\documentclass[sigconf, nonacm]{acmart}
\usepackage[T1]{fontenc}
\usepackage[utf8]{inputenc}
\usepackage{booktabs}
\usepackage{multirow}
\usepackage{balance}
\usepackage{xcolor}
\usepackage{framed}
\usepackage{xspace}
\usepackage{lscape}
\usepackage{color}

\definecolor{shadecolor}{RGB}{235,235,235}

\usepackage{verbatim}
\usepackage{graphicx}
\usepackage{balance}  
\usepackage{caption}
\usepackage{subcaption}
\usepackage{setspace}
\usepackage{float}
\usepackage{tikz}
\usepackage{amsmath}
\usepackage{breqn}
\usepackage{algorithm}
\usepackage[noend]{algpseudocode}
\usepackage{listings}
\definecolor{key-color}{rgb}{0.8, 0.47, 0.196}
\usepackage{relsize}
\usepackage{makecell}

\definecolor{codegreen}{rgb}{0,0.6,0}
\definecolor{codegray}{rgb}{0.5,0.5,0.5}
\definecolor{codepurple}{rgb}{0.58,0,0.82}
\definecolor{backcolour}{rgb}{0.95,0.95,0.92}

\lstdefinestyle{mystyleCpp}{
language=C++,
float=tp,
floatplacement=tbp,
backgroundcolor=\color{backcolour},   
commentstyle=\color{codegreen},
keywordstyle=\color{magenta},
numberstyle=\tiny\color{codegray},
stringstyle=\color{codepurple},
basicstyle=\ttfamily\footnotesize,
breakatwhitespace=false,         
breaklines=true,
captionpos=b,                    
keepspaces=true,                 
numbers=left,                    
numbersep=5pt,                  
showspaces=false,                
showstringspaces=false,
showtabs=false,                  
tabsize=2,
literate={-}{{-}}1,    
morekeywords={MATCH, RETURN, WHERE}
 }

\lstdefinestyle{mystyleSQL}{
language=SQL,
backgroundcolor=\color{backcolour},   
commentstyle=\color{codegreen},
keywordstyle=\color{magenta},
numberstyle=\tiny\color{codegray},
stringstyle=\color{codepurple},
basicstyle=\ttfamily\footnotesize,
breakatwhitespace=false,         
breaklines=true,
captionpos=b,                    
keepspaces=true,                 
numbers=left,                    
numbersep=5pt,                  
showspaces=false,                
showstringspaces=false,
showtabs=false,                  
tabsize=2,
morekeywords={MATCH, RETURN, WHERE, PROJECT, GRAPH, CALL}
 }

\usepackage{enumitem}
\usepackage{hyphenat}

\usepackage{hyperref}
\expandafter\def\expandafter\UrlBreaks\expandafter{\UrlBreaks\do\/\do\-\do\.} 
\definecolor[named]{Purple}{cmyk}{0.55,1,0,0.15}
\definecolor[named]{DarkBlue}{cmyk}{1,0.58,0,0.21}
\hypersetup{bookmarksnumbered,unicode,naturalnames,colorlinks,breaklinks,
	linkcolor=Purple,
	citecolor=Purple,
	urlcolor=DarkBlue,
	filecolor=DarkBlue}

\graphicspath{ {images/} }

\newenvironment{squishedlist}
{
	\begin{list}{$\bullet$}
		{
			\setlength{\itemsep}{0pt}
			\setlength{\parsep}{1pt}
			\setlength{\topsep}{0.5pt}
			\setlength{\partopsep}{0pt}
			\setlength{\leftmargin}{01.0em}
			\setlength{\labelwidth}{1em}
			\setlength{\labelsep}{0.5em}
		}
	}
{
	\end{list}
}

\newcommand{\revision}[1]{{\color{blue} {#1}}}

\newcommand{\Indexname}{NaviX}
\newcommand{\Indexnamecopy}{NaviX-copy}
\newcommand{\Kuzu}{Kuzu}

\newcommand{\bruteforce}{\texttt{bruteforce}}
\newcommand{\onehop}{\texttt{onehop-s}}
\newcommand{\onehopa}{\texttt{onehop-a}}
\newcommand{\directed}{\texttt{directed}}
\newcommand{\blind}{\texttt{blind}}
\newcommand{\adaptiveg}{\texttt{adaptive-g}}
\newcommand{\adaptivel}{\texttt{adaptive-local}}
\newcommand{\sdc}{\texttt{s-dc}}
\newcommand{\tdc}{\texttt{t-dc}}
\newcommand{\CALL}{\texttt{CALL}}

\tikzstyle{io} = [trapezium, trapezium left angle=70, trapezium right angle=110, minimum width=3cm, minimum height=0.7cm, inner sep=1pt, text badly centered, text width=1.2cm, draw=black, fill=blue!10]
\tikzstyle{decision} = [diamond, minimum width=1cm, minimum height=1cm, text width=1.5cm, text badly centered, inner sep=1pt, draw=black, fill=green!10]
\tikzstyle{process} = [rectangle, minimum width=3cm, minimum height=1cm, text width=2.5cm, text centered, draw=black, fill=red!10]
\tikzstyle{arrow}=[draw, -latex]

\newcommand\vldbdoi{XX.XX/XXX.XX}

\newcommand\vldbavailabilityurl{URL_TO_YOUR_ARTIFACTS}
\newcommand\vldbpagestyle{plain} 

\begin{document}
\title{\Indexname: A Native Vector Index Design for Graph DBMSs With Robust Predicate-Agnostic Search Performance}

\newif\ifcv
\cvfalse

\author{Gaurav Sehgal}
\affiliation{%
  \institution{University of Waterloo}
  \streetaddress{200 University Ave W}
}
\email{g3sehgal@uwaterloo.ca}

\author{Semih Salihoğlu}
\orcid{0000-0002-1825-0097}
\affiliation{%
  \institution{University of Waterloo}
  \streetaddress{200 University Ave W}
}
\email{semih.salihoglu@uwaterloo.ca}

\newif\iflong\longtrue     
\longtrue 

\ifcv
\noindent {\textbf{\Huge Cover Letter}}\\

We thank all the reviewers and the meta-reviewer for their constructive feedback. In our revised paper,
we highlight the changes we have made to address 
the review comments in blue font. In the cover letter, we have a section for each reviewer's main comments.
In our responses, we give pointers to the necessary sections to identify the main text changes that address reviewers’ comments. 

Addressing reviewers' comments, which primarily focused on a set of new experiments, required 2.5 additional pages of material to be added to our main text. To make space for these changes,
we had to remove some content from our main text. We decided to scope the main text experiments to cover mainly the baselines that do prefiltering and are predicate-agnostic, which form more direct baselines against Navix. Specifically, we moved our post-filtering experiments against PGVectorScale and VBase to our longer version. We also added new iRangeGraph experiments but also put them in the longer version, as iRangeGraph is not predicate-agnostic. The only exception to this choice is that we kept the new FilteredDiskANN experiments in main text. FilteredDiskANN is also not predicate-agnostic but this experiment is relatively shorter than iRangeGraph. We further edited or removed some previous figures from the background and systems section to make some space.
We further shortened many paragraphs throughout the text. We do not highlight those paragraphs. 
We addressed the presentation issues reviewers have
brought up, e.g., the ones highlighted in the metareview and do not respond to them explicitly in the cover letter. Finally, in our response CL X, refers to Section X of the cover letter.

\section{Meta-Reviewer}

\subsection{Highlight the technical contribution and novelty (Rev2.D1, Rev2.D3)}
\label{meta:contributions}

\begin{snugshade*}
\noindent Rev2.D1: The reviewer is wondering how much the underlying GDBMS contributes to the performance gain of the method. The authors can consider conducting some experiments on a variant of the method where the GDBMS is replaced with the HNSWLib implementation and see how the method would perform compared with baselines and also the proposed method.
\end{snugshade*}

In response to another suggestion by reviwers to compare against ACORN,  we implemented \Indexname\'s \adaptivel\ search procedure in FAISS HNSW. Similar to HNSWLib, FAISS is another in-memory vector library that implements HNSW. We call this version of \Indexname\ as FAISS-\Indexname. We show that
FAISS-NaviX consistently outperforms ACORN, validating that its advanced search heuristics, specifically the \adaptivel\ heuristic, can lead to faster filtered search time in an independent in-memory implementation.
Please see our detailed response to \Indexname's performance on top of FAISS in CL~\ref{meta:acorn}. 
We note that as we explain in the introductory section of our paper, leveraging the underlying GDBMS is desirable primarily for the practicality of the design for the system developers and not  necessarily performance. For example, we can directly leverage
several existing capabilities of the GDBMS instead of re-implementing those  capabilities. 

\begin{snugshade*}
\noindent Rev2.D3: NaviX integrates a variety of techniques, including those from prior work and those newly proposed in this paper. It would be helpful if the authors explicitly highlighted which components are original contributions and which are existing ones to give others proper credits.
\end{snugshade*}
In our revised paper, we added a paragraph at the end of Section~\ref{sec:introduction}, where we summarized our contributions. Briefly, our first main contribution is the design and implementation of Navix, which is a native vector index design that leverages some of the core capabilities of an underlying GDBMS, along with a DBMS-specific optimization to perform distance computations efficiently. Our second contribution is the \directed\ fixed heuristic, which outperforms prior heuristics in medium selectivity levels, and our \adaptiveg\ and \adaptivel\ heuristics, which dynamically make decisions during search to improve the search performance. Finally, our \adaptivel\ algorithm works directly on top of HNSW, making it easier to integrate into most existing systems using the HNSW index.


\subsection{Expand the technical description (Rev3.D1, Rev3.D3)}
\label{meta:clarifications}
\begin{snugshade*}
\noindent Rev3.D1: To make it easy to understand and compare the search strategies, please provide a table to summarize how they work and their pros and cons.
\end{snugshade*}
In our revised paper, we added this table (Table~\ref{tab:heuristics_pro_table}) in the beginning of Section~\ref{sec:hybrid_search}.

\begin{snugshade*}
\noindent Rev3.D3: For adaptive-global, the filtering strategy is decided when traversing each node. However, the paper says that pre-filtering is conducted for all vectors before vector search. This seems a overkill for adaptive-global, and I wonder if NaviX actually conducts filtering when visiting each node.
\end{snugshade*}
We may be misunderstanding this comment correctly but we want to
clarify that every single heuristic in our design space performs the
selection sub-query, which is where filtering happens,
once only. This is done prior to the start of the vector search
to identify the nodes that are in the selected subset $S$ of vectors. This computation is also done in every other prefiltering-based approach in literaute. For example, Weaviate and ACORN also run the selection sub-query before the vector search routine starts only once.
In our implementation. once $S$ is computed, $S$ is stored in a node semimask,
and passed to the vector search routine. Our filtered search heuristics then use $S$
to correctly evaluate the filtered kNN search query (i.e., to only find kNN of a query vector $v_Q$ among the nodes that are in $S$).
Prior to starting the vector search, \adaptiveg\ further checks the size of $S$ to compute what fraction of total number of nodes are in $S$, and based on this global selectivity statistic, decides which heuristic to pick. \Indexname\, i.e., \adaptivel, 
does not compute this global statistic. Instead, for each node $c_{min}$ it explores, it uses $S$ to check what fraction of $c_{min}$'s neighbors are in $S$ to decide what heuristic to
pick for $c_{min}$. That is, it computes a local selectivity
statistic per each node it explores. However, the local selectivity computation does not involve any filtering operations. It is merely checking the bits of $c_{min}$'s neighbors in $S$. We have clarified this more in Sections~\ref{subsec:sys-index-search} and~\ref{subsec:adaptive-algorithms}.


\subsection{Enhance the experimental section (Disk)}
\label{meta:disk}
\begin{snugshade*}
\noindent Rev1.W1: I expected to see more analysis about the disk behavior of NaviX as it is positioned as a disk-resident index, but the experiments only show in-memory performance as they are conducted under warm cache. Please give more details about the disk-based performance of NaviX, e.g. I/O cost of the techniques (blind, directed, onehop, adaptive), as well as comparison with Filtered-DiskANN which offers different storage and search strategies. It may also be interesting to compare with PASE, another disk-resident HNSW index with unique page design, on non-predicated queries or possibly extending it with the proposed adaptive search algorithm for a fair comparison of on-disk behavior.

\noindent Rev1.W3: Why do you not compare with DiskANN?

\noindent Rev3.D4: The experiment machine has 180GB memory while the largest dataset takes about 60GB. With the cache and excluding the warm-up queries, the experiments actually measure the performance for in-memory search rather than disk-based search. Please report some experiments for disk-based search by using smaller memory (e.g., 15\% of the dataset size) and compare with DiskANN.
\end{snugshade*}
We thank the reviewers for these comments.  
We added a new Section~\ref{subsec:disk}, where we evaluate the disk performance of \Indexname\ against DiskANN and FilteredDiskANN. 
DiskANN is a separate disk-based design that differs from \Indexname\ in many aspects, which makes it difficult to have controlled experiments. However, perhaps most importantly, DiskANN, stores 
all of the vectors in quantized (i.e., compressed) manner 
in memory and does not perform
any disk I/O when accessing the vectors. Our new experiments
in Section~\ref{subsec:disk} aim to
progressively demonstrate the performance of \Indexname, from
a completely cold run to a completely warm run in four configurations:
\begin{squishedlist}
\item NaviX-cold: We start by demonstrating the performance of \Indexname\ 
in cold runs, where every access to vectors and adjacency lists result in an I/O. 
\item NaviX-cold-quant: Then we get to a configuration that is closer to DiskANN, where vectors are quantized and cached but
the adjacency list accesses are still completely cold. 
\item NaviX-warm-quant-1GB: Then we give very little, 1GB of buffer manager space and a warm run that caches the adjacency lists partially. 
\item NaviX-warm-quant-10GB: Finally, we give 10GB of buffer manager to NaviX-cold-quant and do a warm run.
\end{squishedlist}
Our main observations is that in a completely cold run, I/Os of the vectors contributes the most to the performance. Then, 
our second configuration closes the gap with DiskANN significantly but is still slower. This is expected as DiskANN does direct I/O and Kuzu performs two I/Os to access adjacency lists (one to read the size and location of the adjacency list and the second to read the actual list). Then starting with partial caching of adjacency lists, \Indexname\ starts outperforming DiskANN. 
We limited these experiments to our most performant heuristic NaviX (i.e., \adaptivel), since the more effective the search heuristic the less candidate explorations it does, which directly translates to less buffer manager and I/O operations. As a reference, Figure~\ref{fig:navix_cold_pins} shows the average number of buffer manager pin operations (or random IOs) done by different heuristics on our Wiki uncorrelated workload for \Indexname-cold configuration.

\begin{figure}[H]
\centering
  \includegraphics[keepaspectratio, height=4.5cm]{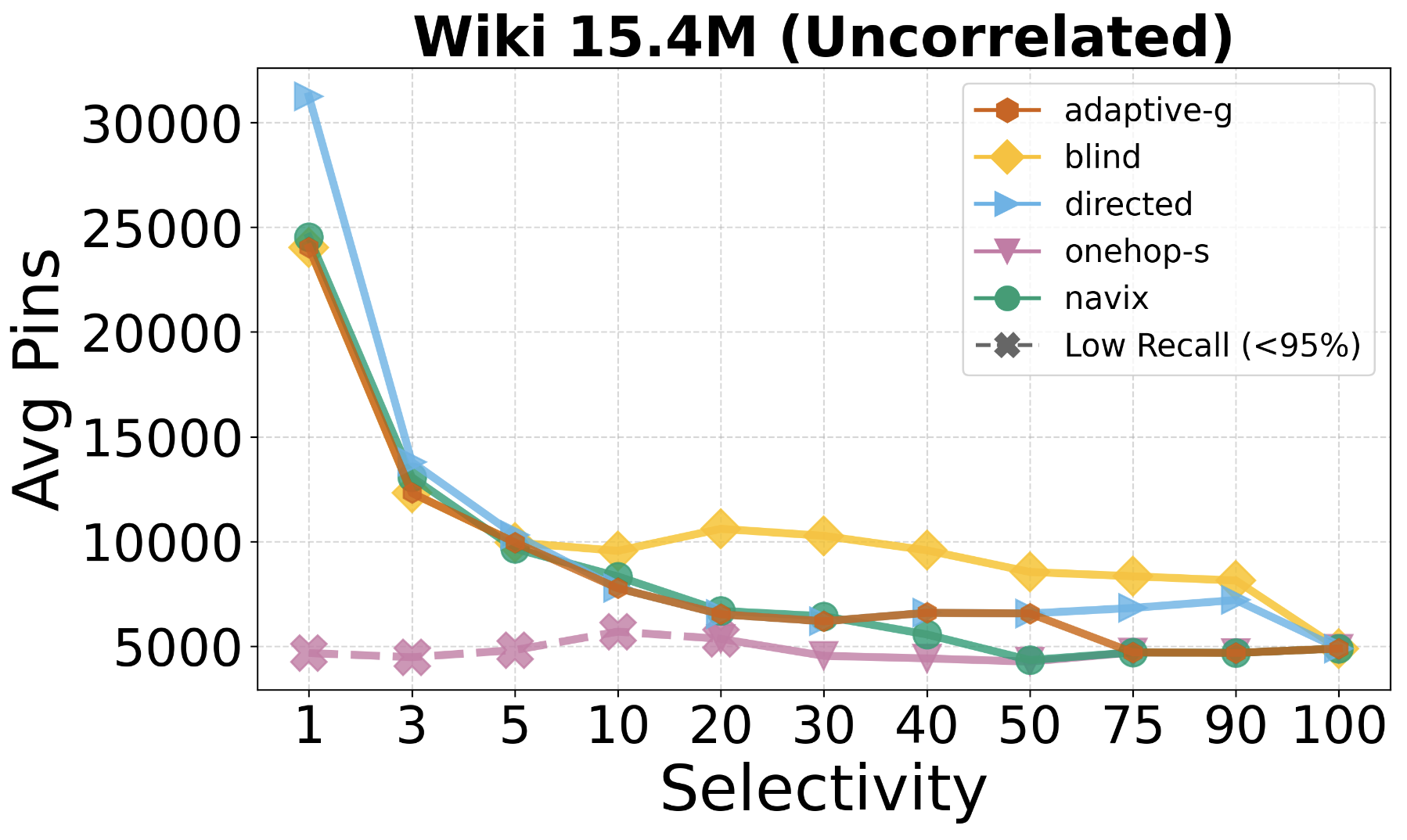}
  \caption{Avg Pins (or random IOs) for different Heuristics in \Indexname-cold config at 95\% recall}
  \label{fig:navix_cold_pins}
\vspace{-10pt}
\end{figure}

We also repeated similar experiments with FilteredDiskANN.
Compared to our DiskANN experiments, in FilteredDiskANN, performance is affected not only due to the differences in disk accesses but also how efficient the filtered search is.
Here, we found \Indexname\ outperforms FilteredDiskANN even 
with cold runs in the \Indexname\ configuration that quantizes and caches vectors. 
Given that ACORN~\cite{acorn} has already shown that FilteredDiskANN's filtered search algorithm is not competitive with ACORN and that \Indexname\ has an even more efficient filtered search algorithm than ACORN, we attribute this to \Indexname\ having a more efficient filtered search algorithm that FilteredDiskANN.

\subsection{Enhance the experimental section (ACORN)}
\label{meta:acorn}

\begin{snugshade*}
\noindent It would also be useful to add a direct comparison between the modified ACORN used by NaviX and the baseline ACORN, clearly reporting latency and recall differences.

\noindent Rev1.W2: Please clarify why ACORN is not included in the main results and instead in the appendix?
\end{snugshade*}

In the longer version of our paper, we had comparisons against ACORN implemented on top of Kuzu. 
In our revised paper, we instead 
implemented \Indexname\ in FAISS HNSW, which is an in-memory vector index library.
We call this version of \Indexname\ FAISS-\Indexname. We added a direct comparison of FAISS-\Indexname\ against ACORN-1 and ACORN-10 in a new Section~\ref{subsec:acorn-ex}. 
First, this allowed us to use the original ACORN implementation from reference~\cite{acorn}. Second, this allowed us to address the comment 
from CL\ref{meta:contributions} to implement \Indexname\ (i.e., \adaptivel\ heuristic) on an in-memory index implementation.
We observe that FAISS-\Indexname\ outperforms both ACORN-1 and ACORN-10
due to its better search heuristics. In addition to these,
we make several other interesting observations here. For example,
we observe that FAISS-\Indexname\ outperforms ACORN with larger factors
in correlated cases. Further, we observe that ACORN-1 is unable to achieve a good recall rate at our default $M$=$32$ configuration, which indicates the importance of the pruning step (ACORN-1 skips the pruning step during vector index building). 

\subsection{Enhance the experimental section (Ablation Study)}
\begin{snugshade*}
\noindent Add an ablation study to better understand the contribution of the different NaviX components, e.g. the graph component returning the bit set of valid points vs. the vector search component built on top of ACORN. In particular, it would be useful to report how substituting Kuzu with another backend affects predicate evaluation latency. 
\end{snugshade*}
In our revised paper, we provide detailed performance breakdowns in several places. First, at the end of our ACORN experiments in Section~\ref{subsec:acorn-ex}, we note that readers can get a sense of the performance difference between implementing 
NaviX inside a DBMS vs a completely in-memory HNSW implementation. This is the point in the paper,
where we have presented NaviX's performance on our datasets both in Kuzu and FAISS HNSW. Second, in a new Section~\ref{subsubsec:prefiltering-vs-vector-search}, we provide the breakdown of NaviX's end-to-end time into the prefiltering time and vector search time. Finally, in our new DiskANN experiments in Section~\ref{subsec:disk}, we also provide performance breakdown of
Kuzu-NaviX due to the amount of I/O it performs (cold vs warm runs).  

\subsection{Enhance the experimental section (Other)}
\label{meta:recall}
\begin{snugshade*}
\noindent Rev2.D2: This paper reports results solely at 95\% recall in its experimental evaluation. Providing additional comparisons at varying recall levels (e.g., 90\%, 99\%) would strengthen the claim of NaviX’s superiority over existing baselines.
\end{snugshade*}
In our revised paper, we provide experiments at different recalls in two places. Figure~\ref{fig:faiss_vs_diff_recall} presents the benchmarks for Faiss-\Indexname\ and ACORN for 90\% and 97\% recall and Figure~\ref{fig:prefilter_diff_recall} presents the benchmarks for Kuzu-\Indexname\ and pre-filtering baselines for 90\% and 97\% recall.

\begin{snugshade*}
\noindent Rev3.D6: Please report the sampling ratio and the size of the high-level index.
\end{snugshade*}

\noindent We added a Section~\ref{subsub:index_size_sampling} where we report the sampling ratio (5\%). We also report the higher level index size for our biggest dataset Wiki (only $\sim$200MB).

\subsection{Cite relevant related works}
\label{meta:rw}
\begin{snugshade*}
\noindent Rev2.D4: The Introduction contains the statement, “This has been proposed in some prior works [] and often 2nd degree neighbors is enough,” but lacks the necessary references.
This was the ACORN reference. We added this reference. We also qualified the latter part of this sentence. We meant that both in the ACORN paper and in our own experiments in this paper, we find 2nd degree to be enough to achieve our high recall targets.

\noindent Rev3.D2: Figure 9 is reproduced from ACORN, please provide a citation.
\end{snugshade*}
We thank the reviewers for pointing out these mistakes. We have fixed these issues.
\section{Reviewer 1}

\subsection{W1 and W3: Disk based evaluation}

\begin{snugshade*}
\noindent W1: I expected to see more analysis about the disk behavior of NaviX as it is positioned as a disk-resident index, but the experiments only show in-memory performance as they are conducted under warm cache. Please give more details about the disk-based performance of NaviX, e.g. I/O cost of the techniques (blind, directed, onehop, adaptive), as well as comparison with Filtered-DiskANN which offers different storage and search strategies. It may also be interesting to compare with PASE, another disk-resident HNSW index with unique page design, on non-predicated queries or possibly extending it with the proposed adaptive search algorithm for a fair comparison of on-disk behavior.

\noindent W3: Why do you not compare with DiskANN?
\end{snugshade*}

For this point, please see our response in CL~\ref{meta:disk}.

\subsection{W2: Acorn}

\begin{snugshade*}
\noindent W2: Please clarify why ACORN is not included in the main results and instead in the appendix?
\end{snugshade*}

For this point, please see our response in CL~\ref{meta:acorn}
\section{Reviewer 2}

\subsection{W1 and D3: specify newly made contributions more explicitly}

\begin{snugshade*}
\noindent W1: The authors should specify the newly made contributions more explicitly (See D3).

\noindent D3: NaviX integrates a variety of techniques, including those from prior work and those newly proposed in this paper. It would be helpful if the authors explicitly highlighted which components are original contributions and which are existing ones to give others proper credits.
\end{snugshade*}

For this point, please see our response in CL~\ref{meta:contributions}.

\subsection{D1: Improvement in experiments}

\begin{snugshade*}
\noindent D1: The reviewer is wondering how much the underlying GDBMS contributes to the performance gain of the method. The authors can consider conducting some experiments on a variant of the method where the GDBMS is replaced with the HNSWLib implementation and see how the method would perform compared with baselines and also the proposed method.
\end{snugshade*}

For this point, please also see our response in CL~\ref{meta:contributions}.

\subsection{D2: Different Recall Benchmarks}

\begin{snugshade*}
\noindent D2: This paper reports results solely at 95\% recall in its experimental evaluation. Providing additional comparisons at varying recall levels (e.g., 90\%, 99\%) would strengthen the claim of NaviX’s superiority over existing baselines.
\end{snugshade*}

For this point, please see our response in CL~\ref{meta:recall}.

\subsection{D4: Writing Comment}

\begin{snugshade*}
\noindent D4. The Introduction contains the statement, “This has been proposed in some prior works [] and often 2nd degree neighbors is enough,” but lacks the necessary references.
\end{snugshade*}

For this point, please see our response in CL~\ref{meta:rw}.

\section{Reviewer 3}

\subsection{W1, R1: Comparison with other STOA baselines for attribute filtering}

\begin{snugshade*}
\noindent W1: The two key contributions of NaviX, i.e., integrating with a graph DBMS and adaptive vector search, are isolated. I think adaptive vector search can be an independent algorithm contribution that works for both memory-based and disk-based scenarios. To establish this point, adaptive search needs be compared with ACORN, Serf, and iRangeGraph. It is OK even if adaptive search may not match Serf and iRangeGraph for range filtering on a single numerical attribute, which they mainly target. However, the trade-offs need to be illustrated.

\noindent R1: Compare adaptive search STOA baselines for attribute filtering. Note that adaptive search is NOT required to outperform these baselines.
\end{snugshade*}
For our new ACORN experiments, please see our response in CL~\ref{meta:acorn}. We have also compared the performance of FAISS-\Indexname, which is the name we gave to our implementation of \Indexname\ on FAISS, with iRangeGraph. We used FAISS-NaviX instead of Kuzu-NaviX because iRangeGraph is also an in-memory implementation. iRangeGraph supports range queries on a single property of the vectors. We used our our uncorrelated workloads
because they have range filters in their selection sub-queries.
We found that for most low selectivity levels iRangeGraph is more performant than FAISS-NaviX, which is expected as iRangeGraph constructs a modified and optimized index for different ranges. For some higher selectivity levels, we found FAISS-NaviX to be more performant. As we discussed in the beginning of this cover letter, addressing reviewers' comments required 2.5 additional pages of material, so we decided to limit the experiments in our main paper to cover the baselines that support arbitrary predicate-agnostic search and prefiltering. That is why we present these experiments in the longer version of our paper~\cite{longerpapernavix}. 

\subsection{W2, D4, R2: Disk-performance and GDBMS integration}

\begin{snugshade*}
\noindent W2: The design space of integrating vector search with graph DBMS is not well discussed. NaviX stores the vectors on disk, and when checking each candidate v, proximity graph search needs to visit all v’s neighbors, which can incur many small random disk accesses for the vectors. In the meantime, NaviX stores in the memory a high-level graph. What is the size of the high-level graph w.r.t. the entire dataset? DiskANN[a] shows that using a memory that is 10\%-20\% of dataset size to store the compressed vectors can avoid most of the disk accesses for vectors during graph traversal. If NaviX already uses considerable memory for the high-level graph, it is unclear whether it is better to use the memory to store the compressed vectors.

\noindent D4: The experiment machine has 180GB memory while the largest dataset takes about 60GB. With the cache and excluding the warm-up queries, the experiments actually measure the performance for in-memory search rather than disk-based search. Please report some experiments for disk-based search by using smaller memory (e.g., 15\% of the dataset size) and compare with DiskANN.

\noindent R2: Discuss the data layout decisions of integrating vector search in a disk-based system and compare with DiskANN when using limited memory for vector search.
\end{snugshade*}

In our revised manuscript, we report the size of the lower and higher levels of NaviX in Section~\ref{subsub:index_size_sampling}. As we report, the storage costs of the high-level index is minimal. The main storage costs are the vectors. This is also shown our new experiments comparing NaviX with DiskANN. For details of these experiments, please see our response in CL~\ref{meta:disk}. 
As hinted by the reviewer, we show that the main contributor to a pure cold/disk-based performance of NaviX is indeed the scanning of vectors. As we show, if we cache the vectors as in DiskANN, the performance difference between a pure cold run of NaviX and DiskANN closes significantly. We also show that with further
partial caching of the adjacency lists, NaviX starts outperforming DiskANN. Then, with full caching, NaviX opens the performance gap with DiskANN.

In our work, we did not change Kuzu's storage structures, and instead leveraged them as they are. Leveraging Kuzu's structures and capabilities has some performance advantages as well as disadvantages. For example, Kuzu's CSR indices are automatically compressed
 and the system has a buffer manager that caches adjacency lists and vectors automatically. At the same time Kuzu does not do direct I/O and requires two I/Os to access any adjacency list (one to read some metadata, which is often cached, and the other to read the actual lists). We agree that more specialized designs can layout the graph and vectors in more optimized way than Kuzu's default layouts. We make this observation in Section~\ref{subsec:sys-index-creation}.

\subsection{D3: Clarification around adaptive-global}

\begin{snugshade*}
\noindent D3: For adaptive-global, the filtering strategy is decided when traversing each node. However, the paper says that pre-filtering is conducted for all vectors before vector search. This seems a overkill for adaptive-global, and I wonder if NaviX actually conducts filtering when visiting each node.
\end{snugshade*}

For this point, please see our response in CL~\ref{meta:clarifications}.

\subsection{D6: Please report the sampling ratio and the size of the high-level index.}

For this point, please see our response in CL~\ref{meta:recall}.

\subsection{W1: I would suggest introduce Adaptive Search as Section 3, and put the current Section 3.1 as Section 4 for implementation details.}

Thank you for this advice, which we agree is a more natural organization of our paper. We have split Section 3 as suggested. In Section~\ref{sec:hybrid_search} we first cover our search heuristics. Then in  Section~\ref{sec:system_implementation}, we cover the implementation of Navix in Kuzu.

\clearpage
\balance

  \setcounter{page}{1}
  \setcounter{section}{0}
  \setcounter{subsection}{0}
  \setcounter{subsubsection}{0}
  \setcounter{figure}{0}
  \setcounter{table}{0}
  \renewcommand{\thesection}{\arabic{section}}
\fi

\begin{abstract}
There is
an increasing demand for extending existing DBMSs
with vector indices to become unified systems that can support modern predictive applications, which 
require joint querying
of vector embeddings and structured  properties and connections of objects.
We present  \Indexname, a {\em {\bf Na}tive {\bf v}ector {\bf i}nde{\bf X}} for graph DBMSs (GDBMSs)
that has two main design goals. First, we aim to implement a 
disk-based  vector index
that leverages the core storage and query processing capabilities of the underlying GDBMS. To this end,
\Indexname\ is a
{\em hierarchical navigable small world} (HNSW) index, which is itself a graph-based structure. 
Second, we aim to evaluate {\em predicate-agnostic} filtered 
vector search queries, where the k nearest neighbors (kNNs) of a query vector $v_Q$
are searched across an arbitrary subset $S$ of vectors that is specified by
an ad-hoc selection sub-query $Q_S$.
We adopt a prefiltering-based approach that evaluates $Q_S$ first
and passes the full information about $S$ to the kNN search operator.
We study how to design a prefiltering-based
search algorithm that is robust under different selectivities as well as correlations of $S$ with $v_Q$.
We propose an adaptive
algorithm that utilizes local selectivity of each vector in the HNSW graph
to pick a suitable heuristic at each iteration 
of the kNN search algorithm. 
We demonstrate \Indexname's robustness and efficiency through extensive experiments against both existing prefiltering- and postfiltering-based
baselines.





\end{abstract}

\maketitle

\pagestyle{\vldbpagestyle}

\ifdefempty{\vldbavailabilityurl}{}{
\begingroup\small\noindent\raggedright\textbf{PVLDB Artifact Availability:}\\
The source code, data, and/or other artifacts have been made available at 
\url{https://github.com/gaurav8297/kuzu}.
\endgroup
}

\section{Introduction}
\label{sec:introduction}

Many modern applications process high-dimensional vector embeddings of objects,
such as images or text. Consider as an example
large language model-based (LLM)  question and answering (Q\&A) systems.
To answer a natural language question $NL_Q$ that requires private knowledge,
these systems need to provide LLMs additional information 
obtained from private documents. A common approach to provide this information is
retrieval augmented generation (RAG)~\cite{rag}. RAG-based systems embed the chunks of 
documents in a high-dimensional vector space. Then, they embed $NL_Q$ into the same space as a vector $v_{Q}$ 
and find chunks whose embeddings are close to $v_{Q}$. These chunks are then given to an LLM to generate an answer.
Other techniques, such as  {\em graph RAG}~\cite{langchainGraphRag,linkedinGraphRag,graphrag}, further connect these chunks with 
structured records, such as a knowledge graph, and  
retrieve chunks based on a mix of these connections and a search in the embedding space. 

A core querying capability these applications require is a vector index~\cite{hnsw,lhs,ivf_pq}, which can find the {\em k-nearest neighbors} (kNN) of
a query vector $v_Q$ in a set of vectors $V$.
In addition to kNN queries, applications also require performing other database querying on their objects, such as filtering them based on other attributes.
Consider an e-commerce recommendation system that recommends products with a price range as well as similarity to another product's image.
Image similarity can be solved by a kNN search on the vector representations of product images, while filtering
on price can be an attribute filter \cite{analyticaldb,multi_meta_rag}. 
Since existing DBMSs already implement advanced querying and storage capabilities, there is immense value 
in implementing a vector index in existing DBMSs. 
This allows users to use a single system 
to build their applications, without the need for an additional specialized vector database.

Implementing a vector index in an existing DBMS also has benefits for system implementers, who can leverage the core capabilities of an existing system  to implement the index, such as the persistent storage structures,  query processor, or the buffer manager. This paper presents {\em \Indexname}, a {\bf Na}tive {\bf v}ector {\bf i}nde{\bf X} for graph DBMSs (GDBMSs)
that has two main design goals. First, we aim to implement a 
disk-based  vector index
that leverages the core storage and query processing capabilities of the underlying GDBMS. Second, we aim to 
efficiently evaluate {\em predicate-agnostic kNN queries}, i.e., kNN queries over arbitrary subsets $S$ of vectors $V$,
that performs robustly under queries that contain both attribute filtering and joins, and whose selected subsets have different
selectivities and correlations to the query vector $v_Q$.
We implemented \Indexname\ in \Kuzu~\cite{kuzu:cidr}, which is a modern columnar GDBMS.

\Indexname\ adopts the {\em hierarchical navigable small worlds} (HNSW) vector index design~\cite{hnsw}. In HNSW, the index itself is a multi-layered graph (henceforth {\em HNSW graph}). 
The lower layer of HNSW contains all vectors $V$ and a set of edges $E_{H}$ that connect close pairs of vectors. The higher layers contain 
progressively fewer nodes that are used to
find good ``entry vectors'' in
the lowest
level that are close to $v_Q$.
With a slight abuse of notation, we refer to the HNSW index
simply as $G_{H}(V, E_{H})$ ignoring the upper layers.
kNNs of $v_Q$ are found by an iterative graph search algorithm on $G_{H}$.
The search starts from an entry vector $v_e$. For each neighbor $w$ of $v_e$, it
computes the distance of $w$ to $v_Q$ and
puts $w$ into a \texttt{candidates} priority queue based on
this distance. At each iteration, the
candidate $c_{min}$ closest to $v_Q$ is extracted from the queue and 
its neighborhood is explored, until the closest $k$
vectors that have been seen so far stop improving.

GDBMSs are uniquely positioned as suitable DBMSs for implementing HNSW-based indices because 
they already contain specialized disk-based graph 
storage structures as well as 
graph-optimized querying capabilities. 
\Indexname\ stores the lower layer of $G_{H}$ 
as a relationship table, which are stored in \Kuzu\ as compressed sparse row-based
graph structures on disk. During kNN search, all accesses  to the lower layer happens through the buffer manager.

To motivate our approach to evaluating
predicate-agnostic kNN queries, consider a graph database
of \texttt{Chunk} nodes, \texttt{Person} nodes, and \texttt{Mention} edges. Suppose \texttt{Chunk} nodes store chunks of text documents and their embeddings in an \texttt{embedding} property. \texttt{Person} nodes have \texttt{name} properties and
there is a \texttt{Mentions} edge from a \texttt{Chunk} node to a \texttt{Person}
node if the text of the chunk mentions a person.
Suppose that a user has created a \Indexname\ index 
called \texttt{ChunkHNSWIndex} on the \texttt{embedding}
properties of \texttt{Chunk} nodes.
Consider 
the following Cypher query:
\begin{lstlisting}[language=SQL,style=mystyleSQL]
MATCH (a:Person)<-[m:Mentions]-(b:Chunk) 
WHERE a.name = "Alice"
PROJECT GRAPH AliceChunks(b); 
CALL QUERY_HNSW_INDEX(AliceChunks, 'ChunkHNSWIndex', k=100,
                      q=[0.1, 0.4, ..., 0.8])
RETURN b, _rank 
\end{lstlisting}

\noindent The query asks for 100 nearest neighbors of $v_Q$ [0.1, 0.4, ..., 0.8] only across
\texttt{Chunks} that mention \texttt{Person} nodes with
name ``Alice''. We refer to the part of the query that selects 
the subset $S$ of vectors among which the kNNs must be found
as the {\em selection subquery}
\iflong
\footnote{Note that the syntax used in the actual open-source Kuzu system~\cite{kuzu-github} has a different syntax to project graphs.}
\fi
.
In this work, we aim to support arbitrary, i.e., predicate-agnostic, selection subqueries.

There are two broad approaches to 
perform predicate-agnostic kNN search.
{\em Prefiltering} approaches~\cite{acorn,milvus,weaviateFilteredSearch,analyticaldb} compute the subset $S$ first and then pass $S$ to the kNN search algorithm, which
finds the kNNs of $v_Q$ only within $S$.
{\em Postfiltering} approaches~\cite{vbase,pase,faiss} continuously find and stream vectors
in the original $V$ from nearest to furthest to $v_Q$, and check if each streamed vector is in $S$, until $k$ such vectors are found.
These approaches offer different tradeoffs about the time they
spend on preprocessing vs on kNN search.
\Indexname\ adopts the prefiltering approach.
The core challenge of prefiltering approaches
is that the original index is constructed on $V$ and not $S$.
Therefore, the search algorithm is performed over nodes that are not in $S$, which can lead
the search in wrong directions and the algorithm may even end up exploring
fewer than $k$ vectors in $S$. 

We next study the problem of {\em how to design a prefiltering-based search algorithm} that is efficient when evaluating 
queries whose selection subqueries have two different properties: (i) different {\em selectivity levels}; and (ii) different {\em correlations} of $S$ with $v_Q$'s close neighborhood in $G_{H}$~\cite{acorn}. In uncorrelated
queries, vectors in $S$ are uniformly selected from
$V$, so the chances of nodes in $S$ being $v_Q$'s nearest neighbors
in $G_{H}$ is
roughly the same as the ``global'' selectivity of $S$ in $V$.
In positively and negatively correlated queries, vectors in $S$ are,
respectively, more and less likely to be in $v_Q$'s nearest neighbors in $G_{H}$. Naive HNSW search algorithm,
which explores
all selected and unselected neighbors of a candidate can 
be very inefficient because it can explore very few selected nodes, or
lead the search away from selected regions in $G_H$.


We propose an adaptive algorithm that is based on
observing the behaviors of a space of 
heuristics that choose which nodes to explore during kNN search.
Our space is based on the observation that prefiltering approaches 
need to make two main decisions during vector search:
\begin{squishedlist}
\item[1.] {\em How much of the neighborhood of $c_{min}$ to explore:}
The two main choices are: 
(i) to explore only the 1st degree neighbors of $c_{min}$ as in the original kNN search algorithm; and (ii) to explore higher degree neighbors, until some predefined number of nodes in $S$ are explored.  This has been proposed in some
prior works~\cite{acorn} and as we will demonstrate in our experiments, in our workloads 2nd degree neighbors are enough.
\item[2.] {\em The order to explore 2nd degree neighbors:} 
We identify two further choices: (i) the {\em blind} order, which is used in prior solutions~\cite{acorn, weaviate}, follows the arbitrary order in which the 1st degree neighbors are scanned; (ii)
the {\em directed} approach, which we propose in this work, orders 1st degree neighbors according to their distance to $v_Q$ and explores 2nd degree neighbors
in this order.
The directed approach has the advantage 
that it can better direct
the search towards regions that are closer to $v_Q$. However, it also pays the upfront cost of computing
the distances to all 1st degree neighbors of $c_{min}$.
\end{squishedlist}
We show that each heuristic has a range of different
selectivity ranges within which they outperform others.
Since in a prefiltering approach, the kNN search
algorithm knows the selectivity of $S$ a priori, one 
can design an {\em adaptive-global} heuristic that picks an appropriate heuristic
after $S$ is computed to get the best of 
all worlds across this space of heuristics.
We further improve the adaptive-global
algorithm by using the (local) selectivity of the neighborhood of each candidate $c_{min}$
to make even more nuanced adaptive decisions. We call this 
the {\em adaptive-local} heuristic and
propose it as a robust 
prefiltering algorithm to evaluate predicate-agnostic
kNN queries. The summary of our main contributions are as follows:
\begin{squishedlist}
    \item We present the design and implementation of a novel native vector index for GDBMSs. Our design leverages the core components of the underlying GDBMS and introduces an {\em in-buffer manager distance computation} optimization that computes distance functions directly on vectors in buffer manager frames.
   \item We present a new \directed\ heuristic for filtered vector search that is efficient at medium to low selectivities, compared to the blind and the default 1st degree heuristics from prior work~\cite{acorn}.
   \item We present two new adaptive heuristics: (i) adaptive-global utilizes the global selectivity of $S$ to pick a fixed heuristic; and (ii) adaptive-local instead utilizes the local selectivity of each candidate vector $c_{min}$ during a single search iteration. While, adaptive-global is able to capture the best of all fixed heuristics, adaptive-local further outperforms any fixed heuristics and adaptive-global, especially when queries correlate with the selected vectors in $S$.
   \item Finally, our adaptive-local algorithm works directly on top of the HNSW index thus making it easier to integrate into most existing systems.
\end{squishedlist}

\noindent Our final design, which we call \Indexname, implements our adaptive-local heuristic. We present extensive experiments evaluating \Indexname\
against several baselines including specialized
vector databases that implement predicate-agnostic queries~\cite{weaviate, milvus},  the ACORN search heuristic for 
predicate-agnostic queries~\cite{acorn}, a disk-based vector index~\cite{diskann}, ``predicate-con\-scious'' vector index designs~\cite{filteredDiskANN, irange}, and
systems that implement the post-filtering approach~\cite{pgvectorscale, vbase}.
Our evaluations demonstrate that \Indexname\ is  efficient and robust under various selectivities
and correlations and that it is possible
build a vector index inside an existing DBMS
whose performance is competitive with specialized systems.

\section{Background}
\label{sec:background}
We first provide background on HNSW indices and the problem of predicate agnostic search.  Then we cover the necessary background on \Kuzu, focusing 
on the components of the system we used in our implementation of \Indexname.

\subsection{HNSW}
\label{subsec:hnsw}
HNSW \cite{hnsw} falls under the class of
spatial indices called {\em approximate proximity graphs}. Similar to other spatial
indices, these indices answer kNN queries for different values of $k$.
Let $V$ be the set of vectors to index and $dist(u, v)$ be a distance function in a vector space. 
In proximity graph indices, the index is a graph $G_H(V, E_H)$. Each vector $v \in V$ is represented as a node
and an edge $(u, v)$ represents closeness of $u$ and $v$ according to the distance function $dist$.
Finding kNNs of a {\em query vector} $v_q$ in proximity graphs involves a search algorithm
in $G$. \footnote{HNSW can be thought of as a relaxation of the {\em sa-tree}~\cite{navarro:sa}
proximity graph index, which is an exact kNN index.
We highly recommend this for readers 
interested in the core ideas and intuitions behind HNSW.}



\noindent {\bf HNSW Construction:} Figure~\ref{fig:hnsw} shows
the overall structure of an HNSW index. HNSW indices have multiple levels.
The lowest level contains all the vectors  and the higher levels progressively
contain fewer vectors.  
The construction algorithm 
is configured with
a parameter $M$ that determines the maximum number of neighbors of each vector,
which for simplicity we assume is the same at each level.
The index is constructed by inserting vectors one at a time.
For each vector $v$, the algorithm first determines the maximum layer ($l$) 
to insert $v$ into.
Then from the top-most level, it first 
finds an entry point $v_{el}$ to level $l$ by finding the closest vector to $v$.
Then within level $l$ it finds the $efConstruction$ ($efC$) NNs $w_1, ..., w_{efC}$ of $v$ using the
search algorithm we describe below (and starting from $v_{el}$). Then, if $efC$ is greater than $M$, it prunes the found neighbors to $M$ using an algorithm from reference~\cite{toussaint:rng}. Then it inserts edges both from $v$ to $w_i$ and vice versa. If adding an edge to $w_i$ increases $w_i$'s neighbors to $b$ $>$ $M$,
$w_i$'s neighbors are pruned using the same algorithm ~\cite{toussaint:rng}. Let  $u_1, ..., u_{b}$
be the neighbors of $w_i$ in increasing
closeness to $w_i$. Each $u_j$ is kept if it is closer to $w_i$
than any of the previously kept $u_{t < j}$.
Then $v$ is inserted into each level below $l$ in the same manner. 
\iflong
Algorithm \ref{lst:hnsw-index-creation} shows the high-level pseudocode.
\fi

\iflong
\begin{figure}[t!]
\centering
  \includegraphics[keepaspectratio, height=3.5cm]{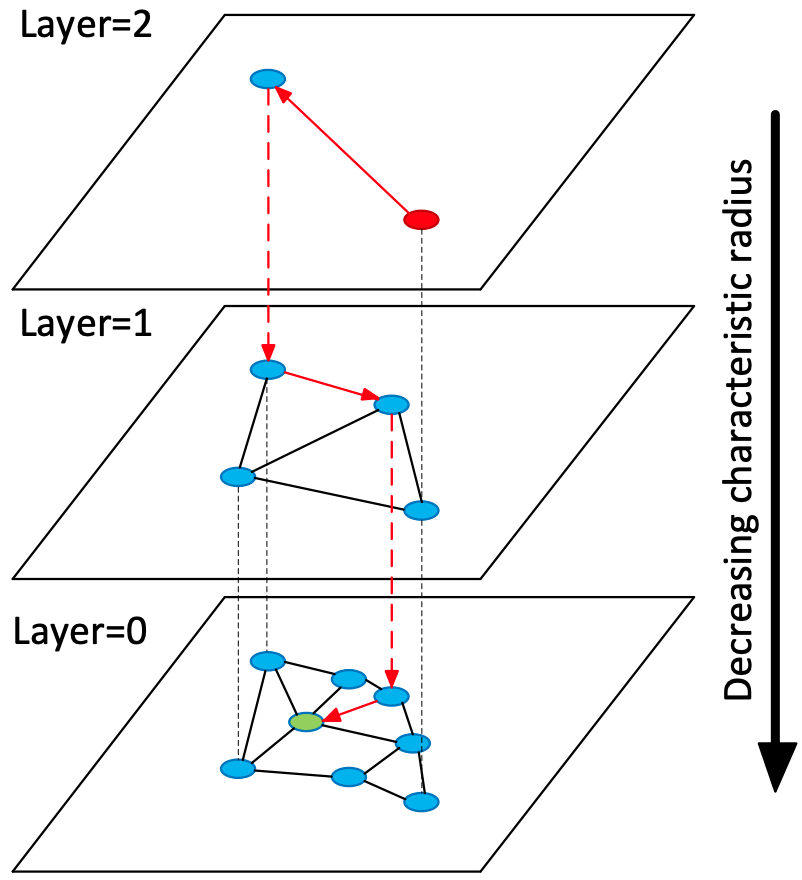}
  \caption{HNSW Index (replicated from reference~\cite{hnsw}).}
    \vspace{-15pt}
  \label{fig:hnsw}
\end{figure}
\fi

\iflong
\begin{lstlisting}[caption={HNSW index creation algorithm},abovecaptionskip=0pt,belowcaptionskip=-10pt,label={lst:hnsw-index-creation},language=C++,style=mystyleCpp,escapechar=^]
Input: dataset V, efConstruction, M
Output: HNSW index graph G
v_e $\leftarrow$ first element of V    
G $\leftarrow$ v_e
maxLevel $\leftarrow$ randomLevel(v_e)               ^\label{line:init-level}^
for each element v in V \ {v_e}:
    L(v) $\leftarrow$ randomLevel(v)                 ^\label{line:assign-level}^
    currEP $\leftarrow$ v_e                       ^\label{line:reset-entry}^
    for ($ly \in maxLevel to L(v)+1$):                 ^\label{line:upper-layer-search}^
         currEP $\leftarrow$ GreedySearch(v, currEP, 1, $ly$)
    for ($ly \in min(maxLevel, L(v)) to 0$):           ^\label{line:layer-insert}^
         candidates $\leftarrow$ SearchLayer(v, currEntry, efConstruction, $ly$)
         selectedNeighbors $\leftarrow$ SelectNeighbors(v, candidates, M)
         // Prune if more than M neighbors are found
         if (|selectedNeighbors| > M):             ^\label{line:rng-check}^
              selectedNeighbors $\leftarrow$ RNGShrink(selectedNeighbors, M) 
         // Add forward edges and shrink
         AddEdgesAndShrink(G, v, selectedNeighbors, $ly$) 
         for each neighbor n in selectedNeighbors:
              // Add backward edges and shrink
              AddEdgesAndShrink(G, n, v, $ly$)
    if L(v) > maxLevel:                              ^\label{line:update-entry}^
         v_e $\leftarrow$ v
         maxLevel $\leftarrow$ L(v)
return G
\end{lstlisting}
\fi

\begin{lstlisting}[caption={HNSW search algorithm in a partilar layer $l$.},abovecaptionskip=0pt,belowcaptionskip=-10pt,label={lst:hnsw-search},language=C++,style=mystyleCpp,escapechar=^]
Input: Query vector $v_q$, $k$, entry point $v_e$, candidate size $efs$
Output: $k$ nearest vectors
min priority queue $C$ $\leftarrow$ {$v_e$} // candidates
max priority queue $R$ $\leftarrow$ {$v_e$} // results 
visited set V $\leftarrow$ {$v_e$}
while(C $\neq$ $\emptyset$):
    $c_{\text{min}}$ $\leftarrow$ Pop-Min($C$) ^\label{line:pop-min}^
    $r_{\text{max}}$ $\leftarrow$ $|R| < efs$ ? Peek-Max($R$) : $\infty$ ^\label{line:peek-max}^
    if ($d(v_q, c_{\text{min}}) > d(v_q, r_{\text{max}})$): break; // convergence criterion  ^\label{line:convergence}^
    for ($n \in neighbours(c_{\text{min}}, l)$): // neighbors in layer l ^\label{line:min-c-explore}^
        if ($n \notin V$):
            $V \gets V \cup \{n\}$
            if ($|R| < efs$ OR $d(v_q, n) < d(v_q, r_{\text{max}})$):
                Insert($C$, $n$)
                Insert($R$, $n$)
                if ($|R| > efs$): Pop-Max($R$)
return closest $k$ vectors in $R$
\end{lstlisting}

\noindent \textbf{HNSW search algorithm} performs a depth-first search-like 
traversal algorithm in each level of the index.
Algorithm~\ref{lst:hnsw-search} shows the pseudocode of 
the search in a particular level. The inputs  
are $v_q$, $k$, an entry vector $v_e$, and an $efs$ value (explained momentarily). The output 
is the approximate $k$ NNs of $v_q$ in a particular level.
The algorithm uses two priority queues: (i) a min priority
queue of {\em candidates}, which represent vectors whose 
neighbors have not been explored; and (ii) a max priority queue of {\em results}, which store the $efs$ closest vectors seen so far. 
At each iteration, the algorithm iteratively explores the
closest candidates' neighbors ($c_{min}$ in Algorithm~\ref{lst:hnsw-search}) on line~\ref{line:min-c-explore}, starting from $v_e$.
For each neighbor $n$, if $n$ has not been visited already, $dist(v_q, n)$ is computed and put into the candidates queue. If $dist(v_q, n)$ improves the closest vectors seen so far, it is also put into 
results. The iterations stop when a $c_{min}$ has a distance larger than the $efs$'th closest vector in results (line~\ref{line:convergence}). The algorithm returns the
$k$ closest vectors in results.

The full algorithm uses this subroutine at each level, starting from the top level
index. At each level $i+1$ except the lowest level, it sets $k=1$ and $efs=1$ to find the closest neighbor $v_{ei}$ of $v_q$, which is used as an entry point for the search at level $i$. At 
the very top level the search starts from a fixed entry point $v_e^*$. At the lowest level, the algorithm finds $k$ NNs using some
$efs$ value greater or equal to $k$. The $efs$ parameter controls the trade-off between search accuracy and latency. The higher the $efs$ value,
the longer it will take for the search to converge, but since a larger
candidate set is considered, the higher will be the recall.

In HNSW search algorithm the dominant cost is the distance
computations, which require expensive floating point operations on vectors.
The distance computations are done when a neighbor $w$ of a $c_{min}$
is explored and put into the candidates queue. When
we will discuss modified versions of this search algorithm, we will focus on minimizing the distance calculations.

\begin{figure}[h]
\centering
  \includegraphics[keepaspectratio, height=3 cm]{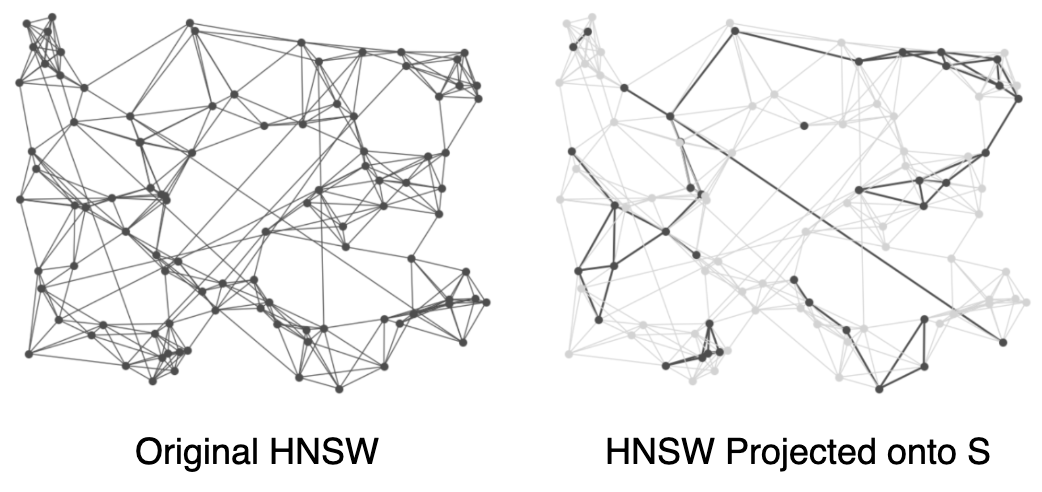}
\vspace{-10pt}
    \caption{Very low selectivity $S$ can disconnect HNSW.}
  \vspace{-15pt}
    \label{fig:disconnected}
\end{figure}

\subsection{Predicate Agnostic Search \& Prefiltering}
\label{subsec:predicate-agnostic-search}
We briefly discuss the challenge of predicate-agnostic vector search,
which is the problem of finding kNNs of $v_Q$ over an arbitrary subset $S$ of the
vectors, that are selected by a selection subquery $Q_S$. 
In this paper we focus on the {\em prefiltering approach}
 to evaluate predicate-agnostic search. In this approach, the system first evaluates $Q_S$
and computes $S$. Then, the entire $S$ is passed to the HNSW search algorithm
which finds the approximate $k$ nearest neighbors
of $v_Q$ in $S$. This is the approach
adopted by some other systems, such as Weaviate~\cite{weaviateacorn} and Acorn~\cite{acorn}.
The core challenge of prefiltering
is that the search is performed on $G_H$, which was constructed by iteratively connecting each vector $v$ with their closest $M$ neighbors in $V$ and not $S$. 
As a result 
$v$'s neighbors in $G_H$ that are actually in $S$ can be very sparse or $v$ can even 
be disconnected from other vectors in $S$. 
Figure~\ref{fig:disconnected} shows this problem pictorially.

\subsection{\Kuzu\ Overview}
\Kuzu~\cite{kuzu:cidr} is a GDBMS that adopts many of the architectural principles of analytical DBMSs, such as adopting disk-based 
columnar storage structures to store the node and relationship records, vectorized query processor~\cite{abadi:ft-columnar},
and morsel-driven multi-core parallelism.
We refer readers to references~\cite{kuzu:cidr, gupta:columnar} 
for a detailed description of \Kuzu's design. Here we provide background on the components 
that are needed to explain the design
and implementation of \Indexname. Other background is provided
in Section~\ref{sec:system_implementation} when describing different components.
 
\subsubsection{Storage Structures}
Node records in \Kuzu\ are stored using a native design that is based on
other columnar designs such as Parquet~\cite{parquet}.
The storage of relationship records are stored in disk-based 
compressed sparse row (CSR) structures. 
This is a highly optimized persistent graph topology design
and an example advantage of leveraging the existing capabilities of GDBMSs to implement
HNSW-based vector indices.

\subsubsection{Node Semimasks}
Query plans in \Kuzu\ are composed into
several subplans $SP_1, ..., SP_t$. Subplans form a directed
acyclic graph and are executed
one after another. Outputs of
one subplan is consumed by the next subplan.
\Kuzu\ frequently uses sideways information passing (SIP) 
by passing {\em node semimasks} 
from one subplan $SP_i$ to another $SP_j$. Node semimasks 
identify a subset of
the nodes whose properties or relationships need to be scanned in $SP_j$.
As we describe in Section~\ref{subsec:sys-index-search}, we use node semimasks to pass the results of selection subquery 
$Q_S$ to the subplan that contains the vector search operator.

\section{Filtered Search Heuristics}
\label{sec:hybrid_search}
We begin by describing our predicate-agnostic filtered search heuristics. The primary advantage of prefiltering is 
that because it pays the upfront cost of computing 
$Q_S$ entirely, it has full information about 
which vectors are in $S$. This information can be utilized 
during kNN search algorithm.
In contrast, post-filtering approaches 
perform the search without any information about what is in $S$.
We next discuss the question of how to use $S$ to design a robust prefiltering algorithm that performs the search
efficiently across different selectivity levels and correlation
scenarios. Throughout this section, we will discuss several existing search heuristics as well as new ones we introduce. For reference, Table~\ref{tab:heuristics_pro_table} summarizes these heuristics.

\begin{table}[t!]
    \centering
    \resizebox{0.45\textwidth}{!}{%
    \begin{tabular}{|c|c|c|}
        \hline
        \textbf{Heuristics} & \textbf{Description} & \textbf{Suitable Selectivity} \\ \hline
        \textbf{\onehop}
          & \makecell{One Hop \\ \& only unfiltered nodes}
          & High \\ \hline
        \textbf{\directed}
          & \makecell{Two Hops up to M \\ in optimal direction}
          & Medium to Low \\ \hline
        \textbf{\blind}
          & \makecell{Two Hops up to M \\ in random direction}
          & Very Low \\ \hline
        \textbf{\texttt{adaptive-global}}
          & \makecell{Use global selectivity \\ to adapt}
          & \makecell{All \\ (Uncorrelated)} \\ \hline
        \textbf{\adaptivel}
          & \makecell{Use local selectivity \\ to adapt}
          & \makecell{All \\ (Uncorrelated \& Correlated)} \\ \hline
    \end{tabular}
    }
    \caption{Summary of search heuristics.}
    \label{tab:heuristics_pro_table}
    \vspace{-15pt}
\end{table}

\begin{figure}[h]
\centering
  \includegraphics[keepaspectratio, height=3.5 cm]{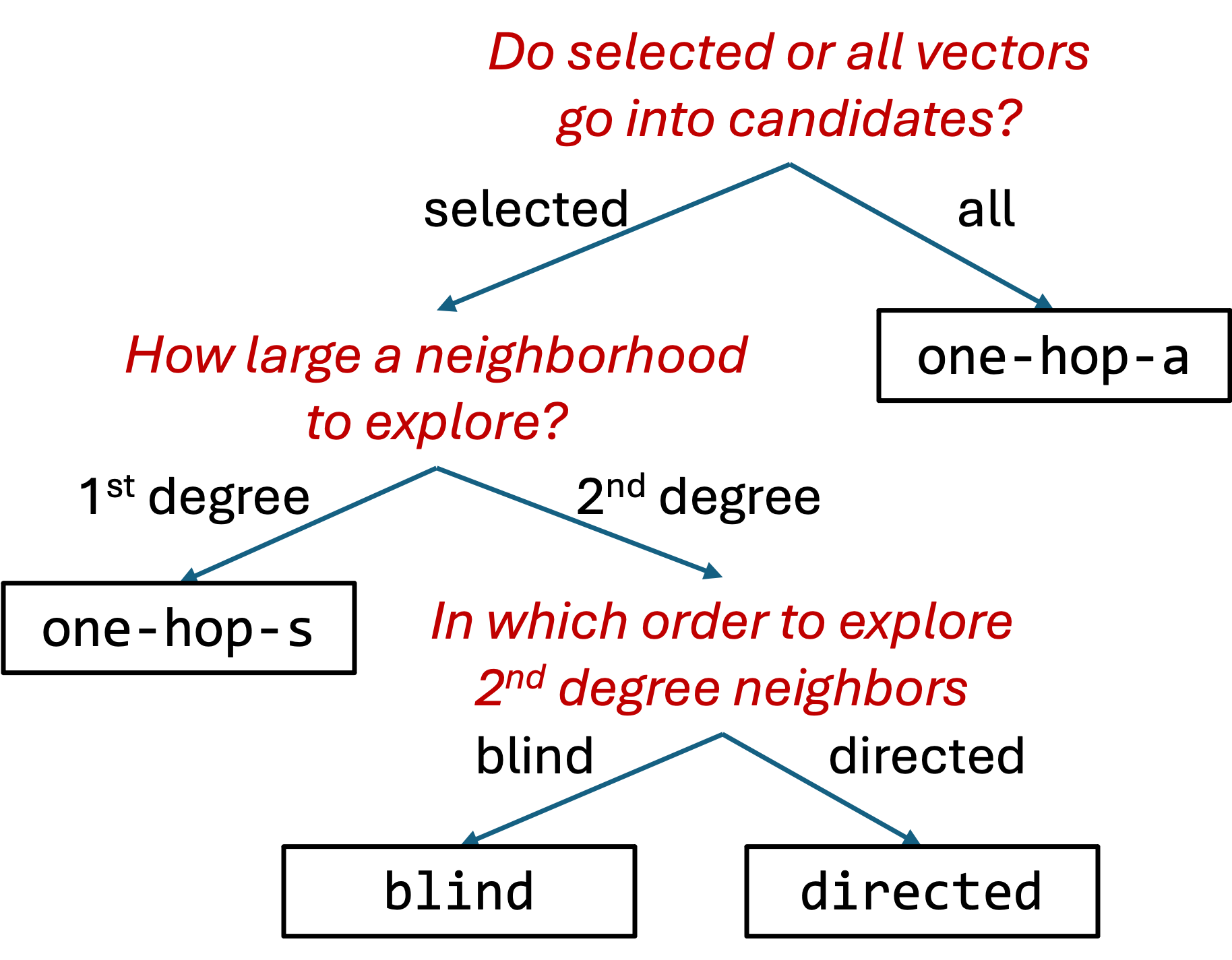}
\vspace{-5pt}
  \caption{Space of heuristics in a modified search algorithm.}
  \vspace{-13pt}
  \label{fig:decision-tree}
\end{figure}

\subsection{Design Space of Fixed Heuristics}
\label{subsec:fixed-design-space}
We begin by outlining a set of fixed heuristic decisions a modified HNSW search 
algorithm can make and discuss the intuitions of the pros and cons of each. 
For reference, Figure~\ref{fig:decision-tree} shows the decision tree that summarizes
the space of heuristics we discuss. We emphasize that
the core step of the original HNSW search algorithm is that
at each iteration, the algorithm explores the {\em closest neighbors of a $c_{min}$ in $V$.} Therefore we are interested in developing 
modified search algorithms that efficiently explore the {\em closest neighbors of a $c_{min}$ in $S$.} Intuitively, heuristics that achieve this {\em primary goal} should
perform the search more efficiently than others. 

The first decision is {\em if all or only selected (or unfiltered) vectors should be put into candidates queue}. The unmodified HNSW algorithm puts all vectors,
which may be inefficient due to exploring unselected vectors. Further,
exploring unselected vectors
can misdirect the search away from regions that contain selected vectors,
slowing down convergence.
We refer to the original
unmodified HNSW as the \onehopa\ heuristic (for one hop
{\bf a}ll vectors).
An algorithm can instead decide to explore only selected vectors. 
We refer to this heuristic as \onehop. 
Intuitively \onehop\ would improve the convergence of \onehopa\ at high selectivity levels 
but creates
a separate problem. Specifically,  at low selectivity levels the
{\em selected projection} of $G_H$, i.e., the subgraph
that contains $S$ and their edges, can be disconnected.  Recall Figure~\ref{fig:disconnected}, which
shows this problem pictorially. 


The second decision  is
{\em how much of each candidate's
neighborhood to explore in the original index $G$}. 
To address the disconnected graph problem at lower selectivity levels, an algorithm can explore higher degree
neighbors of each candidate. Specifically an algorithm
can explore 2nd degree nodes, which we show is adequate
in our evaluations. A version of this heuristic has been explored in ACORN~\cite{acorn}. Given 
a candidate $c_{min}$ and its neighbors $n_1, ..., n_{\ell}$, ACORN's heuristic
takes the first neighbor $n_1$ of $c_{min}$ and explores
the selected vectors among
$n_1$ and $n_1$'s neighbors $n_{11}, ..., n_{1k}$. This is 
then repeated for the second neighbor $n_2$ of $c_{min}$,
until $M$ many selected vectors are explored.
We refer to
this as the \blind\ 2nd degree heuristic (\blind\ for short).
We note that in our evaluations, we implement and evaluate an 
improved version 
of this heuristic than the one used in reference~\cite{acorn}. Specifically,
we first explore all 1st degree neighbors  $n_1, ..., n_{\ell}$
and then start exploring 2nd degree neighbors. This modified version
performs strictly better.

The third decision is {\em the order 
in which the 2nd degree neighbors are explored}.
The \blind\ heuristic is oblivious to how close
each 2nd degree neighborhood it explores is to $v_Q$. In
other words, it does not perform its exploration in a directed manner towards the regions that are closer to $v_Q$.
Alternatively we propose exploring
the neighbors of $c_{min}$ in increasing order of their
distance to $v_Q$. Let $n_1^*, ..., n_{\ell}^*$ be the 1st degree 
neighbors of $c_{min}$ ordered from closest to $v_Q$ to furthest. We refer to the heuristic that explores 
the 2nd degree neighbors of $c_{min}$ in this order
as the \directed\ 2nd degree heuristic (or \directed\ for short). \directed\ has the advantage of prioritizing the search
towards regions that are closer to $v_Q$ but incurs the cost of
computing the distances to each 1st degree neighbor of $c_{min}$.


Within this space, different heuristics have different advantages in different selectivity regions. 
At high selectivity levels,
we expect \onehop, which limits explorations to
selected 1st degree vectors to work well because there is enough selected vectors in $S$ that these heuristics should achieve the primary goal we articulated above.
As selectivity decreases and fewer 1st degree neighbors are selected, \onehop\ should degrade in recall. 
Therefore, heuristics that explore 2nd degree neighbors,
such as \blind\ and \directed\ should
work better. Between these two, \directed\ should converge faster
if we focus on the number of candidates it explores. However,
\directed\ incurs the additional cost of computing the
distance of $v_Q$ to every 1st degree neighbor (selected or unselected) of $c_{min}$. The benefits
of \directed\ can overweigh its overheads 
at medium to slightly low selectivity levels. Eventually however, as
selectivity decreases enough, \directed\ cannot have 
an advantage over \blind. This is because at low-enough selectivity levels,
there is not enough selected nodes even in the 2nd degree neighbors
of $c_{min}$. Therefore both \directed\ and \blind\ effectively explore
the same set of selected vectors, yet \blind\ does not incur the overhead of computing the distances to unselected 1st degree vectors.

\vspace{-5pt}
\subsection{Adaptive Algorithms}
\label{subsec:adaptive-algorithms}
We next describe two adaptive algorithms that utilize the selectivity information in $S$ to pick the best of all worlds across \blind, \directed, and \onehop. 
We first pick an upper bound threshold $ub_{onehop-s}$ above which we pick \onehop. We found 50\% to be a
safe choice here although in some of our evaluations slightly lower thresholds also work. To decide between \blind\ and \directed,
we use the following formula. Recall that under low-enough selectivity,
if there are less than $M$ selected vectors in the 
1st and 2nd degree neighbors of $c_{min}$, \directed\ has no
particular advantage over \blind. So it is guaranteed to be suboptimal.
Therefore, we try to estimate if we are in this safe region. 

\begin{figure}[t!]
\centering
  \includegraphics[keepaspectratio, height=3.5 cm]{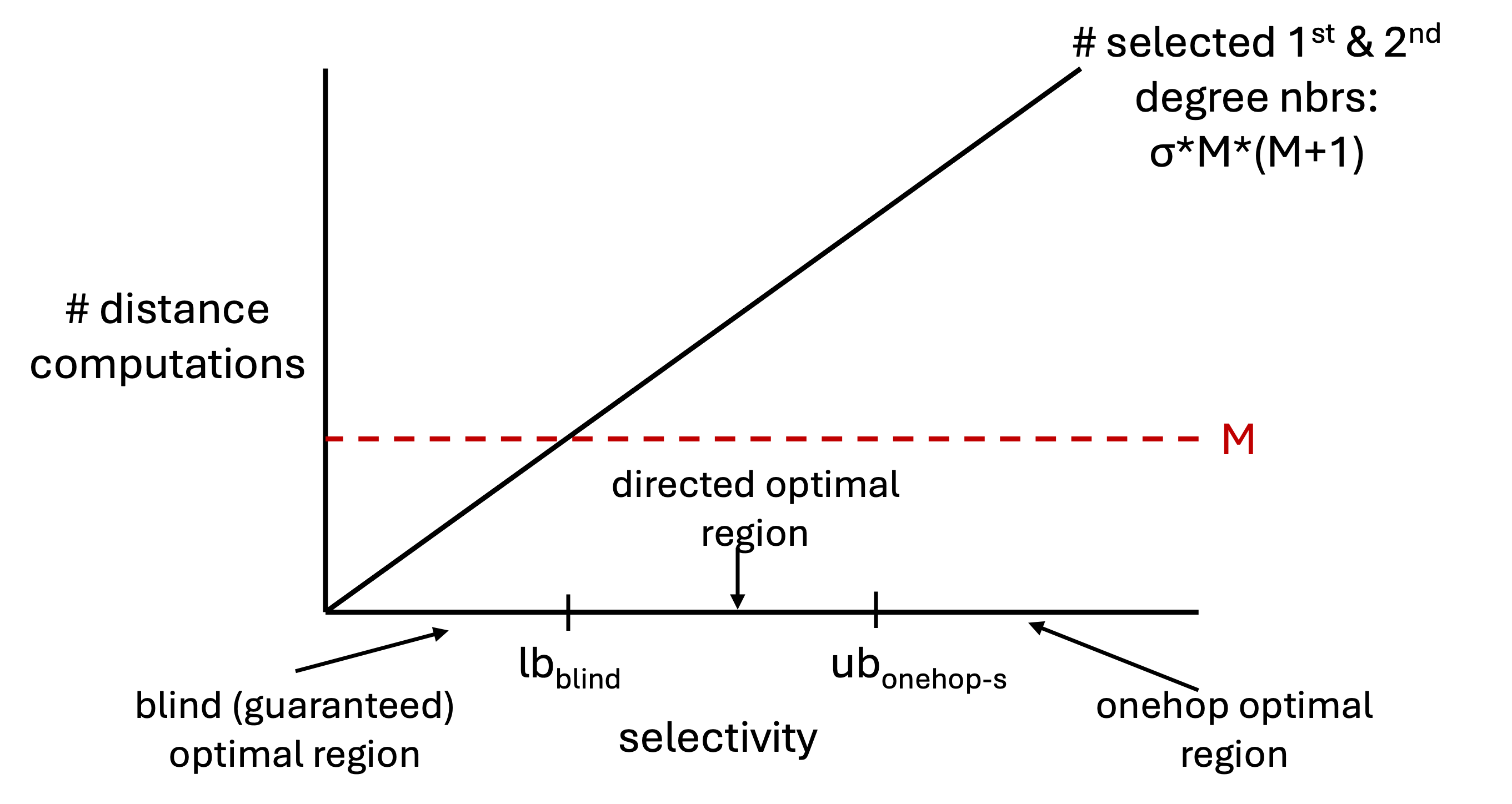}
\vspace{-5pt}
  \caption{Optimal selectivity regions of fixed heuristics.}
  \vspace{-10pt}
  \label{fig:fixed-heuristics-choice}
\end{figure}

Specifically,
we calculate the {\bf e}stimated number of {\bf s}elected {\bf v}ectors ($esv$) in the 1st and 2nd degree neighbors of $c_{min}$ with the formula: $esv=\sigma \times (M+1) \times M$,
Let $\sigma$ be the selectivity of $S$ (more on this later).
The formula multiplies the selectivity with 
the total sizes of the 1st and 2nd degree neighbors of $c_{min}$.
If $esn$ is less than $M$, which is the maximum number 
of selected nodes \blind\ explores, then \directed\ cannot have an advantage over \blind\ by calculating the distances to each 1st degree node. Therefore we default to blind. This safe region is conservative,
so in our implementation we are more lenient and compare $esn$ with $M$*$lf$, where $lf$ (3 by default) is a leniency factor.
Figure~\ref{fig:fixed-heuristics-choice} pictorially shows
the regions where we hypothesize each fixed heuristic to work well.

\begin{figure}[t!]
\centering
  \includegraphics[keepaspectratio, height=3.8 cm]{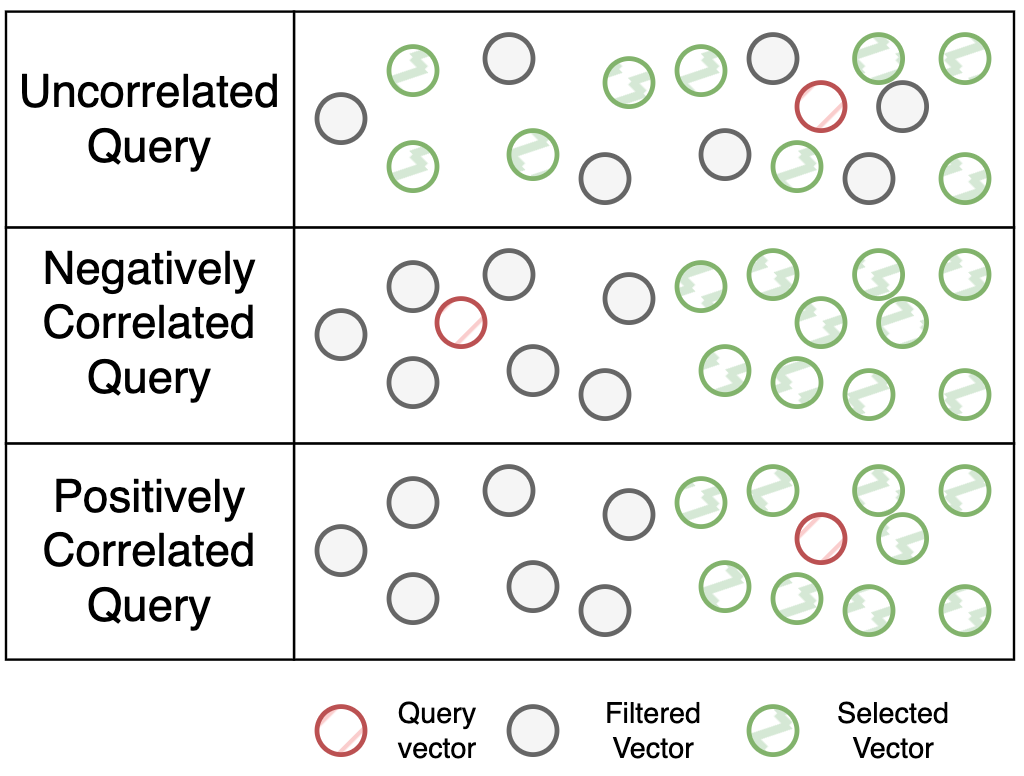}
\vspace{-5pt}
  \caption{Pictorial depiction of different correlations $S$ can have with $v_Q$'s neighborhood. Replicated from reference~\cite{acorn}.}
  \vspace{-10.5pt}
  \label{fig:pictorial-correlation}
\end{figure}
We can configure the above adaptive algorithms in two different ways. First, we can use the global selectivity of $S$ as $\sigma$, $\sigma_g=|S|/|V|$. We refer to this version of the algorithm as \texttt{adaptive-global} (\adaptiveg\ for short). As we demonstrate in our evaluations, this algorithm is able to capture the best of all worlds. However, we can improve this algorithm, and even any
fixed-heuristic in their optimal regions, by choosing a possibly different heuristic for each candidate $c_{min}$. This is useful if $S$ correlates with certain regions in $G$ and is not a random subset of $V$. Figure~\ref{fig:pictorial-correlation} pictorially shows the three different possibilities of $S$ being uncorrelated, positively, or negatively correlated 
with the region around $v_Q$. Especially in correlated scenarios, even though the global
selectivity of $S$ is low enough and we are in the optimal selectivity region for the \blind\ heuristic, if $c_{min}$ is in a region that contains more selected vectors,
it might be better to choose \onehop\ heuristic for $c_{min}$.

We refer to the adaptive algorithm that uses the {\em local selectivity} of $c_{min}$ during each search iteration to pick a heuristic as \adaptivel. 
Specifically, \adaptivel\ uses $\sigma_l=$ \\ $|S(nbrs(c_{min}))|/|nbrs(c_{min})|$, where $nbrs$ $(c_{min})$
is the neighbors of $c_{min}$ and $S(nbrs(c_{min}))$ is the selected 
neighbors of $c_{min}$. $\sigma_l$ calculates the fraction of 
$c_{min}$'s neighbors that are selected. Note that $\sigma_l$ is computed merely by checking if each neighbor of $c_{min}$ in $S$. 
In our implementation, this is done by checking
the bits of these neighbors in a Kuzu node mask (see Section~\ref{sec:system_implementation}).
Importantly, this operation does not require any filtering
or distance computations. Finally, Table~\ref{tab:heuristics_pro_table} presents the summary of these heuristics. 

In this paper, 
we propose \adaptivel\ as a 
predicate-agnostic filtered search algorithm and demonstrate
that it is efficient under both different selectivity levels and
correlation scenarios.

\section{Implementation Details}
\label{sec:system_implementation}
We next describe the design and implementation of \Indexname\ in Kuzu. 

\subsection{Index Creation}
\label{subsec:sys-index-creation}
Index creation is executed as a standalone \CALL\ function
in Cypher. 
Suppose there is a node table with schema `Chunk(id UINT64, docID UINT64, embedding FLOAT[1024])'. Below is an example query
creating an index called \texttt{ChunkHNSW\-Index} on the \texttt{embedding} property of \texttt{Chunk} nodes.

\begin{lstlisting}[language=SQL,style=mystyleSQL]
CALL CREATE_HNSW_INDEX('ChunksHNSWIndex', 'Chunks', 'embedding', M$_U$)
\end{lstlisting}
\Indexname\ is a 2-level HNSW implementation. 
$M_U$ is the maximum degree of vectors in the lower level index.
During construction, two in-memory CSR data structures
$G_L$ (for lower level) and $G_U$ (for upper level) are initiated.
The sizes of these CSRs are 
respectively, $n$ and $n*s$, where $n$ is the number of nodes
in the indexed node table and $s$ is sampling rate 
for $G_U$ (by default 5\%). 
$G_U$ and $G_L$ are initiated with $M_U$
and $M_L=M_U*2$ pre-allocated edges for each node. 
Each worker thread concurrently
scans a morsel of vectors (2048 many) from disk and updates the shared CSR structures
using the HNSW construction algorithm from Section~\ref{sec:background}. Note that
as a thread $T_i$ updates these CSRs,
other threads might be modified by other threads
as well.
However HNSW is already an approximate index and
our evaluations demonstrate that the index quality can tolerate
this possible data race. So, we 
recommend this optimization to obtain better parallelism during construction.
We note that while $G_U$ and $G_L$ are kept in memory during construction, all accesses to the vectors, 
which are much larger than $G_U$ and $G_L$, happen through the buffer manager.
For example, on our Wiki dataset, vectors take 63GB while the lower level of the index only takes 7.8GB.
Once $G_L$ is contructed, we pass it
to Kuzu's CSR construction pipeline.
$G_U$ could also be persisted using \Kuzu's existing disk-based structure.
However,
because it is intended to be very small,
we persist it using a simpler format that writes the offsets and edges
consecutively on disk.


Observe that as per our first design goal, we leverage many of the existing capabilities of the underlying GDBMS, including
its storage structures, which are automatically compressed on disk
by \Kuzu, its automatic query parallelization mechanism, and many
of its operators, such as scans, Node Masker (next subsection), and CSR Construct.
We did not modify any of these core components.
This is an important advantage that simplifies the engineering efforts
of system developers.
From a performance point of view, leveraging Kuzu's existing capabilities has both performance advantages and disadvantages. For example, Kuzu's CSR indices are automatically bit compressed and its semimask creation pipeline is well optimized. Furthermore, by storing vectors and the index (i.e., the adjacency lists) separately, a buffer manager can pack 
multiple adjacency lists in a single 4KB page.
This constrasts with other designs,
such as DiskANN~\cite{diskann}, which packs
the actual vector and the adjacency list of
a vector in a page and caches 
compressed versions of the vectors in memory. Our design helps in caching more adjacency lists with less amount of memory through the buffer manager.
However, at the same time all adjacency list accesses in Kuzu perform two buffer manager (and possibly I/O if not cached) operations: one to read some metadata and the other to read the actual list. Instead, more
specialized designs can do other optimizations that are not implemented in the underlying GDBMS. 
For example, DiskANN performs a single IO to access both the actual vector and its adjacency list. 




\subsection{Index Search Query Syntax and Plan}
\label{subsec:sys-index-search}
\begin{figure}[t!]
\centering
  \includegraphics[keepaspectratio, height=3 cm]{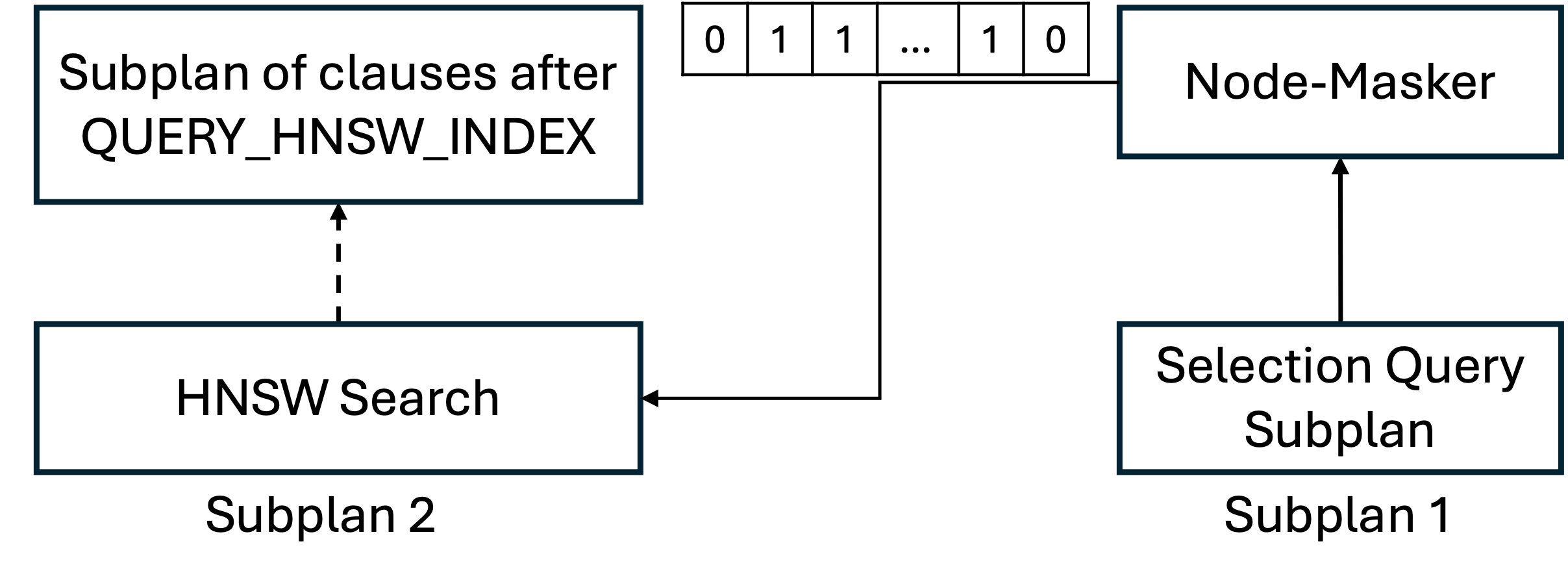}
\vspace{-5pt}
  \caption{\Indexname\ search plan.}
  \vspace{-10pt}
  \label{fig:index-search-plan}
\end{figure}

To describe the index search query syntax, we replicate our query
from Section~\ref{sec:introduction}:
\begin{lstlisting}[language=SQL,style=mystyleSQL]
MATCH (a:Person)-[:Mentions]->(b:Chunk) 
WHERE a.name = "Alice"
PROJECT GRAPH AliceNodes (b); 
CALL QUERY_HNSW_INDEX(AliceGraph, 'ChunkHNSWIndex', k=100,
                      q=[0.1, 0.4, ..., 0.8])
RETURN q, b, _rank 
\end{lstlisting}
Here, $Q_S$ is the subquery specified by the MATCH and WHERE 
clauses and identify the \texttt{Chunk} nodes that are mentioned
by a \texttt{Person} with name Alice. 
The PROJECT GRAPH 
clause is used to identify the selected vectors $S$ that will be passed to QUERY\_HNSW\_INDEX function. 
 Figure~\ref{fig:index-search-plan} shows the plan structure for 
 evaluating predicate-agnostic search queries. The first subplan
 evaluates $Q_S$ and uses \Kuzu's \texttt{Node-Masker} operator
 to create a node semimask. This is then passed to the second 
 subplan that starts with an \texttt{HNSW Search} operator that takes
 in the semimask and runs a modified HNSW algorithm. We note that during \texttt{HNSW Search} computation, no filtering is done. All filtering is performed apriori in the subplan that evaluates $Q_S$.
 All accesses to the persisted adjacency lists of vectors ($G_L$) in the index and to the actual vectors happen through the buffer manager of the system. The upper layer $G_U$, however, remains in memory at all times since it is extremely small compared to the complete index (including vectors). For instance, in our Wiki dataset, $G_U$ occupies only $\sim$200MB while the full index including vectors requires $\sim$70.8GB.

\subsubsection{In-Buffer Manager Distance Computations}
\label{subsec:zero-copy-distance-computation}
We end this section with an optimization that may be of
interest to system developers.
As we discussed above, during both index construction and 
search, vectors are scanned through the buffer manager to improve the scalability
of \Indexname. 
In the standard approach in DBMSs,
each piece of data is first copied from the buffer manager's frames to an operator-local
buffer. Then, the operator operates on this data. We observed that the
copy operations are a performance bottleneck.
To address this, we extended the storage manager
interfaces of \Kuzu\
to directly run a function on the data residing in buffer manager frames
without performing any copies. Specifically, we pass
a distance function to the buffer manager. 
The buffer manager finds and pins a frame, if necessary, scans the vector from disk
to the frame, runs the function, and unpins the frame. 
\iflong
In Appendix~\ref{app:in_bm_distance} 
\else
In the longer version of our paper~\cite{longerpapernavix}
\fi
we show that this optimization can improve vector search latencies by up to 1.6x. Finally,
we used the SimSIMD \cite{simsimd} library to perform distance computations using SIMD instructions.



\section{Evaluation}
\label{sec:evaluation}

We next evaluate the performance of \Indexname, which
refers to our implementation that uses the \adaptivel\ search algorithm. 

\vspace{-5pt}
\subsection{Experimental Setup}
\subsubsection{Baselines} We used the following baseline systems:

\noindent {\bf \Kuzu\ configurations:} We implemented our different heuristics in
\Kuzu: (i) \Kuzu-\onehop; (ii) \Kuzu-\blind; (iii) \Kuzu-\directed; (iv) \Kuzu-ag, which adopts the \adaptiveg\ heuristic; and (v) \Indexname, which adopts the \adaptivel\ heuristic.  

\noindent {\bf ACORN~\cite{acorn} and FAISS-Navix:} ACORN is a
proximity graph implementation that supports
predicate-agnostic vector search using 
a modification of the \blind\ heuristic. ACORN is an in-memory
implementation on top of the FAISS system~\cite{faiss}. 
We implement our \adaptivel\ heuristic also on top of
FAISS to compare against ACORN, which we refer to as FAISS-Navix. 

\noindent {\bf Weaviate (v1.28.2)~\cite{weaviate} and Milvus (v2.5.0)~\cite{milvus}:} These systems serve as baselines of specialized vector databases. Weaviate also serves as baselines of an external implementation of some of the prefiltering heuristics we covered. Milvus instead performs a mix of 
pre- and postfiltering evaluation depending on the selectivities.

\noindent {\bf DiskANN~\cite{diskann}
and FilteredDiskANN~\cite{filteredDiskANN}:} 
DisANN is a disk-based vector index implementation that performs I/O when
 scanning adjacency lists of candidate vectors. 
FilteredDiskANN is an enhancement of DiskANN that support a limited set of filters.

\noindent {\bf  iRangeGraph~\cite{irange}}: 
Similar to FilteredDiskANN, iRangeGraph supports filtered vector search queries in a ``predicate-conscious'' manner. That is, it modifies its index apriori to support limited range queries. However, it is more efficient than FilteredDiskANN and is an in-memory
implementation. 
\iflong
\else
Due to space constraints, we present these experiments in the longer version of our paper~\cite{longerpapernavix}. These experiments demonstrate that search performance levels that are better than our other baselines, including \Indexname, can be achieved with filter-optimized modified index designs.
\fi

\noindent {\bf PGVectorscale (v0.5.1)~\cite{pgvectorscale} and VBase~\cite{vbase}:} These systems are two vector index implementations on Postgres that serve as baselines of vector indices
that are implemented inside an existing DBMS. They also serve
as baselines that implement postfiltering approaches.
\iflong
\else
\revision{Due to space constraints, we also present these experiments in the longer version of our paper~\cite{longerpapernavix}.}
\fi

\noindent {\em Note on brute force search:} Except for VBase, at very low selectivity levels, our baselines adopt the \texttt{brute-force} heuristic, which computes distances to every vector in $S$ and returns the kNNs with 100\% accuracy. This will be relevant in our experiments in Section~\ref{subsec:weaviate_milvus}.



\begin{table}[h]
    \centering
        \resizebox{0.35\textwidth}{!}{%
    \begin{tabular}{|c|c|c|c|}
        \hline
        \textbf{Dataset} & {\bf \# Vectors} & {\bf Dimension} & {\bf Distance Fn}\\ \hline
        GIST~\cite{ivf_pq} & 1M & 960 & L2\\ \hline
        Tiny~\cite{tiny5M} & 5M & 384 & L2 \\ \hline
        Arxiv~\cite{ann_filtered_dataset} & 2.1M & 384 & Cosine \\ \hline
        Wiki & 15.4M & 1024 & Cosine\\ \hline
    \end{tabular}
    }
    \caption{Datasets}
        \label{tab:datasets}
        \vspace{-20pt}
\end{table}

\subsubsection{Datasets}

We used the four datasets in Table~\ref{tab:datasets}.

\noindent {\bf GIST, Tiny, and Arxiv:} GIST and Tiny are two datasets
that contains embeddings of images that have been
embedded using local GIST descriptors~\cite{gist_descriptor}, which is a feature representation technique for images.
GIST uses local INRIA Holidays images~\cite{hinra_image} and Tiny contains images from the TinyImages dataset~\cite{tiny_images}. 
Arxiv is a recent open-source dataset that embeds titles from the arXiv~\cite{arxiv} paper dataset using the all-MiniLM-L6-v2~\cite{all-MiniLM-L6-v2} text embedding model.



\begin{figure}[h]
  \centering
    \begin{subfigure}[b]{0.2\textwidth}
    \centering
    \includegraphics[width=\textwidth]{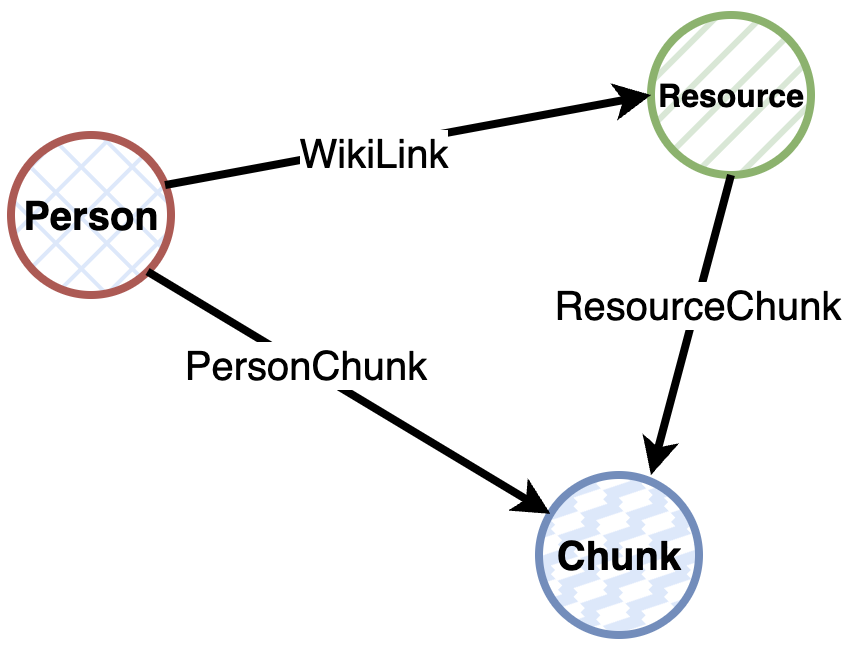}
    \subcaption{Property Graph Schema}
    \label{fig:wiki_graph_schema}
  \end{subfigure}
  \begin{subfigure}[b]{0.2\textwidth}
    \centering
    \includegraphics[width=\textwidth]{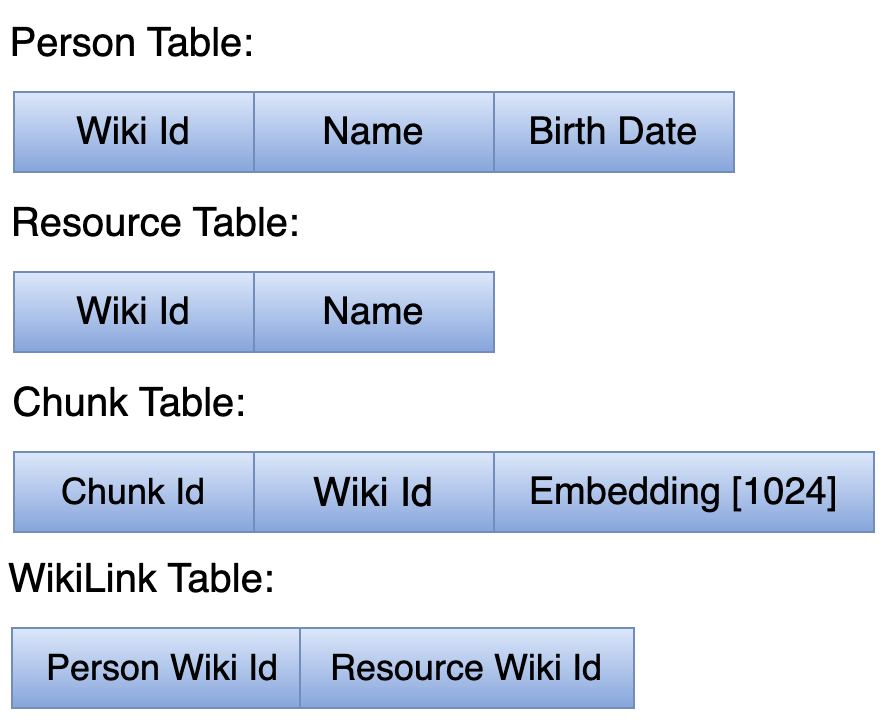}
      \subcaption{Relational Schema}
     \label{fig:wiki_relation_schema}
  \end{subfigure}
      \vspace{-5pt}
      \caption{Wiki Schema}
    \vspace{-10pt}
  \label{fig:wiki_schema}
\end{figure}

\noindent {\bf Wiki:} The above datasets 
contain objects but no connections between objects.
Therefore we can use them with predicate-agnostic queries where
the selection subqueries contain simple filters
but not joins. Moreover, these filters are uncorrelated with the 
query vectors (see Section~\ref{subsec:query-workloads}). To extend our evaluation to
selection subqueries with joins and different correlations, we prepared a new dataset from DBPedia latest version~\cite{dbpedia} and the Wikipedia dump~\cite{wikimedia}.
DBPedia is an RDF graph of (subject, predicate, object) triples.
The dataset schema is shown in Figures~\ref{fig:wiki_graph_schema} and~\ref{fig:wiki_relation_schema} both as a property graph and a relational database.
\begin{squishedlist}
\item {\em Person(pID) nodes:} We modeled each resource with a dbo:\-birth\-Place and dbo:birthDate predicate
as a Person. 
\item {\em Resource(rID) nodes:} In the DBPedia graph, we do a 2-hop traversal around Persons along the dbo:wikiPageWikiLink predicates and model each resource we visit as a Resource node.
\item {\em Chunk(cID, embd) nodes:} We chunked the Wikipedia articles
of each Person and Resource node into 1028 tokens and created
a Chunk node. We embedded each chunk into 1024 
dimensional vectors using the stella_en_400M\-_v5~\cite{stella_en_400M_v5} model on Hugging Face~\cite{huggingface} and stored it as an embedding property. 
\item {\em PersonChunk, ResourceChunk, and
WikiLink relationships:} We connected each Person
and Resource to the Chunk nodes of their corresponding Wikipedia articles, respectively,
as PersonChunk and ResourceChunk relationships. We also added
WikiLink relationships from Person nodes to their first degree Resource nodes. 
\end{squishedlist}

\begin{table}[t!]
    \centering
    \resizebox{0.48\textwidth}{!}{%
    \begin{tabular}{|c|c|c|}
        \hline
        \textbf{Correlation} & \textbf{Query} & \textbf{Filter} \\ \hline
        \textbf{Uncorrelated} & \makecell{Where is \\ Amazon located?} & \makecell{Random filtering \\ on Chunks} \\ \hline
        \textbf{Positively Correlated} & \makecell{Which company did \\ Jeff Bezos start?} & \makecell{Chunks of \\ person nodes} \\ \hline
        \textbf{Negatively Correlated} & \makecell{Where is the \\ Eiffel Tower located?} & \makecell{Chunks of \\ person nodes} \\ \hline
    \end{tabular}
    }
    \caption{Examples of Correlated Queries in Wiki Dataset}
        \vspace{-20pt}
    \label{tab:example_correlation}
\end{table}

\begin{table}[t!]
    \centering
    \resizebox{0.47\textwidth}{!}{%
    \begin{tabular}{|c|c|*{9}{c|}}
        \hline
        \textbf{Selectivity} 
            & \textbf{90\%} & \textbf{75\%} & \textbf{50\%} 
            & \textbf{40\%} & \textbf{30\%} & \textbf{20\%} & \textbf{10\%} & \textbf{5\%} & \textbf{3\%} &\textbf{1\%} \\ \hline
        Wiki
            & 0.98 & 0.98 & 0.98 & 0.98 & 0.99 & 0.98 & 1.00 & 1.02 & 1.12 & 1.12 \\ \hline
        Tiny
            & 1.00 & 1.00 & 1.00 & 1.00 & 0.98 & 0.99 & 1.02 & 1.00 & 0.98 & 0.90 \\ \hline
        Arxiv
            & 0.99 & 0.99 & 1.03 & 1.05 & 1.06 & 1.07 & 1.10 & 1.10 & 1.12 & 1.14 \\ \hline
        GIST
            & 0.90 & 0.99 & 0.99 & 1.00 & 1.01 & 1.01 & 1.01 & 0.96 & 1.04 & 1.18  \\ \hline
    \end{tabular}%
    }
    \caption{Correlation ratios of uncorrelated workloads.}
    \vspace{-20pt}
    \label{tab:un_correlations}
\end{table}

\begin{table}[t!]
  \centering
  \resizebox{0.47\textwidth}{!}{%
    \begin{tabular}{|c|*{5}{c|}*{5}{c|}}
      \hline
      \textbf{Correlation} & \multicolumn{5}{c|}{\textbf{Negatively Correlated}} & \multicolumn{5}{c|}{\textbf{Positively Correlated}} \\ \hline
      \textbf{Selectivity} & \textbf{22.9\%} & \textbf{15\%} & \textbf{9.9\%} & \textbf{5.1\%} & \textbf{1\%} & \textbf{22.9\%} & \textbf{15\%} & \textbf{9.9\%} & \textbf{5.1\%} & \textbf{1\%} \\ \hline
      Wiki & 0.055 & 0.050 & 0.055 & 0.064 & 0.037 & 2.65 & 2.90 & 2.64 & 2.57 & 2.90 \\ \hline    
    \end{tabular}%
  }
  \caption{Correlation ratios of for Wiki negatively and positively correlated workloads.}
  \vspace{-20pt}
  \label{tab:neg_pos_correlation}
\end{table}

\subsubsection{Query Workloads}
\label{subsec:query-workloads}
Our main workloads contain predicate-agnostic queries 
on our datasets. Each query $Q$ consists of two main components:
(i) $Q_S$ is a selection subquery that selects a subset $|S|$ 
of the vectors from the entire indexed vectors $V$; and (ii)
$v_Q$ is a query vector. $Q$ finds kNNs of $v_Q$
within $S$. 
We refer to the  
global selectivity of $Q_S$ as $\sigma = |S|/|V|$.
In each $Q$, $v_q$ can be ``uncorrelated'', ``positively'', or ``negatively'' correlated with $S$, which we define as follows.
Let $\text{knn}^{v_Q}_{V}$ be the kNNs of $v_Q$ in $V$. Let 
$\text{knn}^{v_Q}_{S} \subseteq \text{knn}^{v_Q}_{V}$ be the subset of these neighbors 
that are in $S$. We use $\sigma_{v_q} = \text{knn}^{v_Q}_{S}/\text{knn}^{v_Q}_{V}$ as a metric
for the selectivity of $v_Q$'s kNNs in $V$. Our
correlation metric measures if  $\sigma_{v_q}$ and $\sigma$ are correlated: 
$$\text{ce} = \sigma_{v_q}/\sigma$$

\noindent If ce $\approx$ 1, then $v_Q$ is uncorrelated with $S$.
If ce $\gg$ 1 (ce $\ll$ 1), then  $v_Q$ is positively (negatively) correlated with $S$, as the
neighbors of $v_Q$ are more (less) likely to be in $S$ than a random node.

\noindent{\bf Uncorrelated Workloads:} For each dataset,
our uncorrelated queries have a $Q_S$ that filter embedded objects on their IDs:
\begin{lstlisting}[language=SQL,style=mystyleSQL]
MATCH (c:Chunk) WHERE c.cid < (MAX_CHUNK_ID * $\sigma$)
PROJECT GRAPH S(c);
CALL QUERY_HNSW_INDEX(S, HNSWIndex, v$_Q$, k) RETURN c.cID;
\end{lstlisting}

We populate this template with different $\sigma$, $v_Q$, and $k$. For GIST, Tiny, and Arxiv, we randomly selected 50 queries from their own query sets. For Wiki, we generated 50 queries for uncorrelated and positively correlated cases as follows. We chose a random Person node  $p$, sampled its chunks and chunks of connected Resource nodes, and used OpenAI’s o1 model~\cite{openaio1} to generate questions about $p$ using these chunks as context. We provided chunks until reaching the 128K token limit. We embedded o1's question as $v_Q$ also using stella\_en\_400M\_v5. To get an uncorrelated $S$, we selected Chunks based solely on chunkIDs as in the above Cypher query, which
filters chunks uniformly.
\noindent {\bf Wiki Positively Correlated Workload:}
We used the same $v_Q$'s of the uncorrelated Wiki workload but
changed $Q_S$ as follows: 

\begin{lstlisting}[language=SQL,style=mystyleSQL]
MATCH (p:Person)-[e:PersonChunk]->(c:Chunk)
WHERE p.birth_date >= {s_date} AND p.birth_date < {e_date}
PROJECT GRAPH S(c)
CALL QUERY_HNSW_INDEX(S, HNSWIndex, v$_Q$, k) RETURN c.cID;
\end{lstlisting}
\noindent The selection subqueries are 1-hop queries that select a subset of Person
nodes based on their birthdates, which we
expect are more likely
to be in the neighborhood of $v_Q$'s than a random Chunk.

\noindent {\bf Wiki Negatively Correlated Workload:} 
For $Q_S$, we use the 1-hop selection subqueries that select Chunks of 
Person nodes. For $v_Q$ we prompt $o1$
to generate questions about non-people entities,
such as cities, monuments, and companies.

Table~\ref{tab:example_correlation} shows an example of natural language queries for each correlation scenario. Tables~\ref{tab:un_correlations} and ~\ref{tab:neg_pos_correlation} report the average correlations (ce) of these queries for different selectivity levels and correlated scenarios. Note that our 
selectivity levels go up to 100\% in uncorrelated 
workloads as we can use predicates that select every object.
For Wiki correlated workloads, selectivities are at most 22.9\% because
the 1-hop queries select Chunks that come from articles of Person nodes,
which constitute 22.9\% of all Chunks.\footnote{See our repo~\cite{repo} 
for questions generated by o1, our o1 prompts, and SQL versions of the selection subqueries we used.}

\subsubsection{Evaluation Metric}
For most experiments, we measure baseline latency on a dataset/workload under varying selectivity levels when searching for the $k$=100 nearest neighbors of $v_Q$. Since kNN is approximate, our main experiments target 95\% recall and adjust the efs parameter to stay within 1\% of it. If a baseline gives higher recall than 95\% with the minimum efs value, then we match that for all other baselines. 
If a baseline fails to reach 95\% recall even at efs 1000, we mark that point with a cross sign in our figures.  
\iflong
VBase does not have
any knobs to tune its recall, so we use it in its default configuration.
\fi

\begin{table}[h]
    \centering
        \resizebox{0.47\textwidth}{!}{%
    \begin{tabular}{|c|c|c|c|c|}
        \hline
        \textbf{Systems} & {\bf GIST 1M} & {\bf Arxiv 2.1M} & {\bf Tiny 5M} & {\bf Wiki 15.4M} \\ \hline
        PGVectorScale & 336.34 & 456.63 & 3153.52 & 5932.81 \\ \hline
        VBase & 54.84 & 71.32 & 294.12 & 942.78 \\ \hline
        Weaviate & 19.14 & 22.80 & 57.44 & 187.65 \\ \hline
        Milvus & 19.85 & 23.66 & 50.15 & 182.42 \\ \hline
        DiskANN & NA & NA & NA & 130.78 \\ \hline
        FilteredDiskANN & NA & NA & NA & 142.89 \\ \hline
        iRangeGraph & NA & 40.90 & 162.42 & NA \\ \hline
        ACORN-1 & 1.69 & 1.75 & 4.99 & 23.85 \\ \hline
        ACORN-10 & 11.55 & 10.97 & 38.55 & 162.66 \\ \hline
        FAISS-Navix & \textbf{3.29} & \textbf{3.78} & \textbf{12.22} & \textbf{42.05} \\ \hline
        \Indexname & \textbf{5.13} & \textbf{7.09} & \textbf{18.25} & \textbf{55.22} \\ \hline
    \end{tabular}%
    }
    \caption{Indexing Time (mins). All systems use 32 threads except for PGVectorScale and VBase, which are single threaded.}
        \label{tab:indexing_time}
        \vspace{-25pt}
\end{table}

\subsubsection{Index Configurations}
\label{subsub:index_config}
We built the HNSW indices with maximum connection $M$ parameter set to 32 in upper layers and 64 in lowest layer and $efC$ set to 200 across all systems. Similarly, we built the other proximity graph-based indices such as DiskANN and ACORN using similar configurations.
We used Milvus in its single segment/partitioned configuration as all other baselines and \Indexname\ stores the index in a non-partitioned way. Table~\ref{tab:indexing_time} reports the index-building
times when using 32 threads on our hardware (see below), except
PGVectorScale and VBase which have single-threaded index construction implementation. Although our focus is not on index construction time,
\Indexname's construction times are always faster than disk-based baselines.

\begin{figure*}[t!]
  \centering
    \begin{subfigure}[b]{0.24\textwidth}
    \centering
    \includegraphics[width=\textwidth]{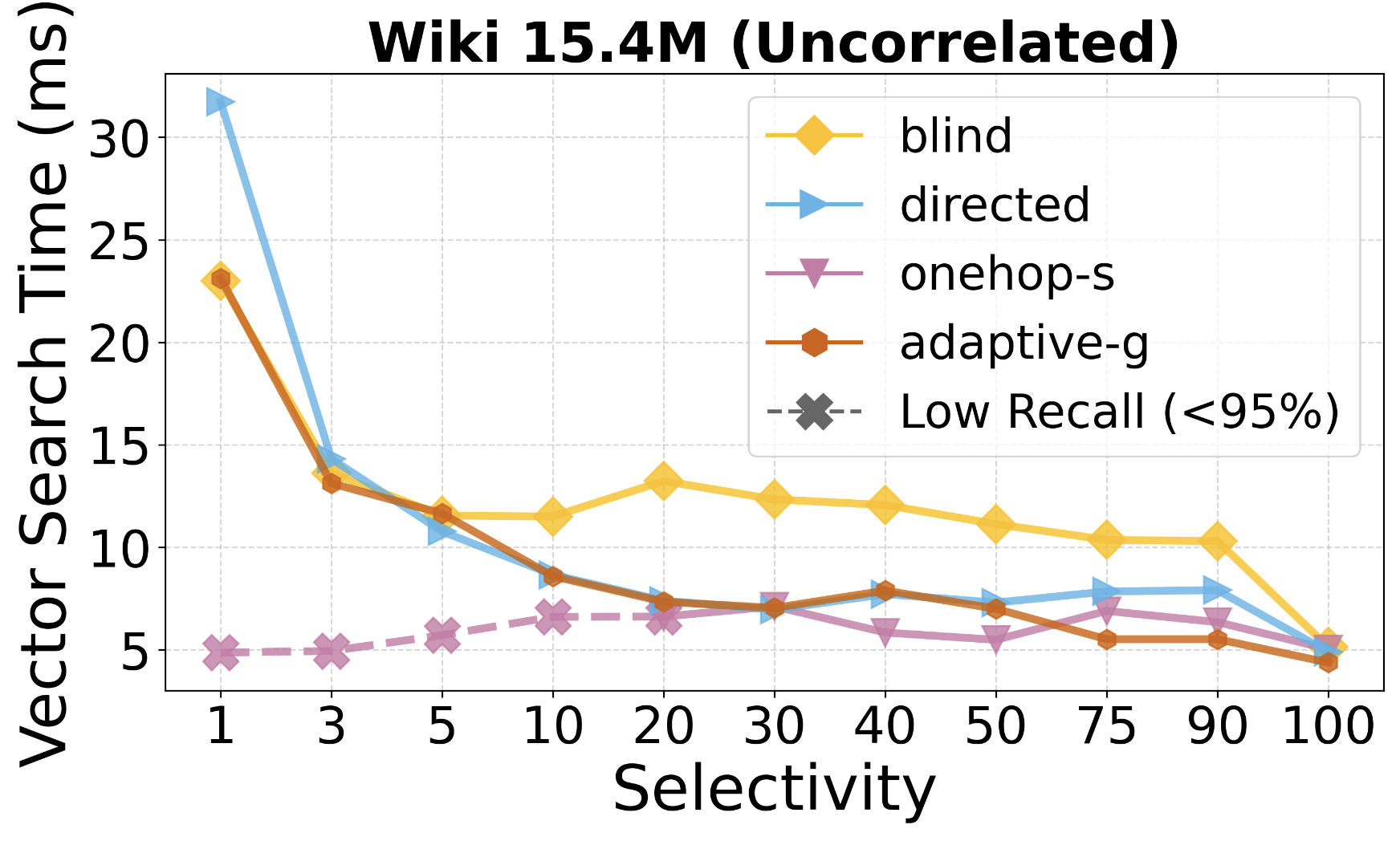}
  \end{subfigure}
  \begin{subfigure}[b]{0.24\textwidth}
    \centering
    \includegraphics[width=\textwidth]{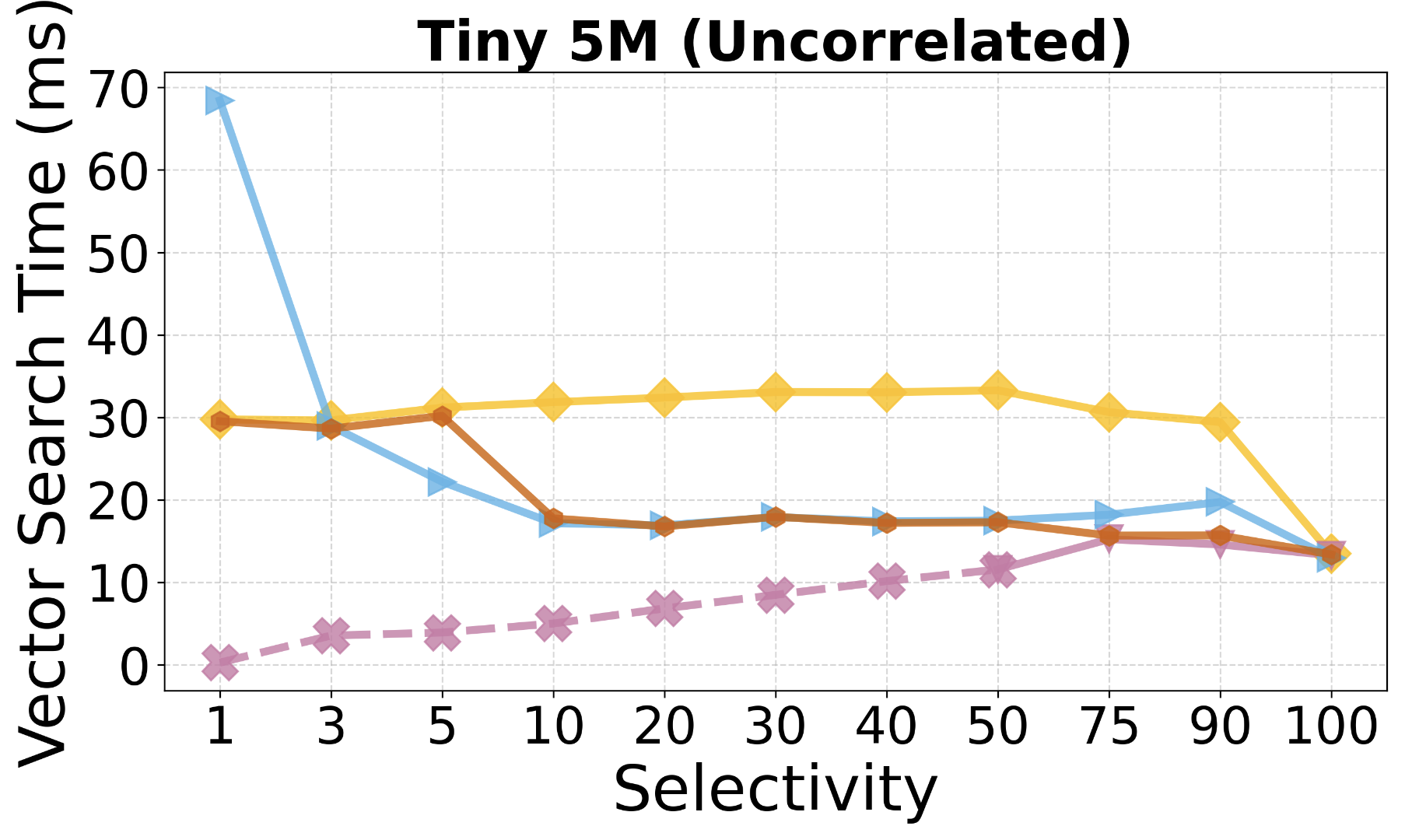}
  \end{subfigure}
   \begin{subfigure}[b]{0.24\textwidth}
    \centering
    \includegraphics[width=\textwidth]{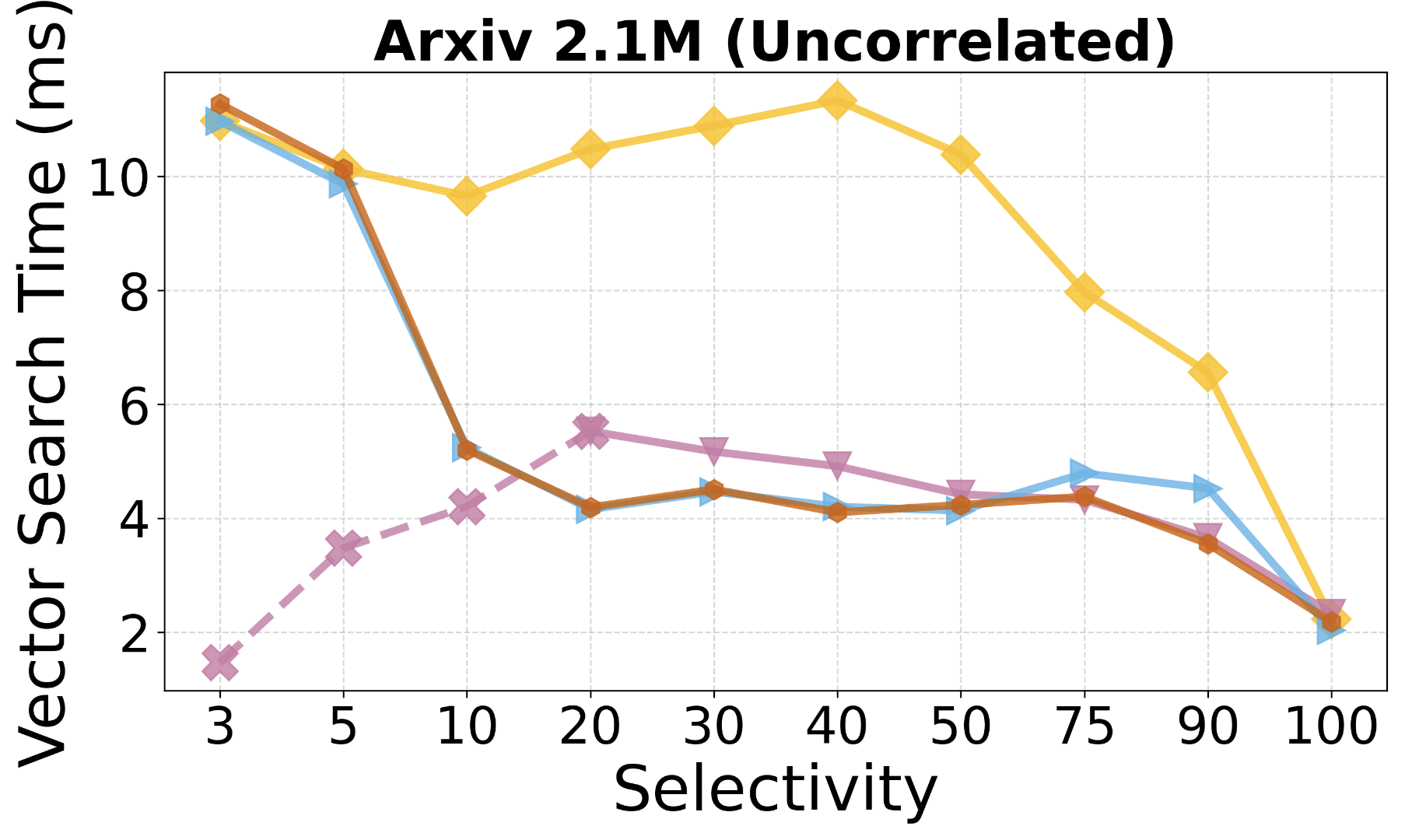}
  \end{subfigure}
   \begin{subfigure}[b]{0.24\textwidth}
    \centering
    \includegraphics[width=\textwidth]{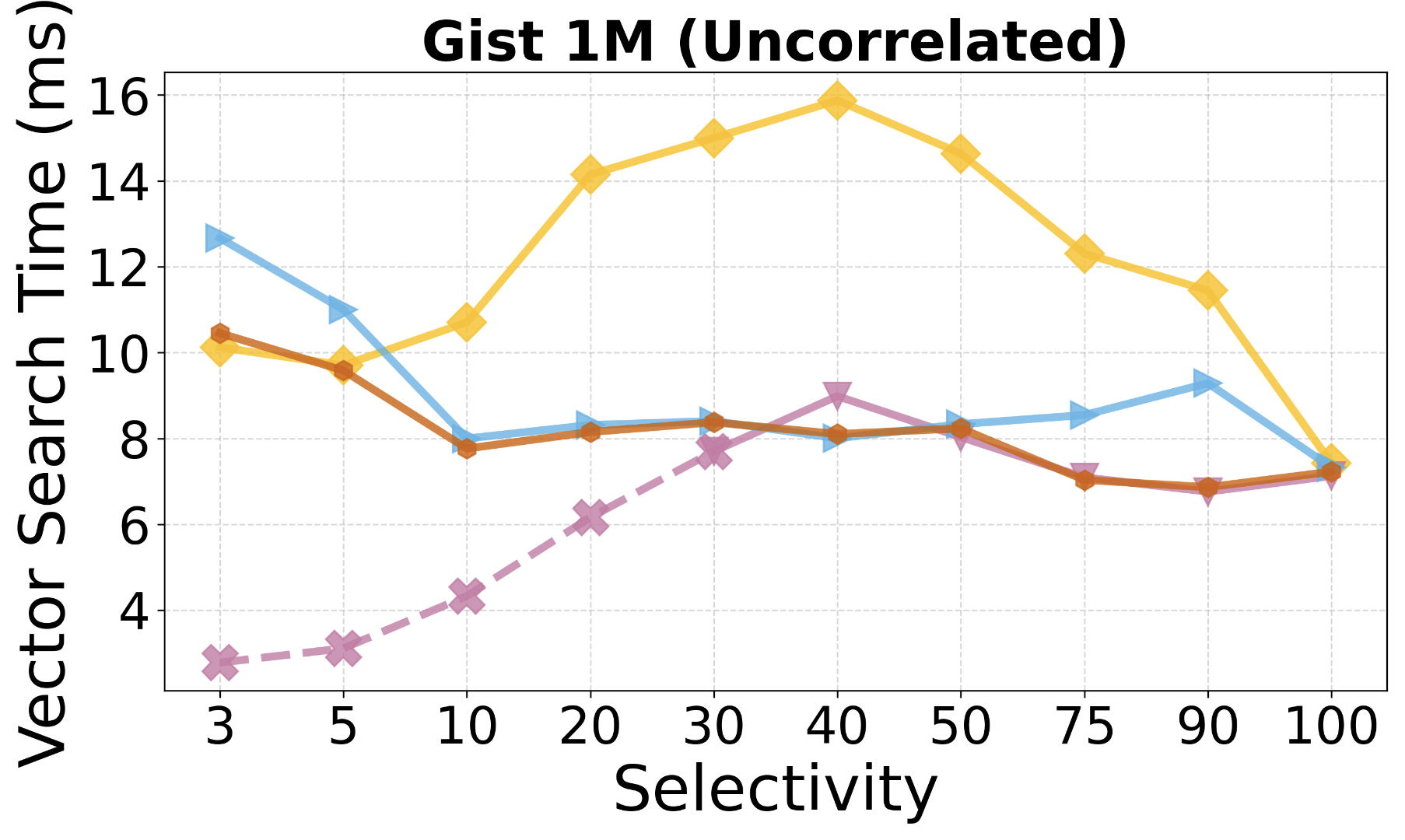}
  \end{subfigure}
      \vspace{-10pt}
  \caption{Vector search time vs Selectivity for different heuristics within 95\% to 95.5\% recall}
    \vspace{-10pt}
  \label{fig:heuristic_vs}
\end{figure*}

\begin{figure}[h]
  \centering
  \begin{minipage}{0.48\textwidth}
    \centering
    \begin{minipage}[b]{0.5\textwidth}
      \centering
      \includegraphics[width=\textwidth]{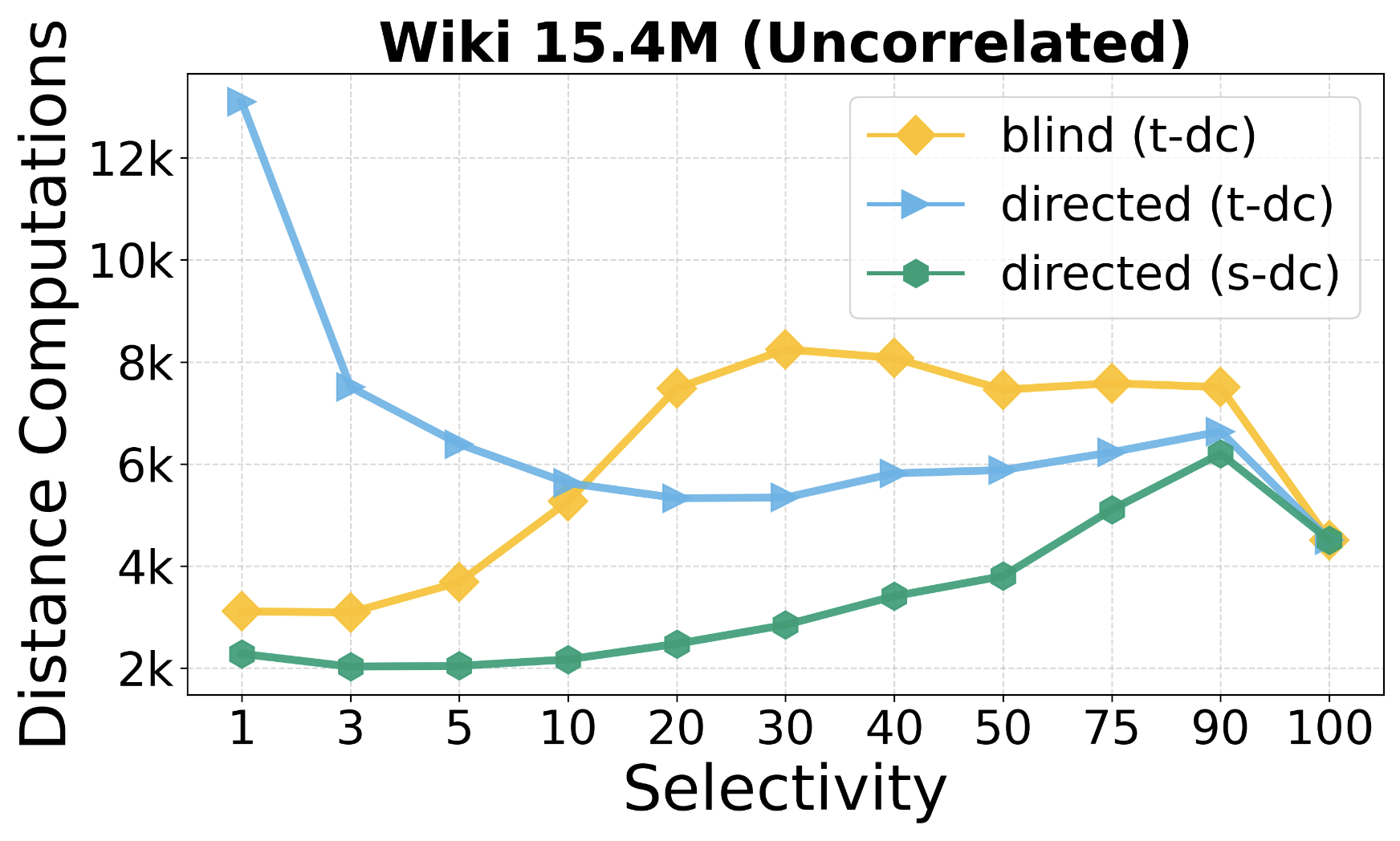}
    \end{minipage}\hfill
    \begin{minipage}[b]{0.5\textwidth}
      \centering
      \includegraphics[width=\textwidth]{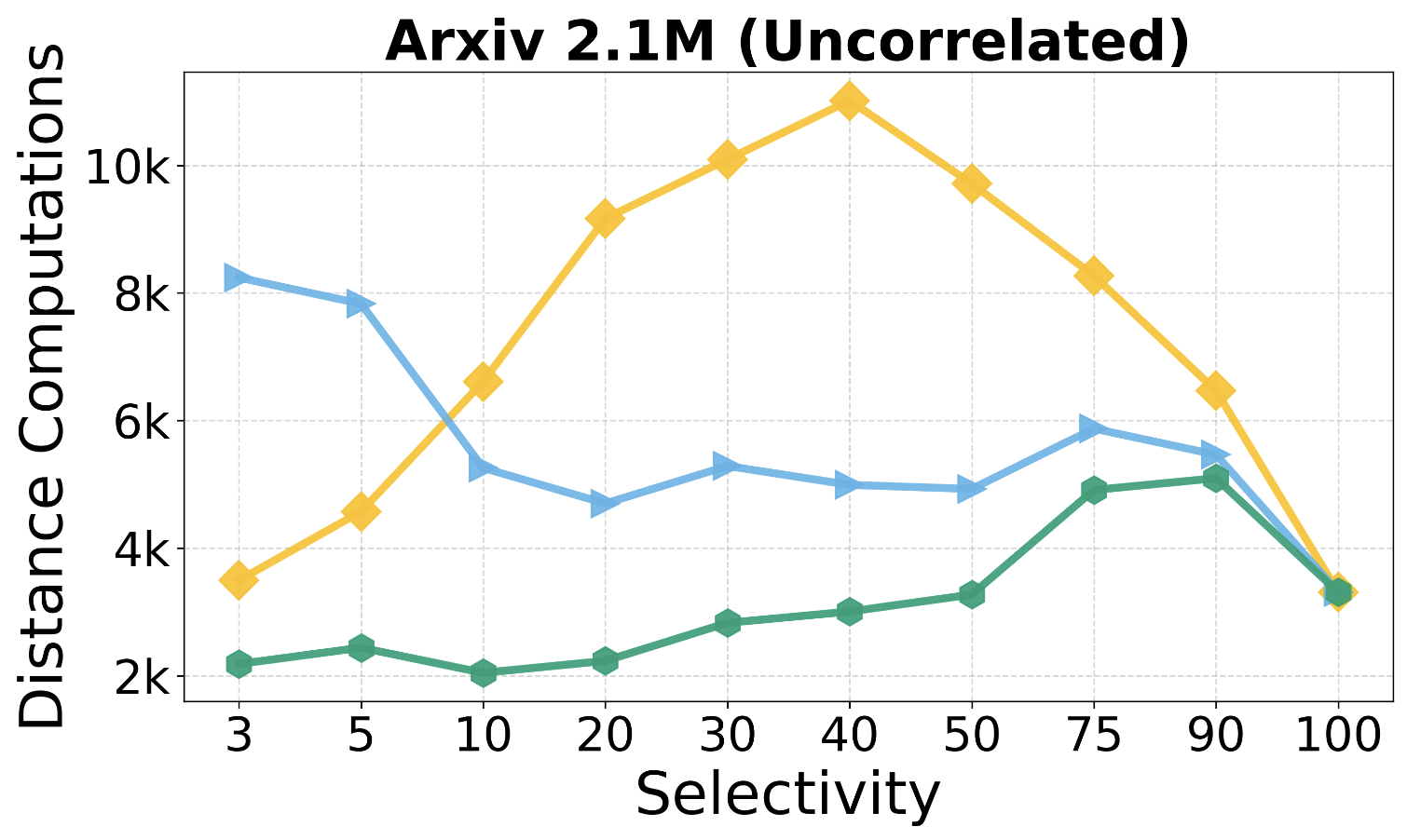}
    \end{minipage}\hfill
             
        \captionsetup{margin=20pt,skip=10pt} 
    \vspace{-10pt}
    \captionof{figure}{\texttt{t-dc} vs \texttt{s-dc} of \blind\ and \directed.}
  \label{fig:total_selected_dc}
  \end{minipage}
  \hfill
\vspace{-10pt}
\end{figure}

\subsubsection{Index Size and Sampling Ratio}
\label{subsub:index_size_sampling}

In \Indexname\, we set the sampling ratio to 5\%. This results in a high-level in-memory index size of at most $\sim$200MB (on our largest dataset Wiki). This requirement is a tiny fraction of the storage requirements for vectors and the lower level graph. For example, on Wiki,
vectors take  $\sim$63GB, while the lower
level graph takes $\sim$7.8GB.

\subsubsection{Other Experimental Setup Details}
Each query workload contains 50 different $v_Q$ and all numbers in our experiments are averages across these queries. 
Except in our disk-based experiments in
Section~\ref{subsec:disk}, for each query $Q$, we first warm up the buffer manager by running $Q$. Then we run each query 5 times and report the average 
latency of $Q$. Unless otherwise specified, all latency
numbers measure end-to-end latency, i.e., including both the time to execute the selection subquery and kNN search.
We ran our evaluation on Compute Canada~\cite{compute-canada} on a single x86 machine with 180 GB of memory and 32 v-CPUs using an Intel(R) Xeon(R) Platinum 8260 processor. Because we could not install DiskANN on this machine, for DiskANN and FilteredDiskANN benchmarks we used a different machine with 132GB of memory, 32 v-CPUs and 2.5T of SSDs.



\begin{figure*}[t!]
  \centering
  \begin{minipage}[t]{0.74\textwidth}
    \centering
    \begin{minipage}[b]{0.33\textwidth}
      \centering
      \includegraphics[width=\textwidth]{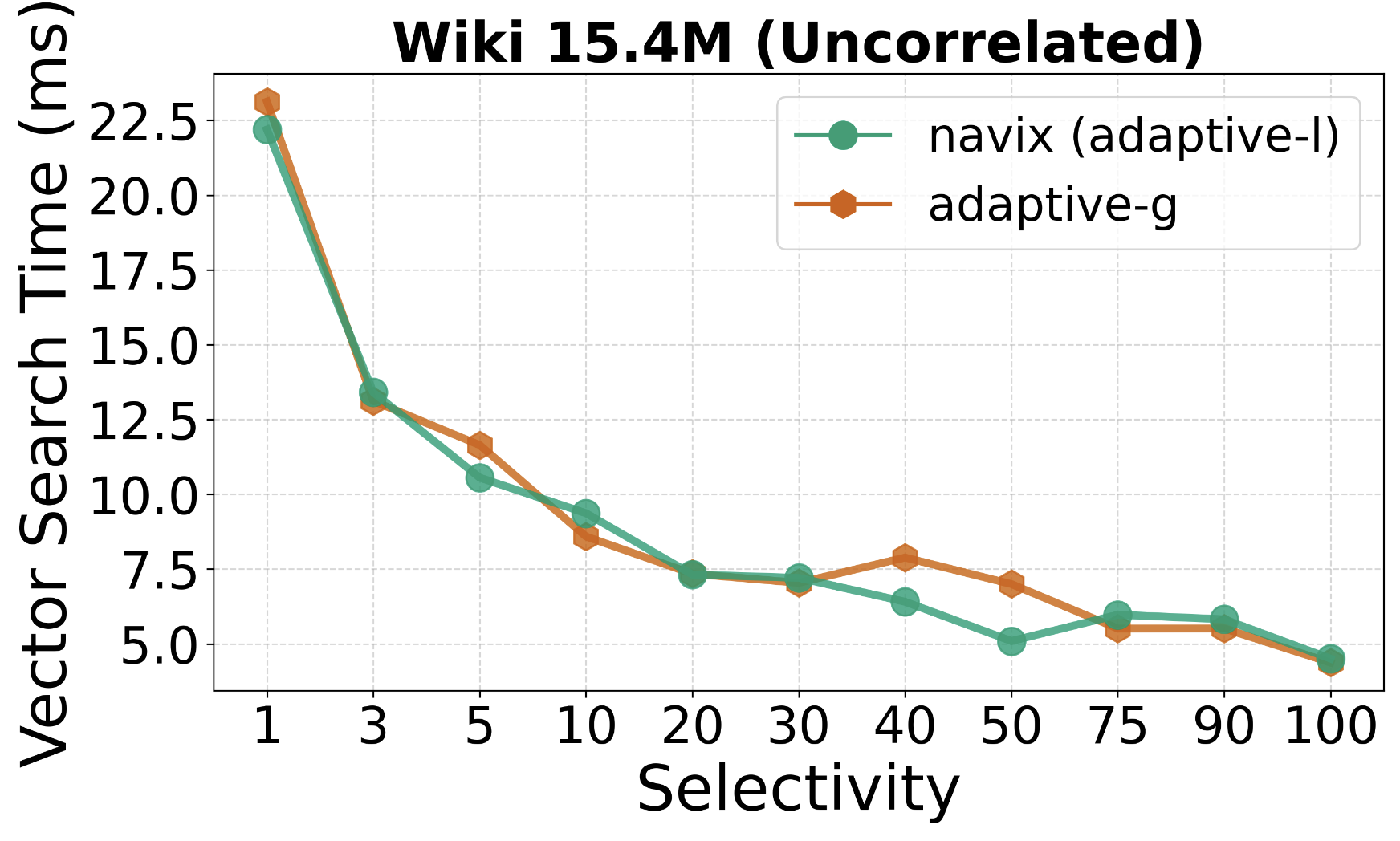}
    \end{minipage}\hfill
    \begin{minipage}[b]{0.33\textwidth}
      \centering
      \includegraphics[width=\textwidth]{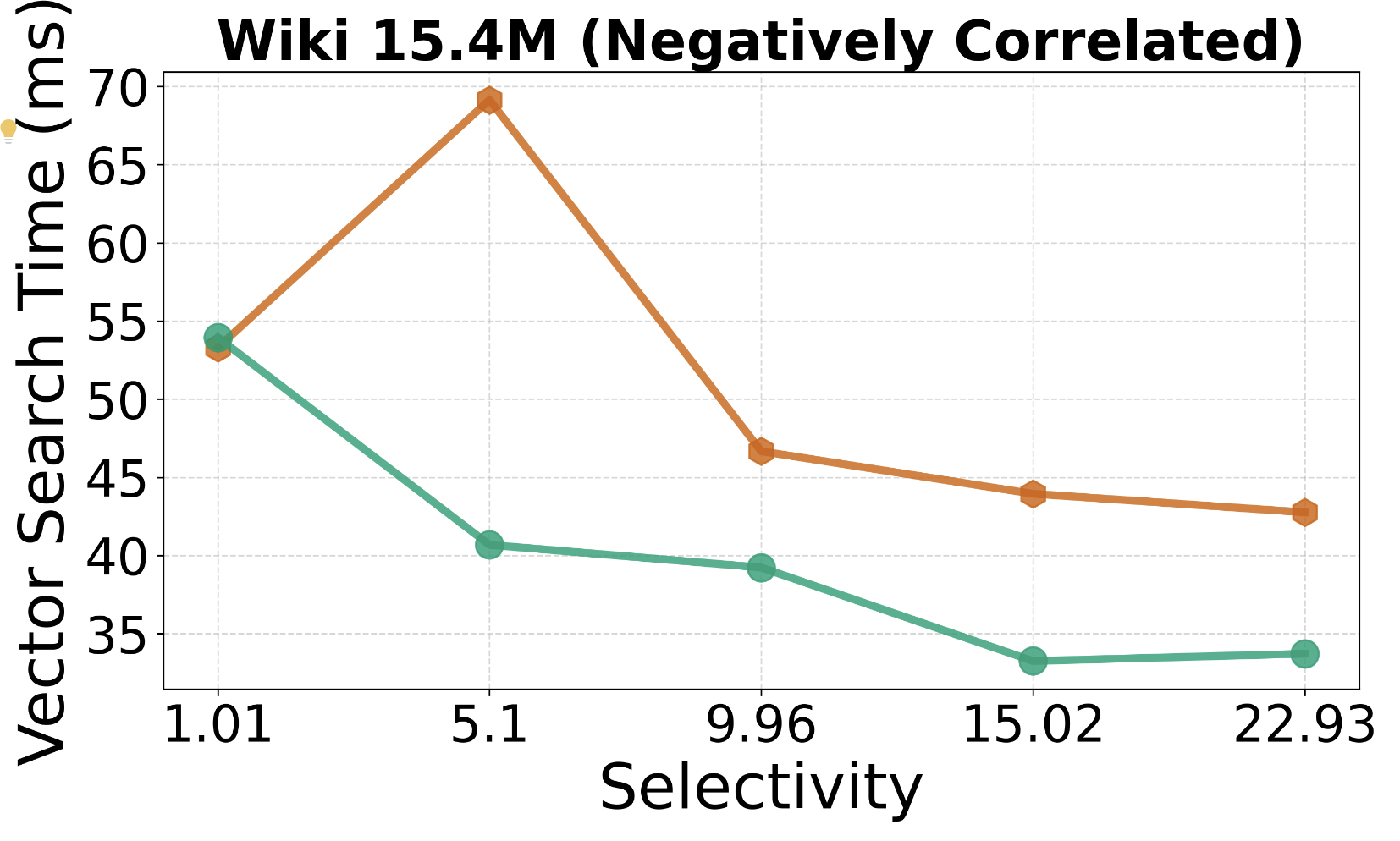}
    \end{minipage}\hfill
    \begin{minipage}[b]{0.33\textwidth}
      \centering
      \includegraphics[width=\textwidth]{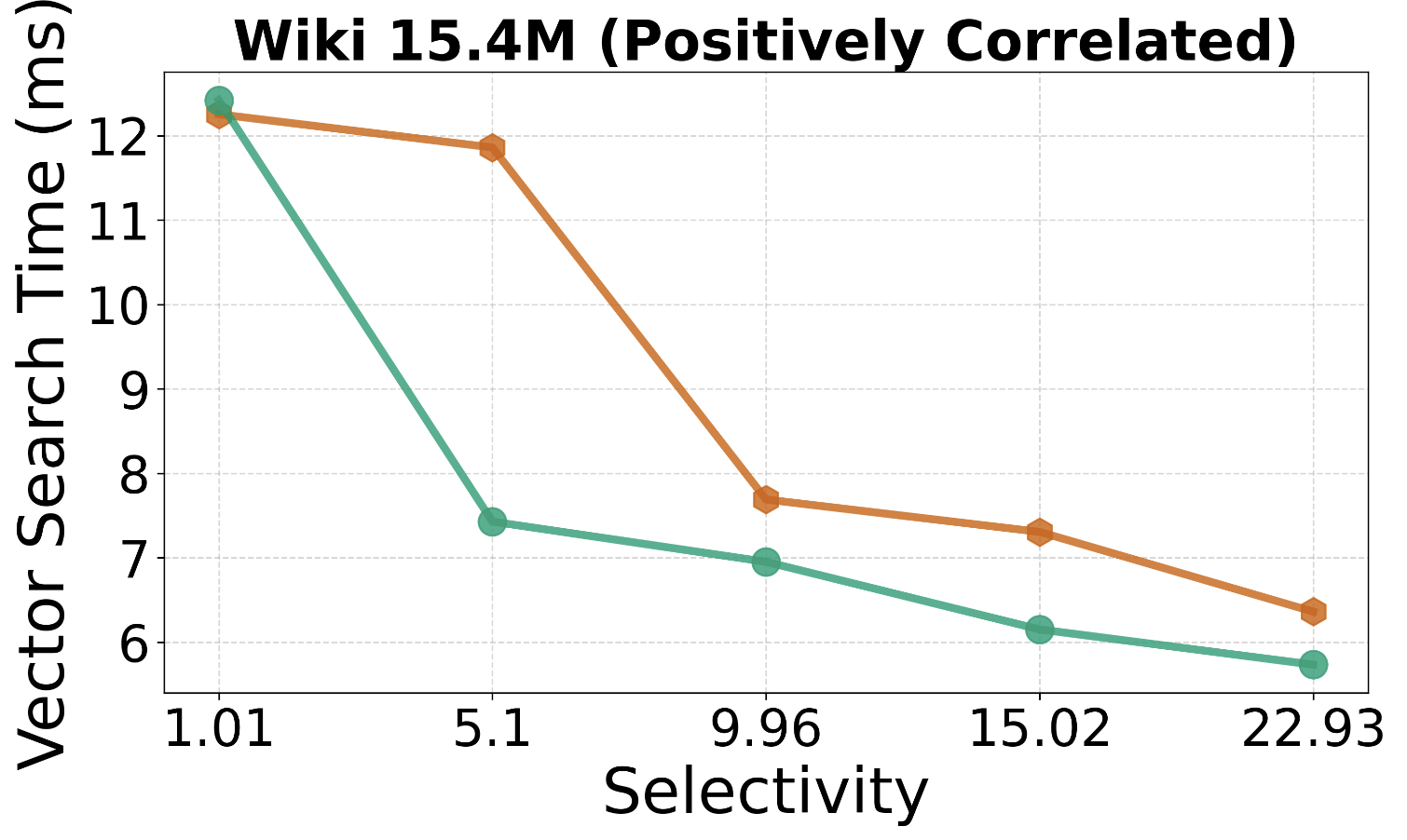}
    \end{minipage}
    \captionsetup{margin=20pt,skip=10pt} 
        \vspace{-5pt}
    \captionof{figure}{Vector search time vs selectivity for \Indexname\ and \adaptiveg.}
        \vspace{-5pt}
    \label{fig:navix_vs_adaptive_g}
  \end{minipage}\hfill
  \begin{minipage}[t]{0.25\textwidth}
    \centering
    \includegraphics[width=\textwidth]{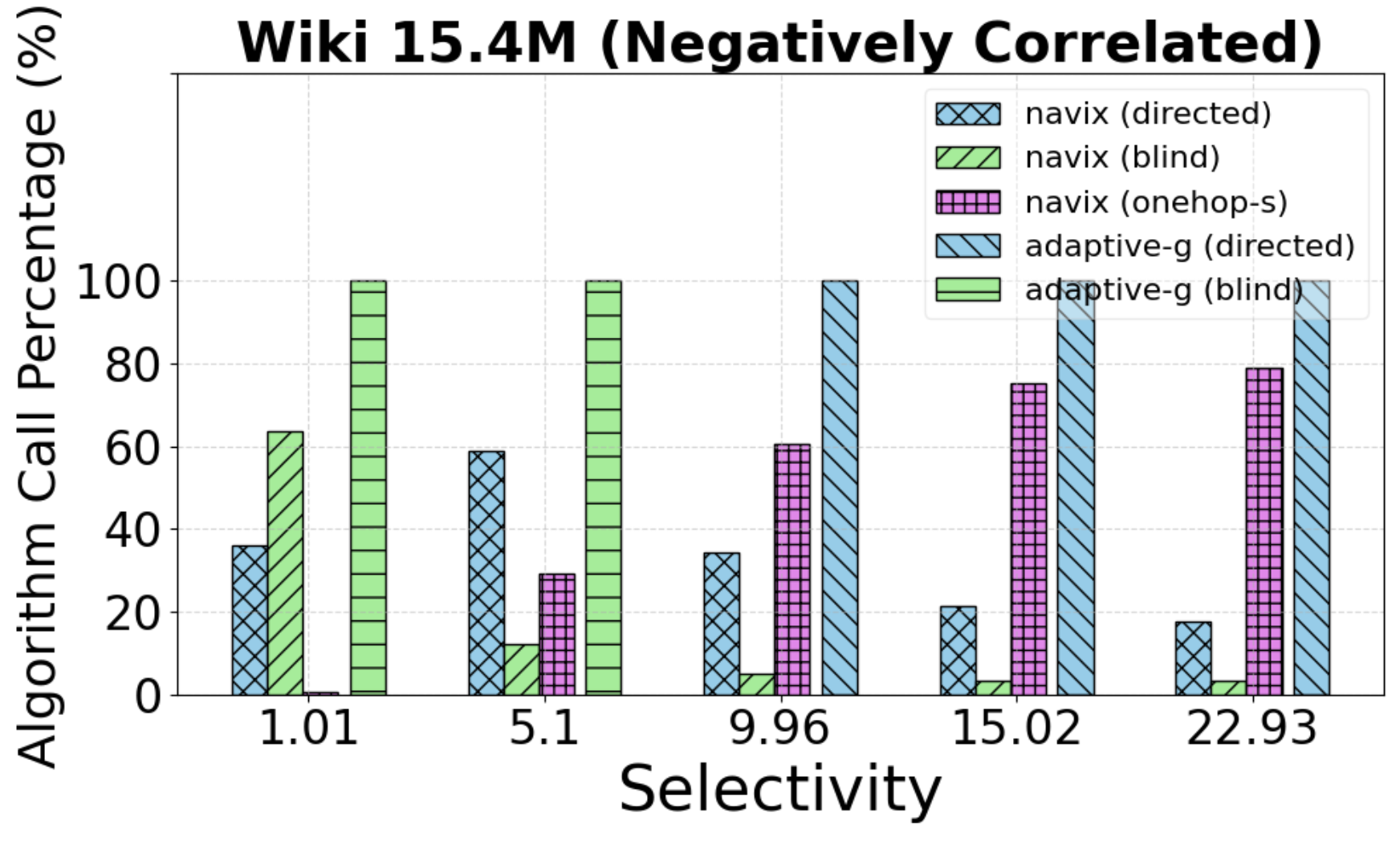}
            \vspace{-17pt}
    \captionof{figure}{Heuristic calls.}
    \label{fig:neg_correlated_heuristics}
  \end{minipage}
      \vspace{-10pt}
\end{figure*}

\subsection{Prefiltering Fixed Heuristics and Adaptive-G}
\label{subsec:prefilter_heuristics}
We begin by studying the behaviors of different fixed prefiltering heuristics and \adaptiveg.
We used each dataset and their uncorrelated workloads.
Since each of the \Kuzu\ configurations have
the same prefiltering cost for evaluating the selection subqueries,
we only report vector search times.
Figure~\ref{fig:heuristic_vs} presents our results.

 First, as hypothesized in
Section~\ref{sec:hybrid_search}, across all datasets, 
\onehop\ consistently has lower latency than other heuristics 
as it always limits its distance calculations to only the selected vectors in first degree neighbors of candidates.
However, its recall drops significantly below 50\% selectivity for  Tiny5M and 30\% for Wiki (Figure \ref{fig:heuristic_vs}). 
Therefore, above 50\% selectivity,
\onehop\ is the most performant of the fixed heuristics across all datasets.

Second, observe that at very high selectivity levels,
\blind\ and \directed\ perform very similarly.
This is because at very high selectivity levels, say 100\%, 
since each first degree neighbor is selected, 
both directed and blind explore only the first degree neighbors. 
Then as selectivities decrease, \directed\ starts outperforming
\blind. Once the selectivities are low enough, \blind\ starts
outperforming \directed. 

To explain this behavior, we measured two further metrics. 
First, we measure the number of distance calculations
these heuristics perform across the selected vectors (\sdc). 
This number is 
equivalent to the number of vectors a heuristic puts into the
candidates priority queue.
This is a measure of how {\em effective} the search is.
Second, we measure the total distance calculations a heuristic makes (\tdc).
For \blind, \tdc\ is always equal to \sdc. 
However, for \directed, \tdc\ can be larger than \sdc\ as \directed\ computes
distances to unselected vectors in the 1st degree neighbors of candidates. 
We use the difference of \tdc\ and \sdc\ 
as a measure of {\em \directed's overhead}. 
Figures~\ref{fig:total_selected_dc} show the \sdc\ numbers for \blind\ and \directed\, and \tdc\ numbers for \directed.




As selectivity levels decrease, a larger fraction of the 1st degree neighbors are unselected, so 
\directed's overhead of 
computing the distances to them increase. 
Further, irrespective of the selectivity level,
\directed\ performs the search as effectively
or better than \blind.  
However, \directed's effectiveness edge over \blind\ is low at very high and very low selectivity levels because at these levels
they perform very similar explorations.
As we already observed above, at very high selectivity levels
these heuristics perform very similar searches. 
At very low selectivity levels, since very few vectors
are selected, both heuristics again behave similarly because both explore every 2nd degree neighbor. That is also why at very low selectivity levels,
\blind\ outperforms \directed\ in latency, because \directed's overhead are
highest and it does not have an important advantage in terms of its search effectiveness.
\directed\ however has an edge over
\blind\ at medium selectivity ranges of 50\% to 5\%
where its search edge is larger and overheads are low enough. At roughly these ranges, \directed\ outperforms
\blind\ by up to 2x in latency. Overall, 
between 50\% and 5\% selectivity, \directed\ is the 
most performant fixed heuristic.
Below and at 5\% selectivity, \blind\ outperforms both \directed\
and \onehop. This matches the regions we outlined in Figure~\ref{fig:fixed-heuristics-choice}.

Next, observe that, except for a few small regions, \adaptiveg\ follows the lowest-latency fixed heuristics
in almost all ranges,
indicating that using global selectivity information is enough
to capture the best of all fixed heuristics.
The only exception is within ranges below but close to 50\% on some datasets. This is because we choose 50\% as a safe
threshold in \adaptiveg\ to switch to \onehop, although in some datasets
even between 30\% and 50\%, \onehop's recall is high enough. 


\vspace{-5pt}
\subsection{Adaptive-G and \Indexname}
\label{sec:adaptiveg-navix}
In our next set of experiments we focus on comparing
\adaptiveg\ and \Indexname\ (adaptive-l). Our goal is to show that 
\Indexname\ is a strict improvement over \adaptiveg\ and 
the advantage of \Indexname\ is especially visible  when
queries are correlated.
We used the Wiki dataset and each of its workloads and 
measured the vector search times of \adaptiveg\ and \Indexname. 
We also used the uncorrelated workloads of our datasets.
Our observations for these experiments are similar to Wiki uncorrelated
dataset and are presented
\iflong
in Appendix~\ref{app:adaptive_g_navix}.
\else
in the longer version of our paper~\cite{longerpapernavix}.
\fi

Figure~\ref{fig:navix_vs_adaptive_g} shows our results. Observe that in the Wiki uncorrelated workload, \adaptiveg\ and \Indexname\ 
perform very similarly. However in both the negatively and positively correlated
cases, the local selectivity of candidate nodes during search can be different than the global selectivity of $S$. Therefore we see 
\adaptiveg\ and \Indexname\ behaving differently. Overall we observe that \Indexname\ does 
better decisions and outperforms \adaptiveg\ consistently, in some cases
up to a factor of 1.7x. For example in the negatively
correlated benchmark at 5\% selectivity level, the average vector search
latency of \adaptiveg\ is 69.13 ms while that of \Indexname\ is 40.67 ms.

To verify that  \Indexname\ and \adaptiveg\ make different choices in terms
of the heuristics they pick, we calculated the distribution of 
each heuristic they pick for different workloads. Figure~\ref{fig:neg_correlated_heuristics} shows
these distributions for the Wiki negatively correlated workload.
The figure reports what fraction of times \Indexname\ and \adaptiveg\
picks each heuristic when exploring a candidate's neighborhoods for each
selectivity level. Observe that since \adaptiveg's choice is based on the
global selectivity, at each selectivity level, it commits to using one heuristic.
Instead, \Indexname\ makes more nuanced decisions.
For example, at 22.9\% selectivity, while \adaptiveg\ commits to
\directed, \Indexname\ picks \onehop\ 80\% of the time,
indicating that it has performed the majority of the search in
a region that contain vectors with very high local selectivity.

\begin{figure*}[t!]
  \centering
    \begin{subfigure}[b]{0.23\textwidth}
    \centering
    \includegraphics[width=\textwidth]{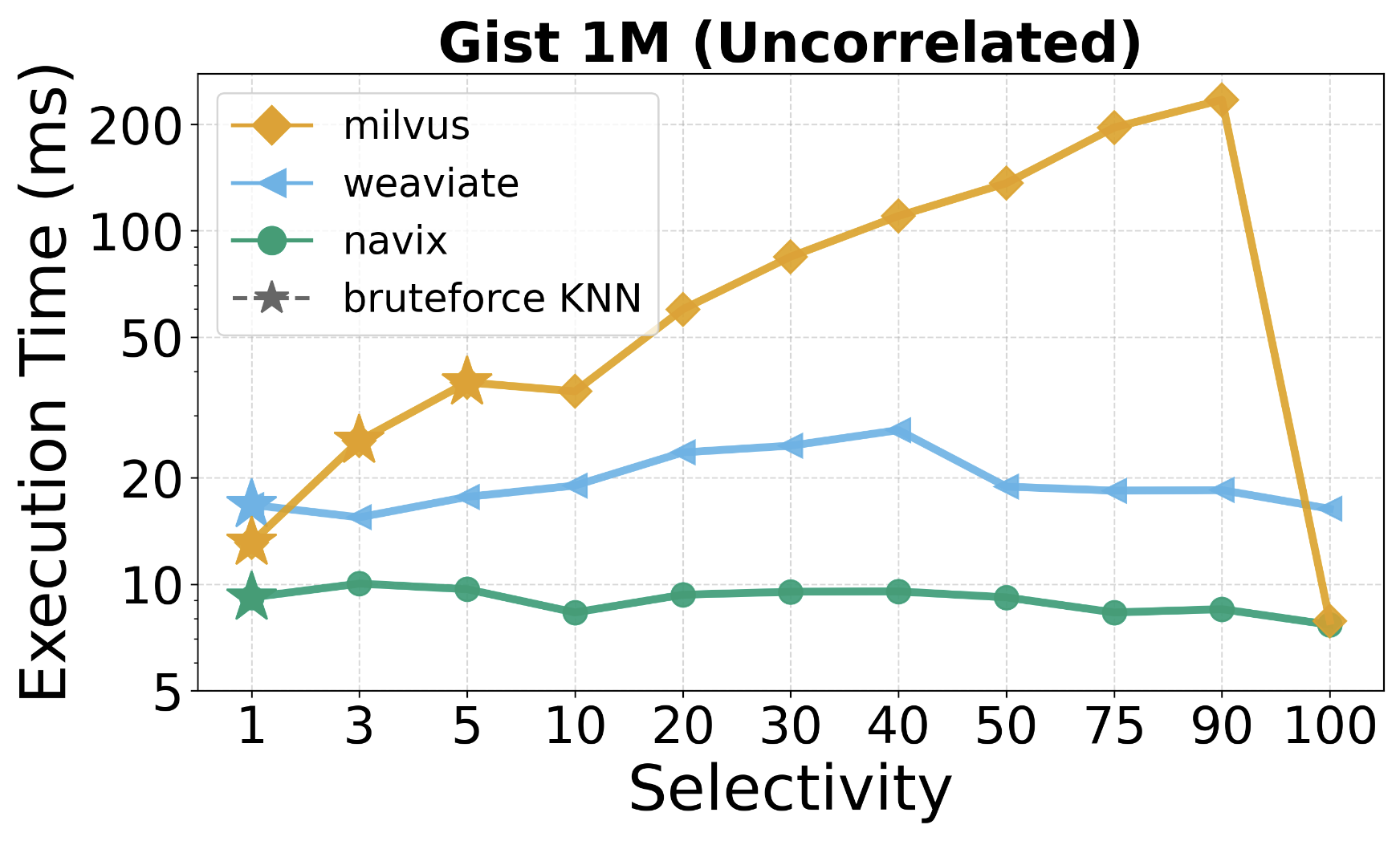}
  \end{subfigure}
  \begin{subfigure}[b]{0.23\textwidth}
    \centering
    \includegraphics[width=\textwidth]{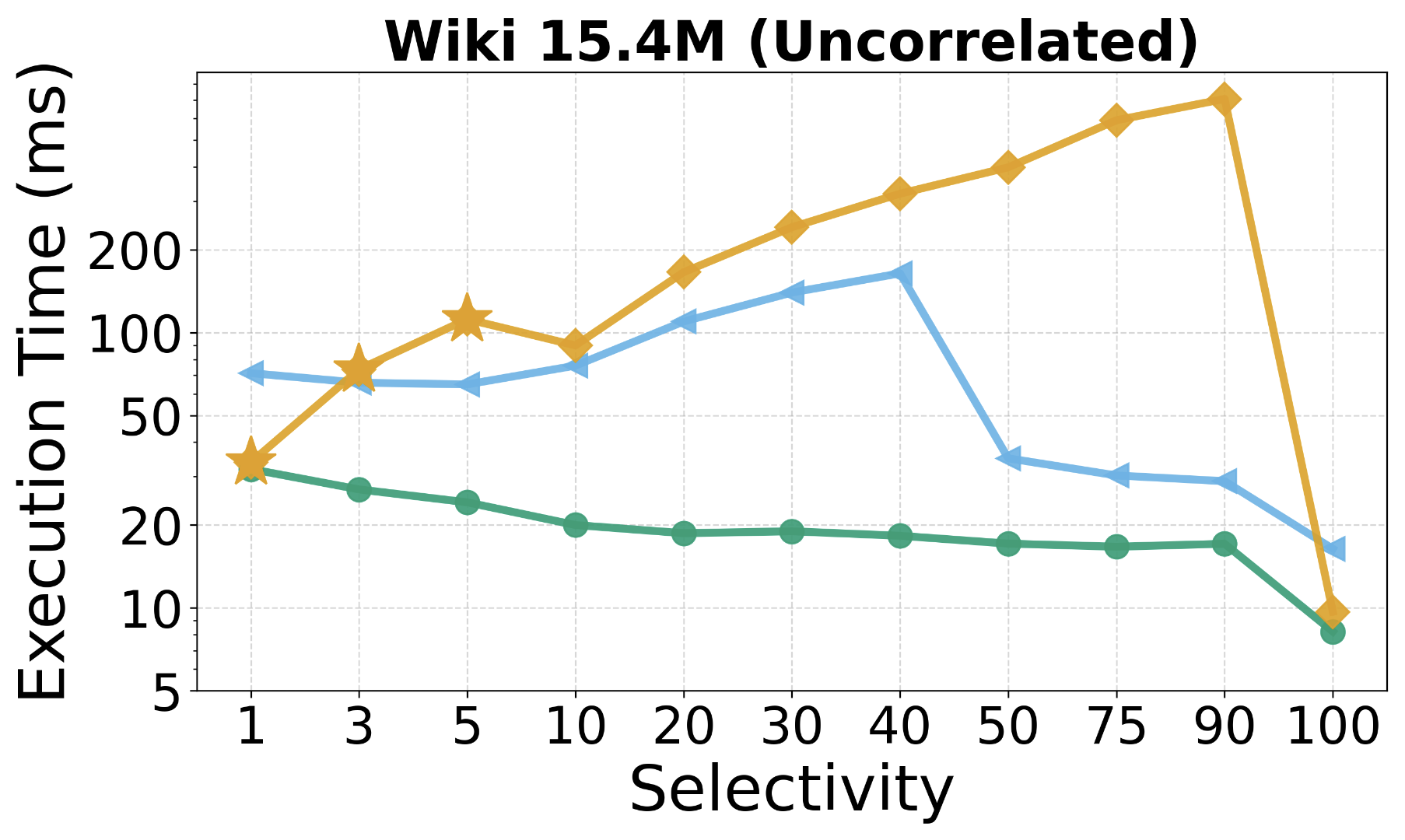}
  \end{subfigure}
   \begin{subfigure}[b]{0.24\textwidth}
    \centering
    \includegraphics[width=\textwidth]{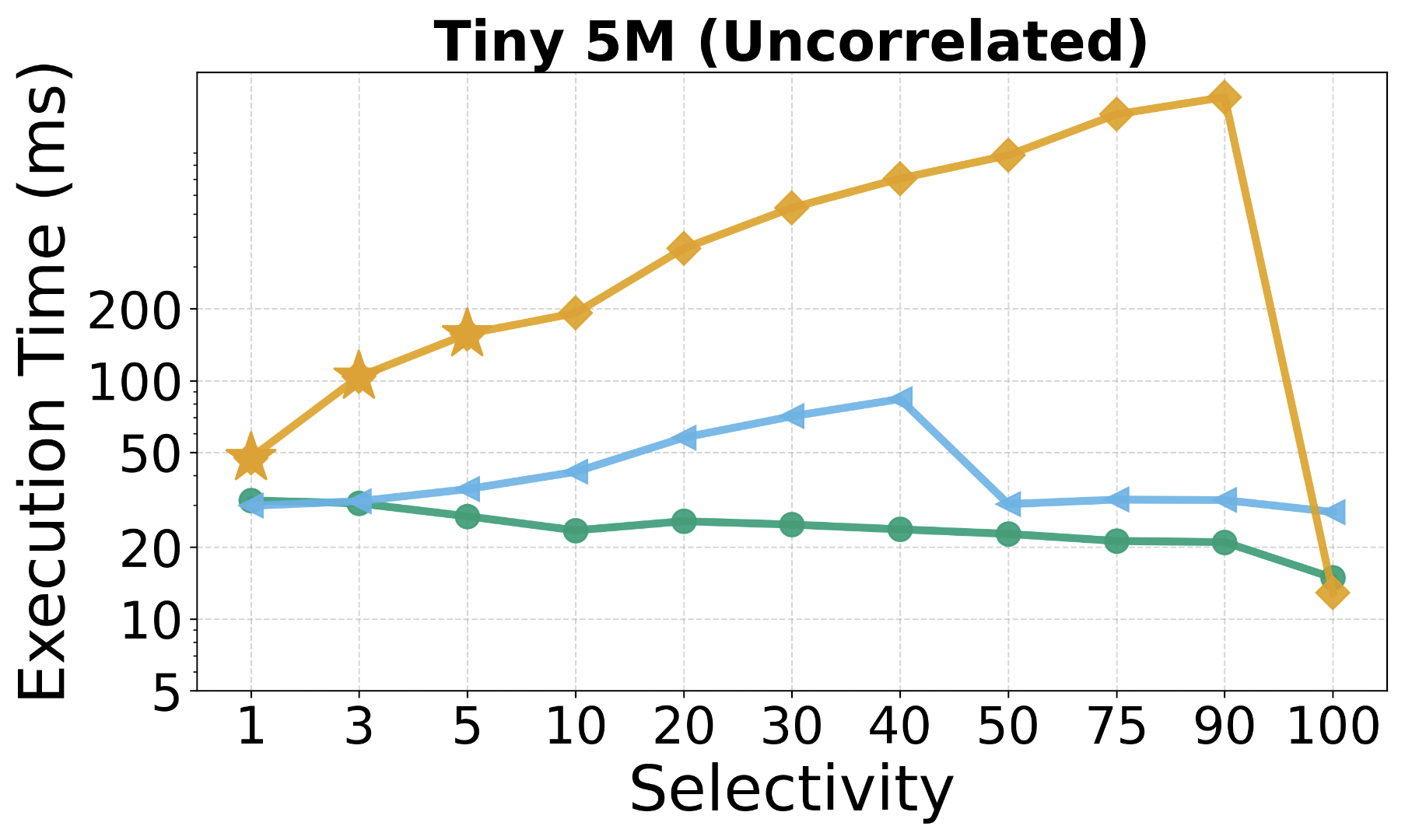}
  \end{subfigure}
   \begin{subfigure}[b]{0.24\textwidth}
    \centering
    \includegraphics[width=\textwidth]{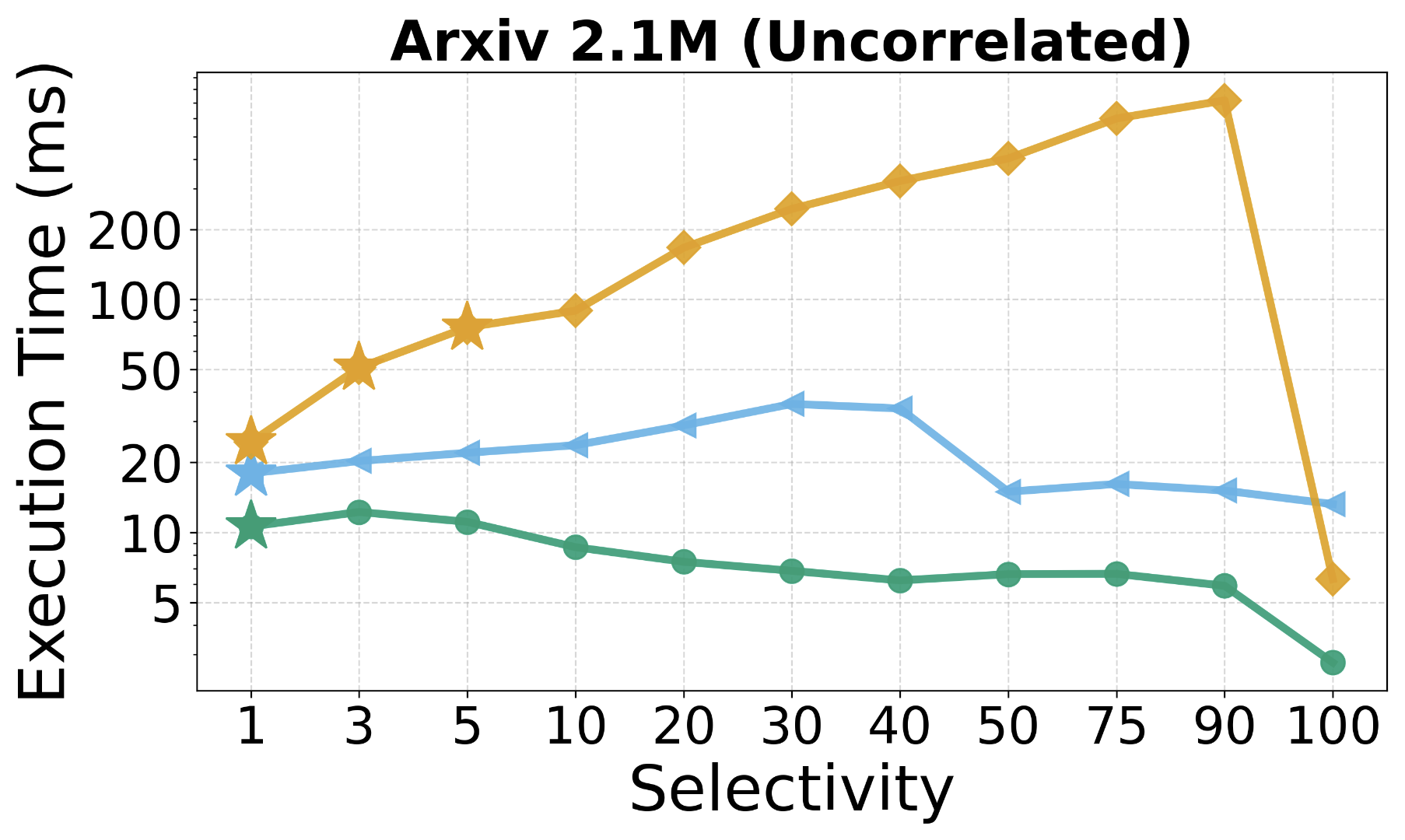}
  \end{subfigure}
              \vspace{-10pt}
  \caption{Execution Time vs Selectivity for prefiltering baselines at 95\% recall}
                \vspace{-10pt}
  \label{fig:prefiltering_baselines}
\end{figure*}

\begin{figure*}[t!]
  \centering
    \begin{subfigure}[b]{0.23\textwidth}
    \centering
    \includegraphics[width=\textwidth]{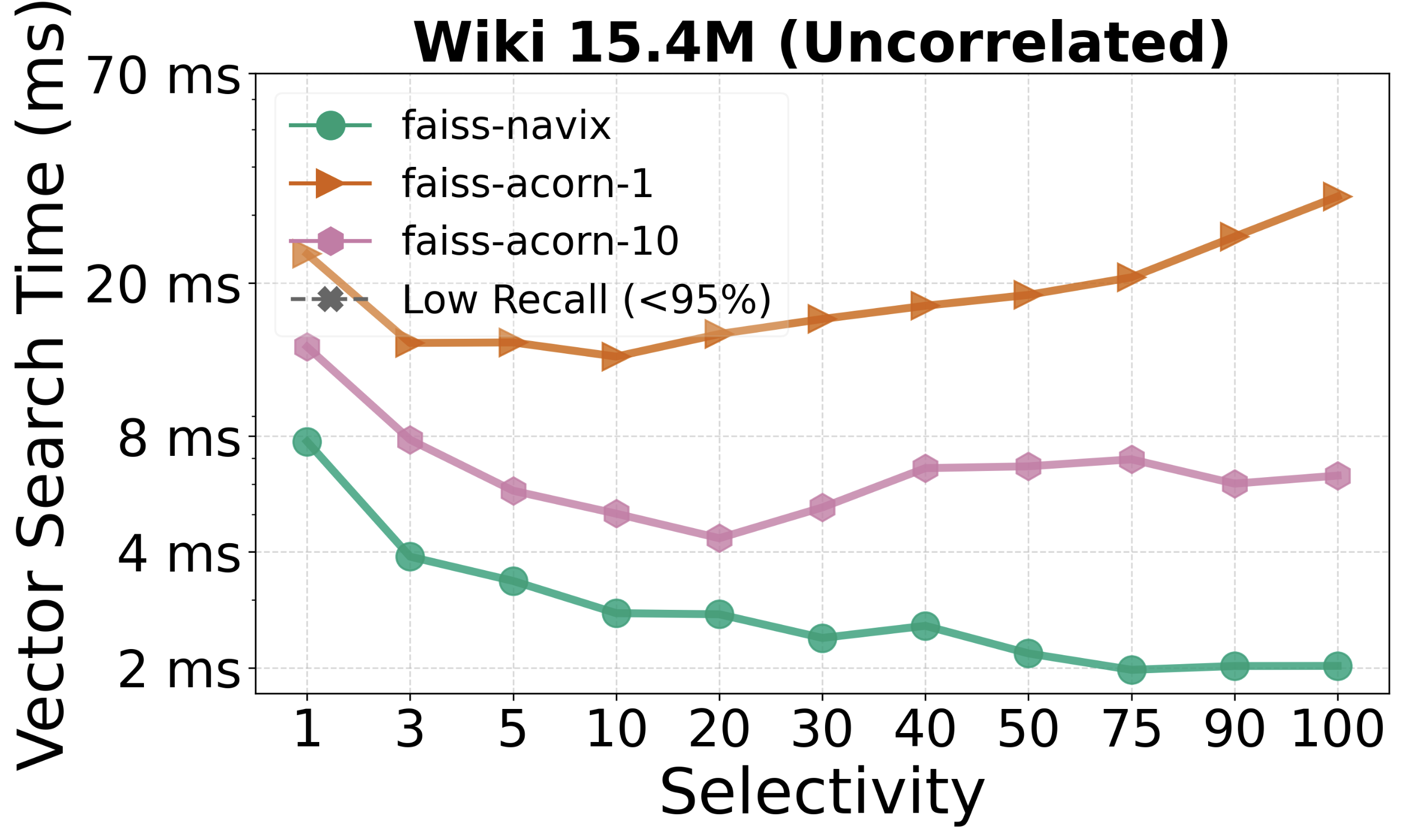}
  \end{subfigure}
  \begin{subfigure}[b]{0.23\textwidth}
    \centering
    \includegraphics[width=\textwidth]{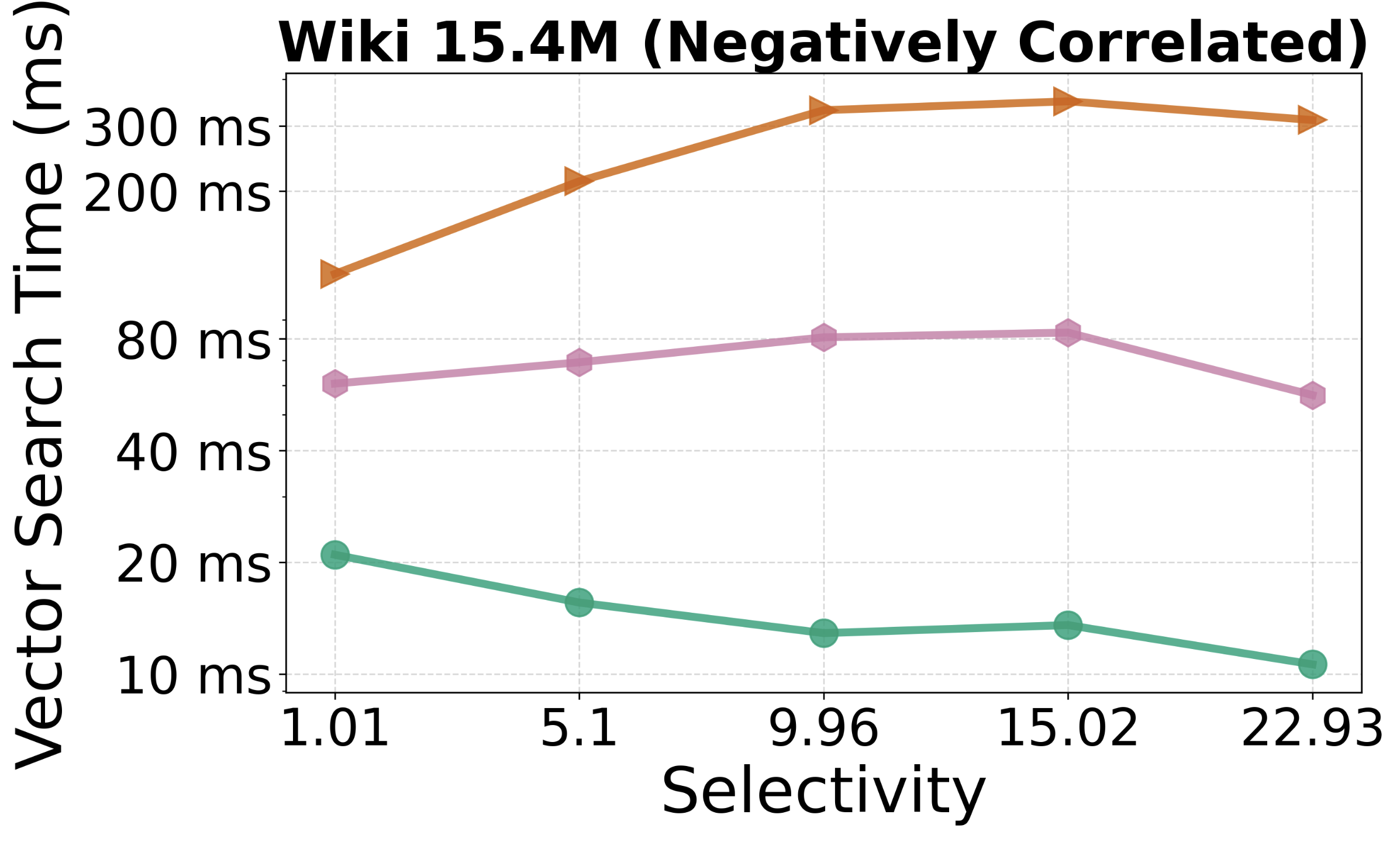}
  \end{subfigure}
   \begin{subfigure}[b]{0.24\textwidth}
    \centering
    \includegraphics[width=\textwidth]{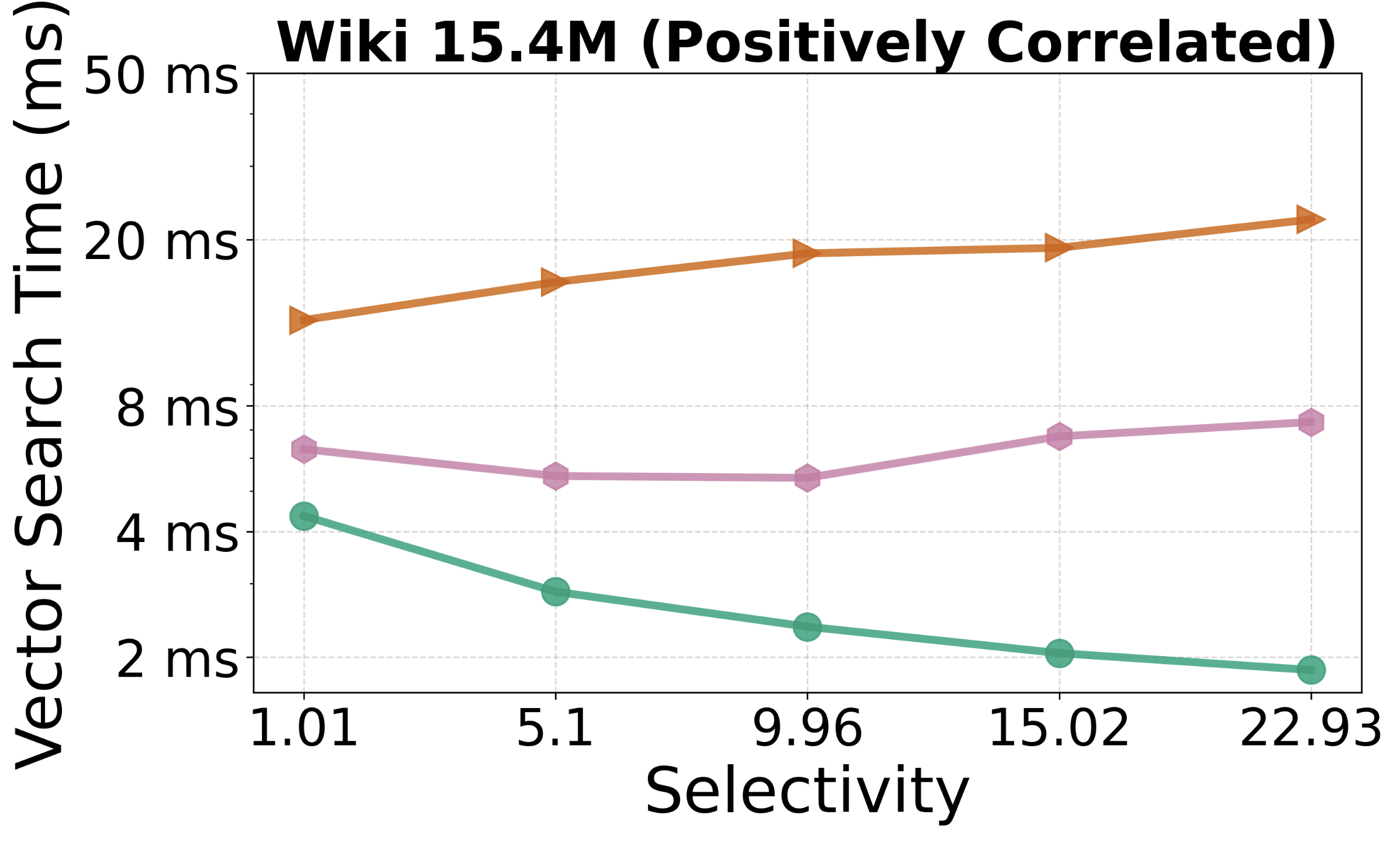}
  \end{subfigure}
  \iflong
     \begin{subfigure}[b]{0.24\textwidth}
    \centering
    \includegraphics[width=\textwidth]{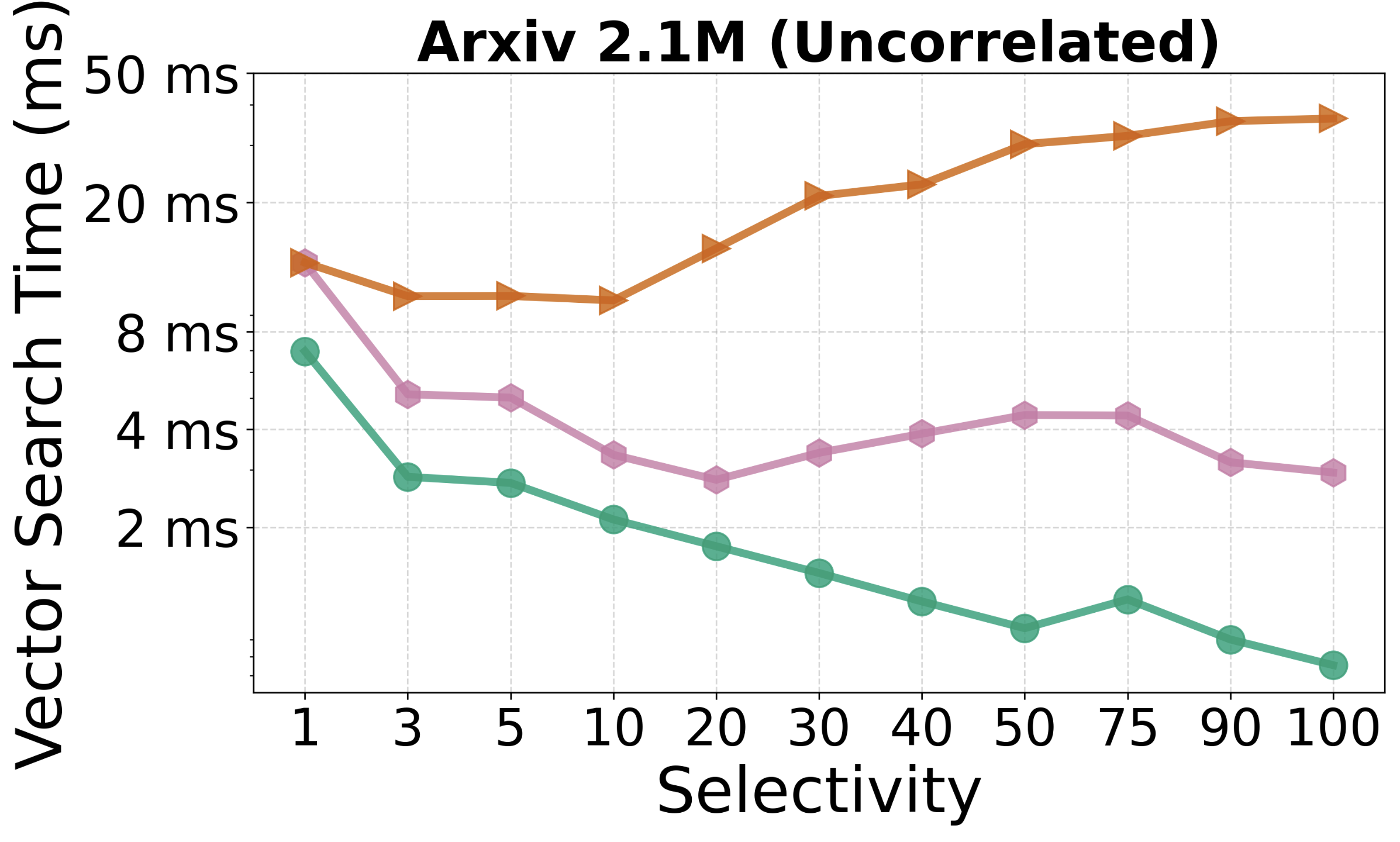}
      \end{subfigure}
  \else
   \begin{subfigure}[b]{0.24\textwidth}
    \centering
    \includegraphics[width=\textwidth]{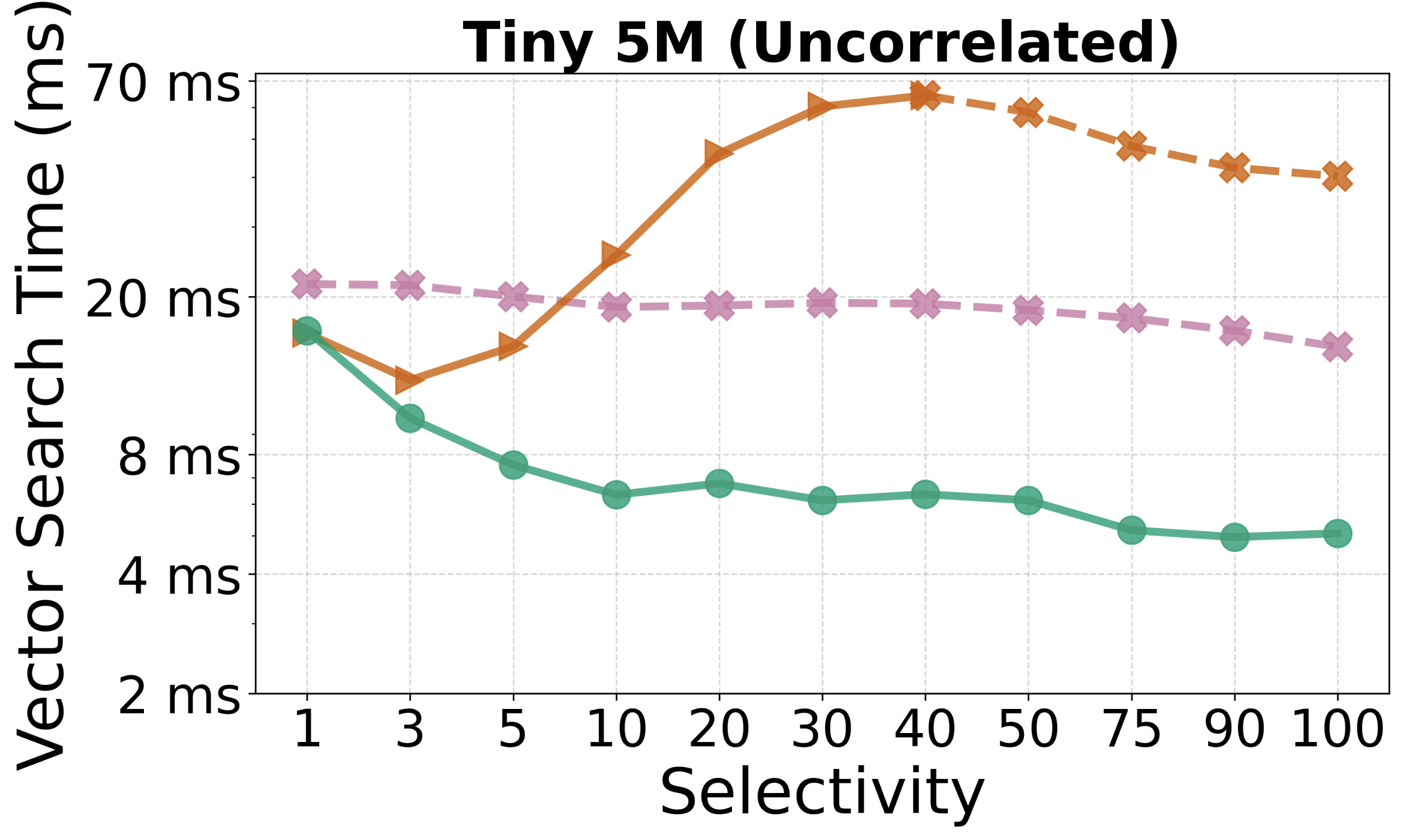}

  \end{subfigure}
      \fi
              \vspace{-10pt}
  \caption{Vector Search Time vs Selectivity for ACORN and Faiss-Navix baselines at 95\% recall}
                \vspace{-10pt}
  \label{fig:prefiltering_faiss_baselines}
\end{figure*}

\begin{figure*}[t!]
  \captionsetup{margin=0pt,skip=0pt}
  \centering
  \begin{minipage}[t]{0.49\textwidth}
    \centering
    \begin{minipage}[b]{0.5\textwidth}
      \centering
      \includegraphics[width=\textwidth]{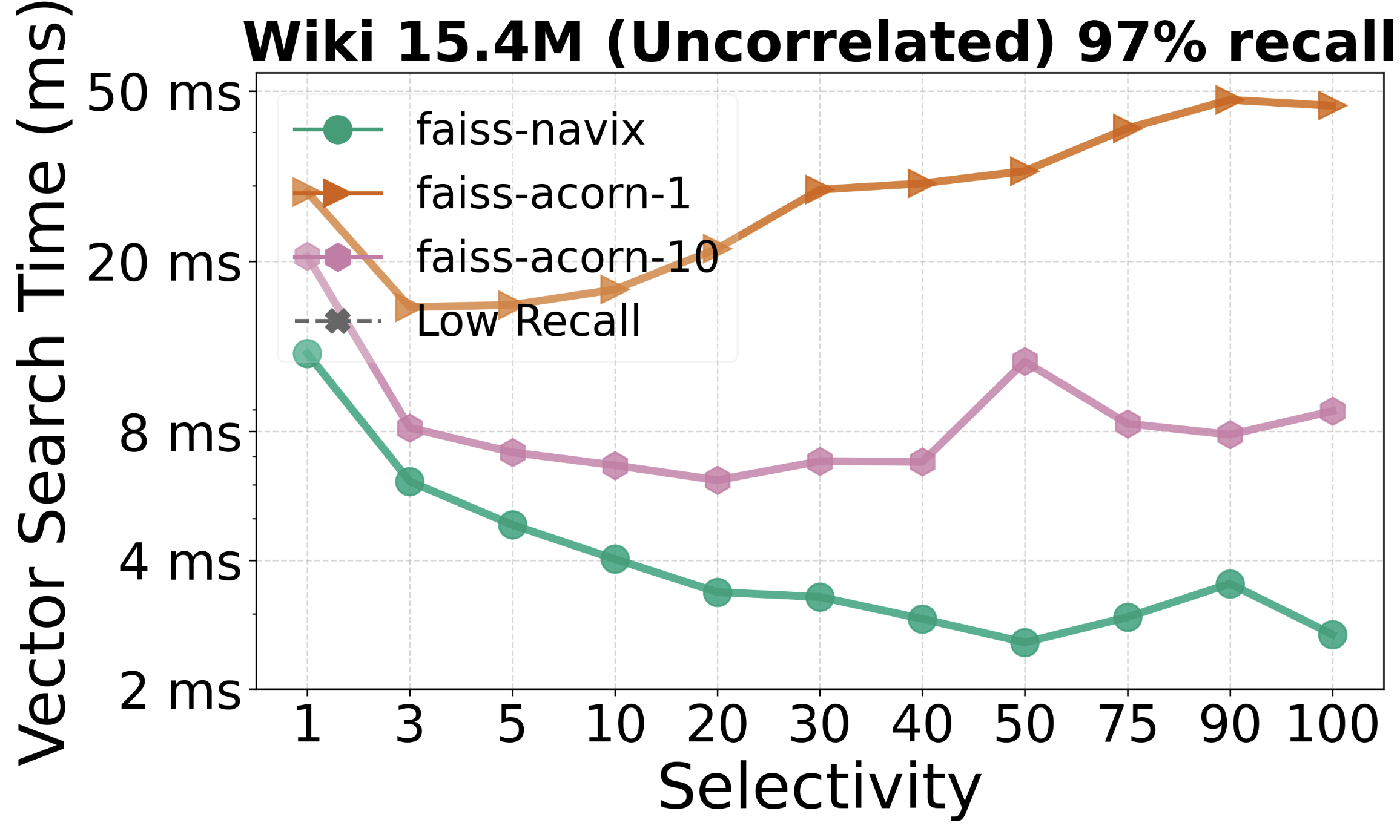}
    \end{minipage}\hfill
    \begin{minipage}[b]{0.5\textwidth}
      \centering
      \includegraphics[width=\textwidth]{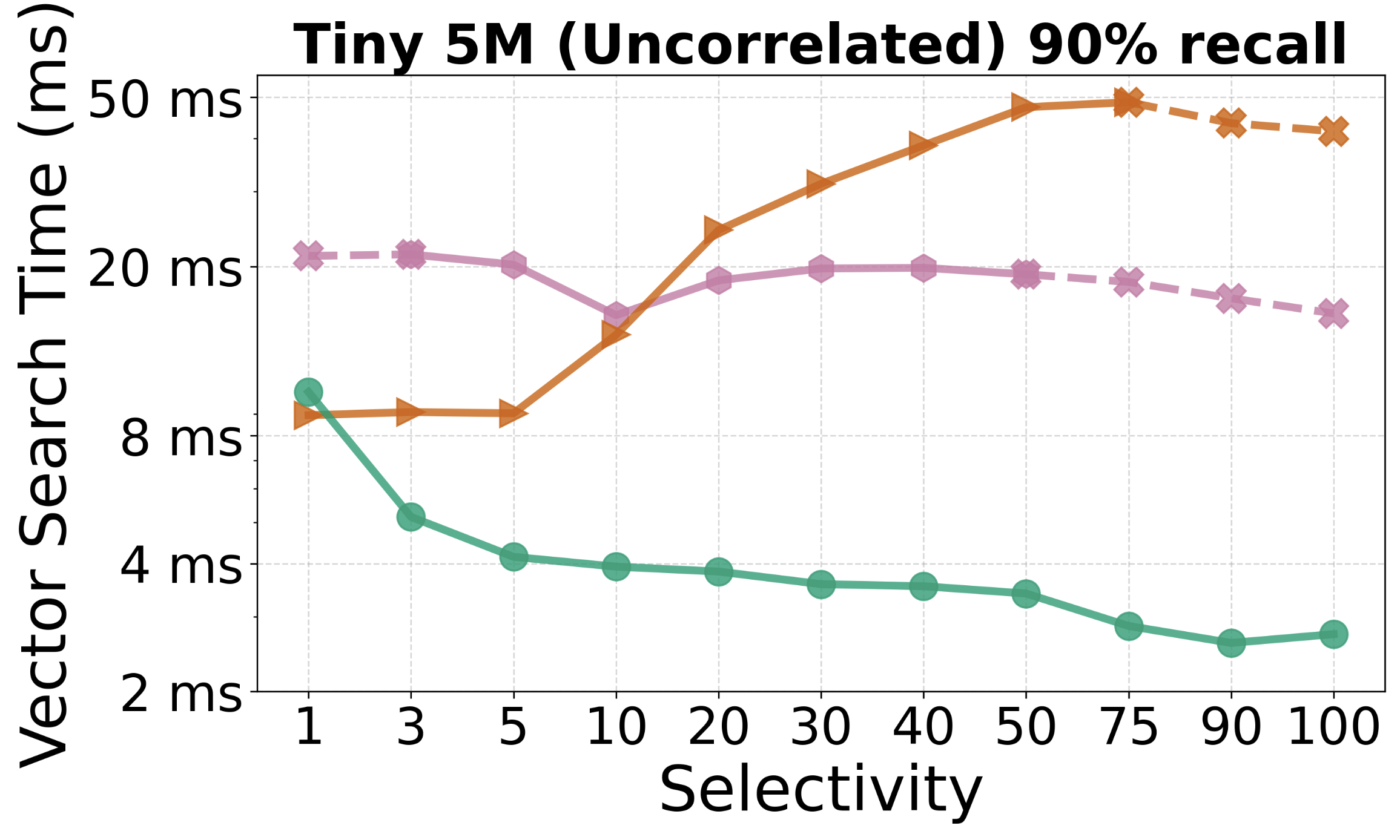}
    \end{minipage}\hfill
    \captionof{figure}{Vector Search Time vs Selectivity for ACORN and Faiss-Navix at different recalls.}
                  \vspace{-10pt}
    \label{fig:faiss_vs_diff_recall}
  \end{minipage}\hfill
  \begin{minipage}[t]{0.49\textwidth}
      \begin{minipage}[b]{0.5\textwidth}
      \centering
      \includegraphics[width=\textwidth]{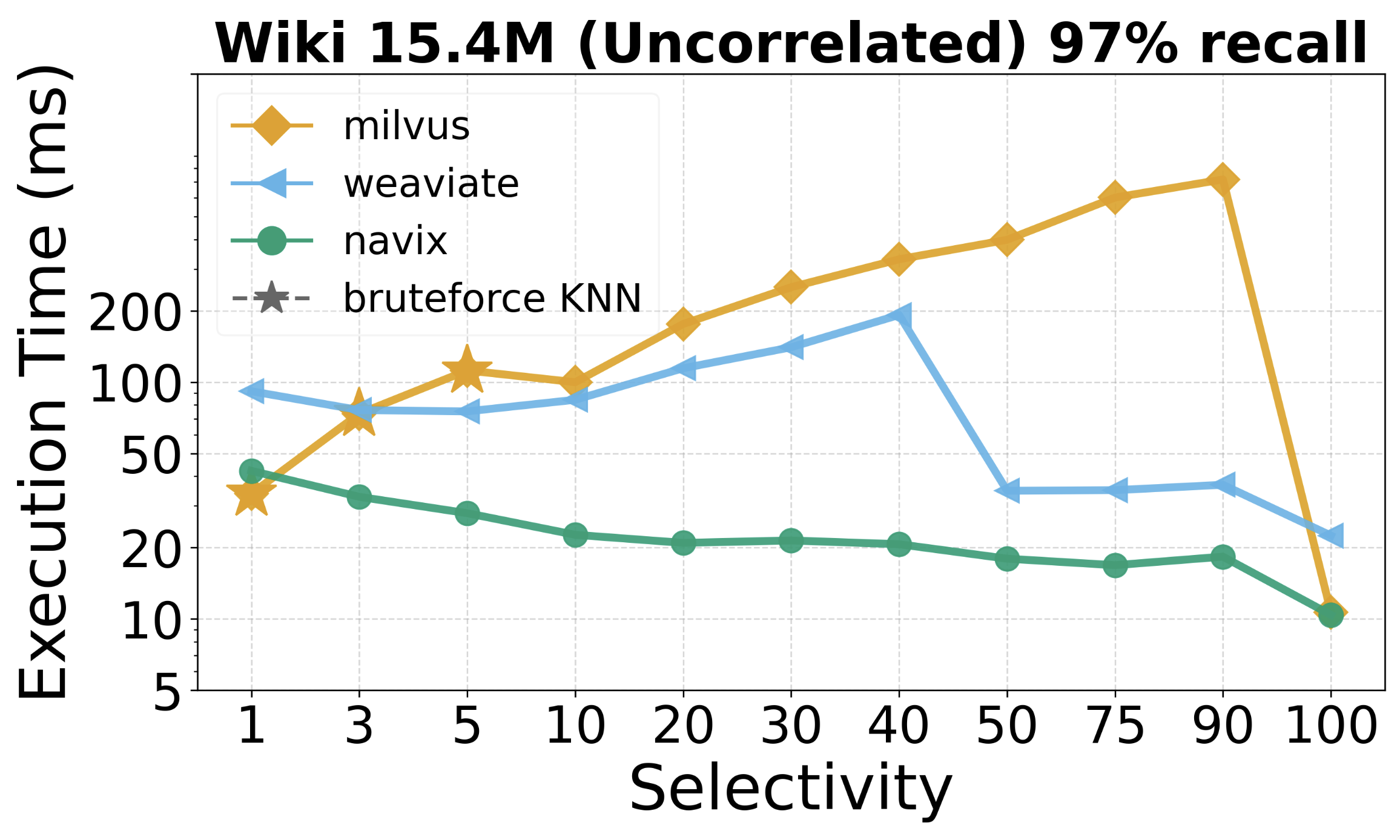}
    \end{minipage}
        \begin{minipage}[b]{0.5\textwidth}
      \centering
      \includegraphics[width=\textwidth]{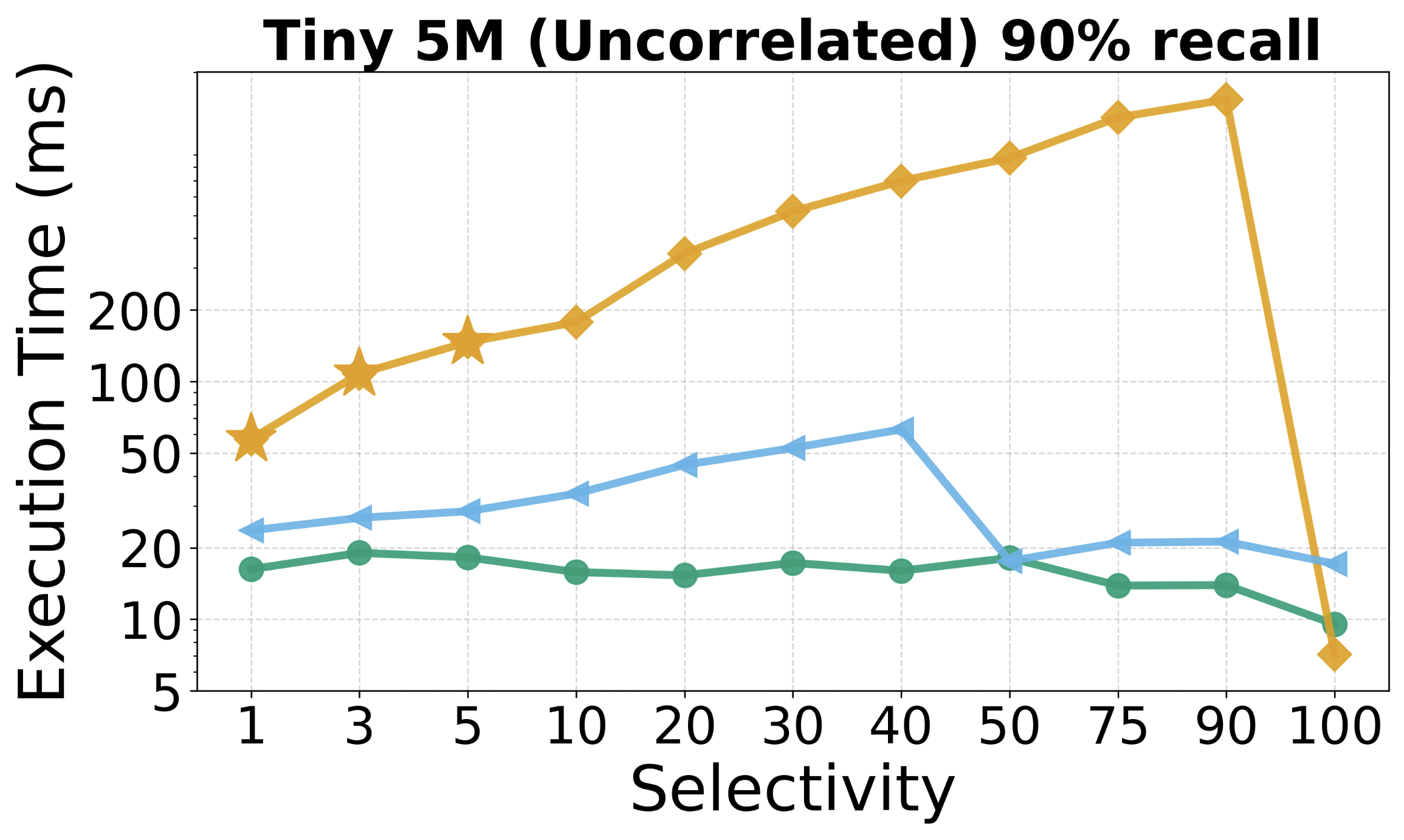}
    \end{minipage}
                  \vspace{-10pt}
    \captionof{figure}{Execution Time vs Selectivity for Weaviate and Milvus baselines at different recalls.}
                  \vspace{-10pt}
    \label{fig:prefilter_diff_recall}
  \end{minipage}
      \vspace{-1pt}
\end{figure*}

\begin{table}[h]
    \centering
    \resizebox{0.47\textwidth}{!}{%
    \begin{tabular}{|c|*{5}{c|}*{5}{c|}}
    \hline
      & \multicolumn{5}{c|}{\textbf{Wiki Uncorrelated}} & \multicolumn{5}{c|}{\textbf{Wiki Negatively Correlated}} \\ \hline
        \textbf{Selectivity} 
            & \textbf{90\%} & \textbf{50\%} 
            & \textbf{30\%} & \textbf{10\%} &\textbf{1\%} & \textbf{22.9\%} & \textbf{15\%} & \textbf{9.9\%} 
            & \textbf{5\%} & \textbf{1\%} \\ \hline
        Prefiltering
            & 11.28 & 12.03 & 11.78 & 10.64 & 10.23 & 32.15 & 26.27 & 21.99 & 18.72 & 11.76 \\ \hline
        Vector Search
            & 5.82 & 5.09 & 7.20 & 9.36 & 22.19 & 33.71 & 33.25 & 39.23 & 40.67 & 53.94 \\ \hline
        Prefiltering \%
            & 65\% & 70\% & 62\% & 53\% & 31\% & 48\% & 44\% & 35\% & 31\% & 17\% \\ \hline

    \end{tabular}%
    }
    \caption{Vector search vs Prefiltering (ms) for Navix.}
    \vspace{-25pt}
    \label{tab:vs_prefilterint_time_uncorrelated}
\end{table}

\subsubsection{Prefiltering vs vector search time:}
\label{subsubsec:prefiltering-vs-vector-search}
The runtime numbers we presented in Figure~\ref{fig:navix_vs_adaptive_g}
focused only on vector search time, since the time spent on prefiltering
is same across \adaptiveg\ and \Indexname. 
We next perform a drill-down analyses into how much time is spent 
by \Indexname\ on prefiltering vs vector search using our largest dataset Wiki. Recall 
from Section~\ref{subsec:query-workloads} that our uncorrelated workloads 
have a simple selection sub-query that puts a range filter on the IDs of embedded objects. 
In contrast, Wiki correlated workloads have selection sub-queries that contain 1-hop joins from a subset of nodes. This can be more expensive,
especially if the selectivities are higher, so the join processes 
a large number of nodes. 
Table~\ref{tab:vs_prefilterint_time_uncorrelated} presents the times spent by \Indexname\ on prefiltering and vector search. We see that 
on Wiki uncorrelated, the prefiltering time is relatively
constant and its contribution to the total execution time decreases
as selectivities decrease and vector search becomes more expensive.
In contrast, we see that for Wiki negatively correlated, the prefiltering times increase as selectivities increase. This is because the time required to perform the join
gets more expensive. However, the contribution of prefiltering to the total time (last row in the table) in both cases decrease as selectivities decrease, and vector search becomes more challenging.



\vspace{-5pt}
\subsection{Weaviate and Milvus}
\label{subsec:weaviate_milvus}
We next compare \Indexname\ against Milvus and Weaviate, which are our disk-based prefiltering
 baselines.  These systems do not support joins, so we compare their behavior  only on
 uncorrelated workloads. Figure~\ref{fig:prefiltering_baselines} shows our results.
Weaviate adapts between
two search heuristics. Above 40\% threshold they use the \onehopa\ heuristic which explores all selected and unselected nodes.
Below and at 40\%, they use the simpler version of \blind\ from ACORN~\cite{acorn} that 
is configured
to explore between 32 and 8$\times$32 many vectors in the 2nd degree. This explains the latency
hikes at 40\%. 
Milvus is overall the slowest system in our experiments. Milvus does \onehopa\ heuristic at 100\% and \bruteforce\ below and at 5\% selectivity, where we start observing latency drops. It performs postfiltering at the higher selectivities and \onehopa\ based prefiltering at the medium to lower selectivities \cite{milvus_paper}. 
Observe further that \Indexname\ outperforms
both of these systems. Specifically, when Weaviate switches to \blind\ heuristic at 40\% selectivity, the difference is quite drastic, 18.28 ms for \Indexname\ compared to 164.19 ms for Weaviate on Wiki dataset. We attribute these performance differences 
partly due to \Indexname's better 
choice of search heuristics and partly to its fast zero-copy distance computations. 
These experiments indicate that vector indices inside DBMSs can be competitive with specialized vector databases.

We further repeated these experiments to verify that our
observations remain the same in other
target recall rates. For Wiki and Arxiv, each system
we used already achieves our target rate of 95\% in the lowest possible efs across most selectivity levels. We therefore increased our target rate to 97\%. For GIST and Tiny, we set a target
rate of 90\%. 
\iflong
Figure~\ref{fig:prefilter_diff_recall} shows our results for Wiki-uncorrelated and Tiny. The remaining results are shown in Appendix~\ref{appsubsec:weaviate_different_recall}.
\else
Figure~\ref{fig:prefilter_diff_recall} shows our results for Wiki-uncorrelated and Tiny. The longer version of our paper shows the rest of our results.
\fi
Observe that the performance patterns of Navix,
Milvus, and Weaviate are similar in these datasets as in Figure~\ref{fig:prefiltering_baselines}.

\subsection{ACORN}
\label{subsec:acorn-ex}
We next compare \Indexname\ against ACORN~\cite{acorn},
which also supports predicate-agnostic vector search on an HNSW-like index. 
Regardless of the selectivity level, 
ACORN implements a variant of the \blind\ heuristic on top of
the in-memory FAISS HNSW index~\cite{faiss}.
ACORN is also configured with a parameter $\delta$, which
changes the original
index by removing (for $\delta$=$1$) or modifying (for $\delta$ $>$ $1$) the pruning routine during index construction, which makes the graph denser for any fixed $M$ value.
Recall that \Indexname\ does not change the underlying HNSW index.

Because ACORN is implemented on top of FAISS, we also implemented \adaptivel\ on top of FAISS HNSW.
We call this version of \Indexname\ as {\em FAISS-Navix}.
We compare FAISS-Navix against ACORN with $\delta$=$1$ and $\delta$=$10$ using all of our workloads.
For $\delta$=$1$, we used $M$=$64$ instead of $M$=$32$ because
$M$=$32$ setting was not able to reach our recall rate. 
This is an important observation: when $\delta$=$1$,
ACORN skips the pruning step, which leads to an index
with actually more edges than Faiss-Navix ($\sim$1.2x).
Therefore the removal of pruning, despite making the HNSW
graph denser in fact degrades recall. This indicates
the pruning routine is essential for creating diverse edges 
within proximity graphs. Table~\ref{tab:indexing_time} shows the indexing times of ACORN and FAISS-\Indexname. 
Observe that ACORN-10 takes significantly more time to build the index as it builds a very dense index. For example, for the Wiki dataset, ACORN-10 takes 3.87x more time compared to Faiss-Navix (162.66 mins vs 42.05 mins).

Figure~\ref{fig:prefiltering_faiss_baselines} shows 
our results on Wiki and 
\iflong
Arxiv, and 
\else
Tiny, and  
\fi
\iflong
Figure~\ref{fig:acorn-irangegraph} shows 
our results on GIST and Tiny.
\else
our longer version of paper~\cite{longerpapernavix}
shows our results on GIST and Arxiv.
\fi 
Our first observation is that
FAISS-Navix consistently outperforms both ACORN configurations
and ACORN-10 consistently outperforms ACORN-1. The latter 
observation is consistent 
with the observations in reference~\cite{acorn}. 
Recall that our previous experiments showed that \blind\ is a good choice
in very low selectivity levels but
is suboptimal in medium and high selectivity levels.
Since ACORN always uses \blind, 
we see that FAISS-Navix outperforms ACORN with bigger margins
at higher selectivity levels.
For lower selectivity levels, we attribute FAISS-Navix's superior
performance to two advantages it has
over ACORN: (i) recall from Section~\ref{subsec:fixed-design-space} that FAISS-Navix's \blind\ is a strict improvement over ACORN's blind; and (ii) FAISS-Navix uses the local selectivities to adaptively select heuristics other than \blind.

Our second observation is that Faiss-Navix generally outperforms ACORN with larger factors in correlated cases.
For example, at around 20-22\% selectivity, while
Faiss-Navix outperforms Faiss-ACORN-10 and Faiss-ACORN-1, respectively,
by factors 1.65x (2.7ms vs 4.44ms) and 5.45x (2.7ms vs 14.72ms),
in the negatively correlated case, the differences
are 5.31x (10.61ms vs 56.41ms) and 29.29x (10.61ms vs 310.8ms). 
We also attribute this to Navix's more advanced search heuristics.

Similar to our experiments with Weaviate and Milvus, we further repeated these experiments
at different recall rates. 
\iflong
Figure~\ref{fig:faiss_vs_diff_recall} shows our results for Wiki-uncorrelated and Tiny. The remaining results are shown in Appendix~\ref{appsubsec:acorn_different_recall}.
\else
Figure~\ref{fig:faiss_vs_diff_recall} shows our results for Wiki-uncorrelated and Tiny. The longer version of our paper shows the rest of our results.
\fi
Observe that the performance patterns of FAISS-Navix, ACORN-1, and ACORN-10 are similar in these datasets as in Figure~\ref{fig:prefiltering_faiss_baselines}
\iflong
 and \ref{fig:acorn-irangegraph}
\fi
.

Finally, our FAISS-Navix experiments give a sense of the performance difference between implementing
Navix inside a DBMS, where accesses to both neighborhoods of vectors and the actual vectors happen through the buffer manager, vs a completely in-memory HNSW implementation.
Specifically, we can compare the Wiki runtimes in Figure~\ref{fig:navix_vs_adaptive_g}
and~\ref{fig:prefiltering_faiss_baselines}, which use the same workloads. 
For example, on the Wiki-uncorrelated benchmark,
the runtimes of Navix in Kuzu change between 4.5-22.19ms, while in FAISS, the runtimes are between 2.02-7.74ms.

\iflong
\begin{figure*}[t!]
  \centering
    \begin{subfigure}[b]{0.23\textwidth}
    \centering
    \includegraphics[width=\textwidth]{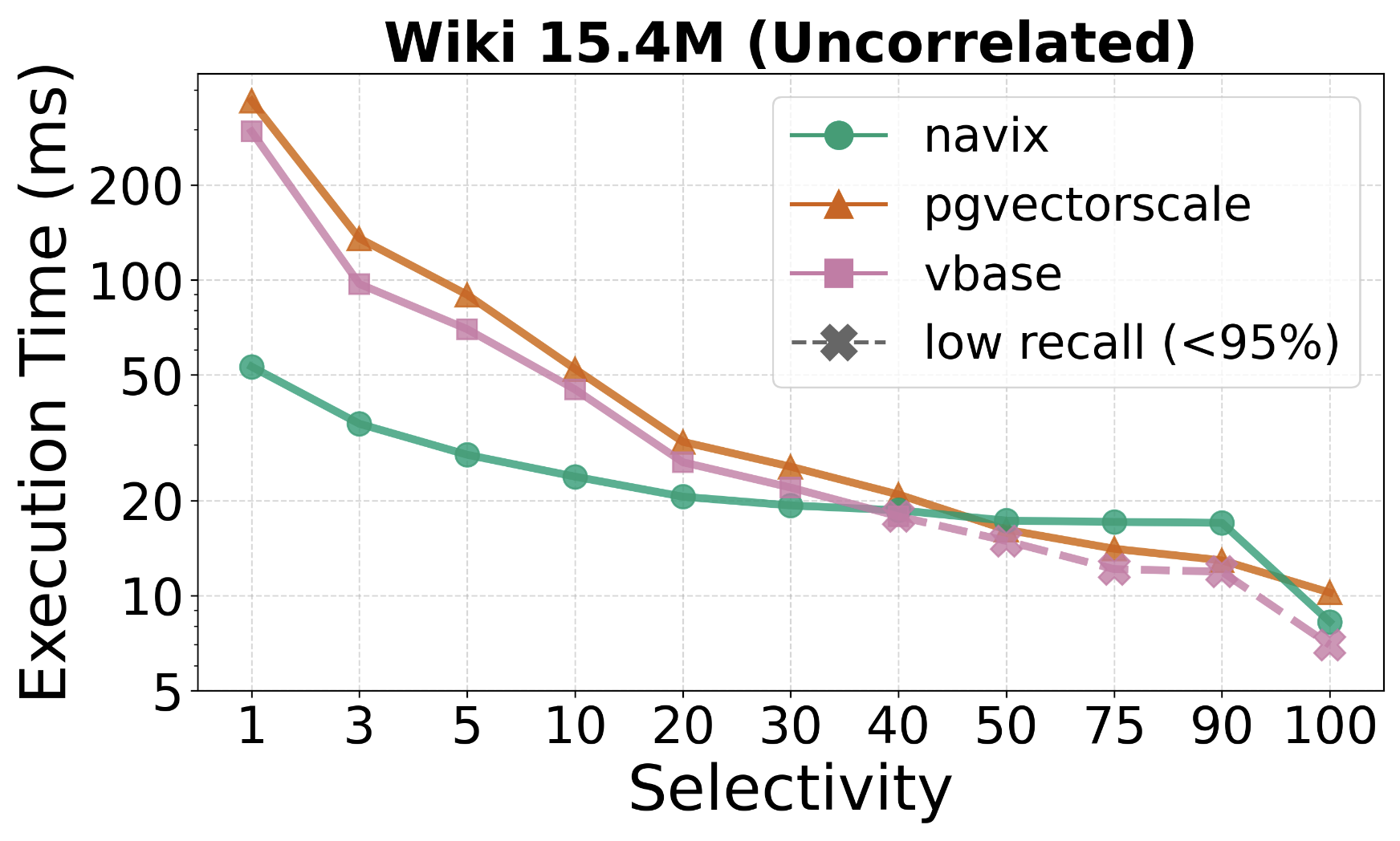}
  \end{subfigure}
  \begin{subfigure}[b]{0.23\textwidth}
    \centering
    \includegraphics[width=\textwidth]{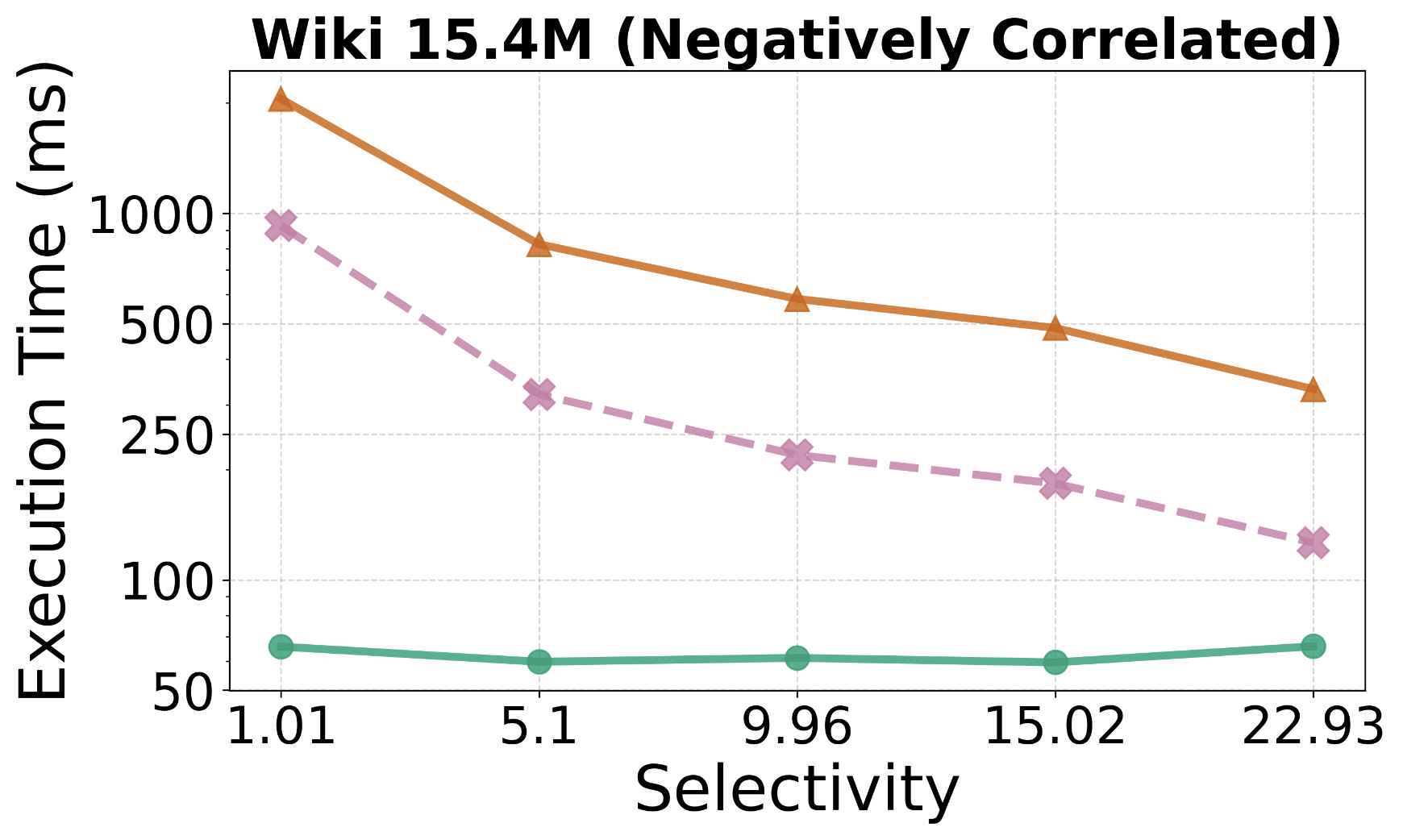}
  \end{subfigure}
   \begin{subfigure}[b]{0.24\textwidth}
    \centering
    \includegraphics[width=\textwidth]{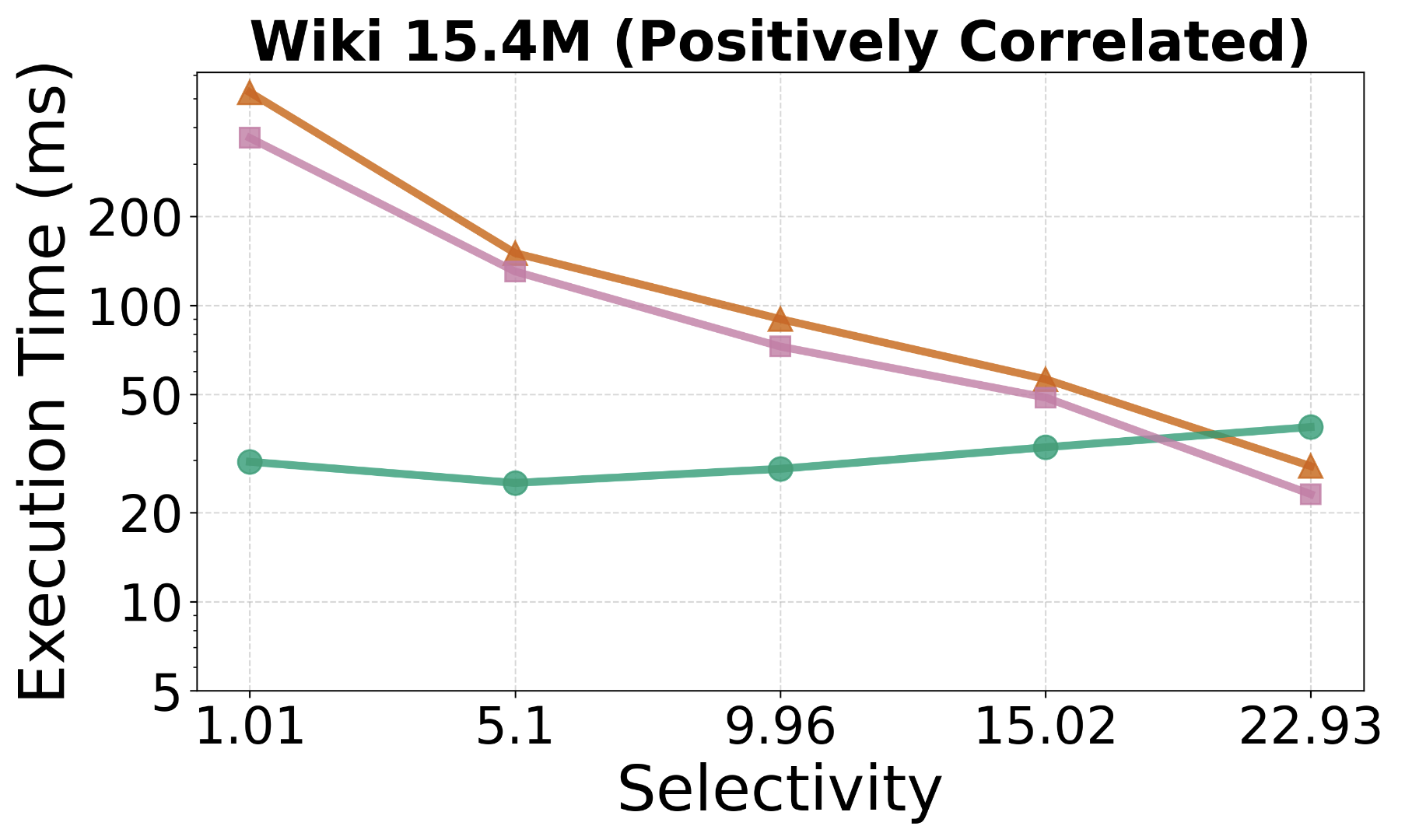}
  \end{subfigure}
   \begin{subfigure}[b]{0.24\textwidth}
    \centering
    \includegraphics[width=\textwidth]{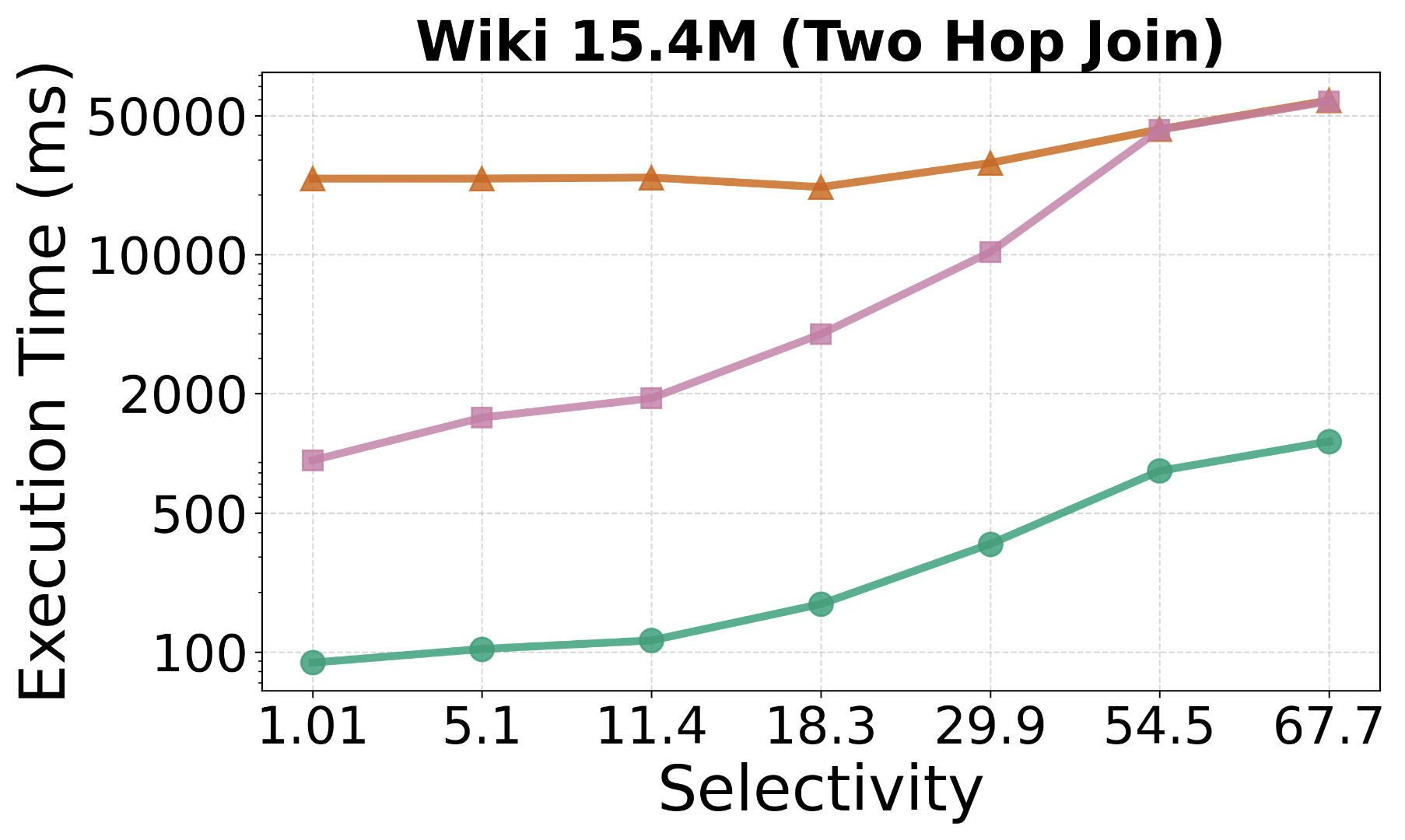}
  \end{subfigure}
                \vspace{-10pt}
  \caption{Execution Time vs Selectivity for postfiltering baselines with >95\% recall and within 1\% of each other}
                \vspace{-10pt}
  \label{fig:postfilter_baselines}
\end{figure*}

\begin{figure}[h]
  \centering
    \begin{subfigure}[b]{0.23\textwidth}
    \centering
    \includegraphics[width=\textwidth]{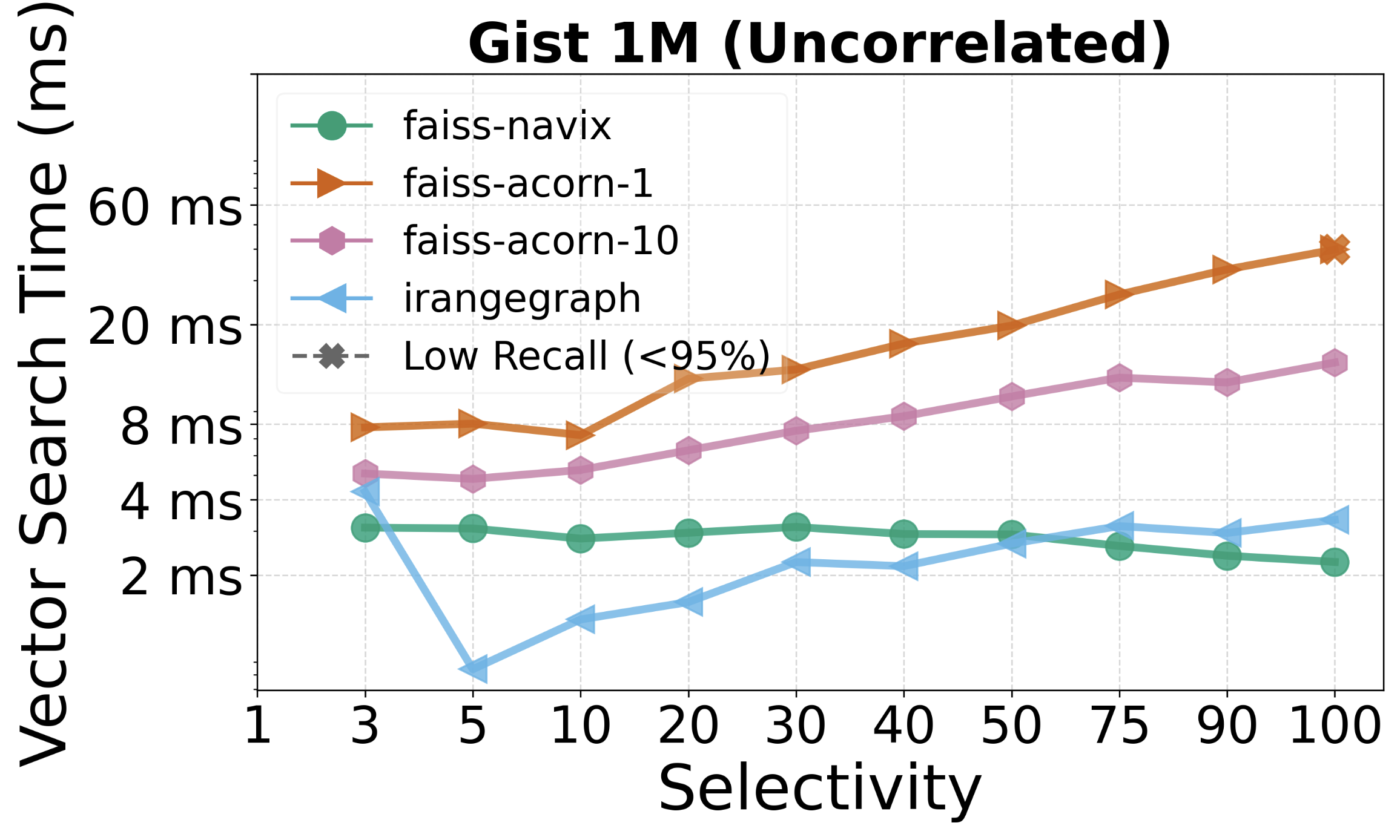}
  \end{subfigure}
  \begin{subfigure}[b]{0.23\textwidth}
    \centering
    \includegraphics[width=\textwidth]{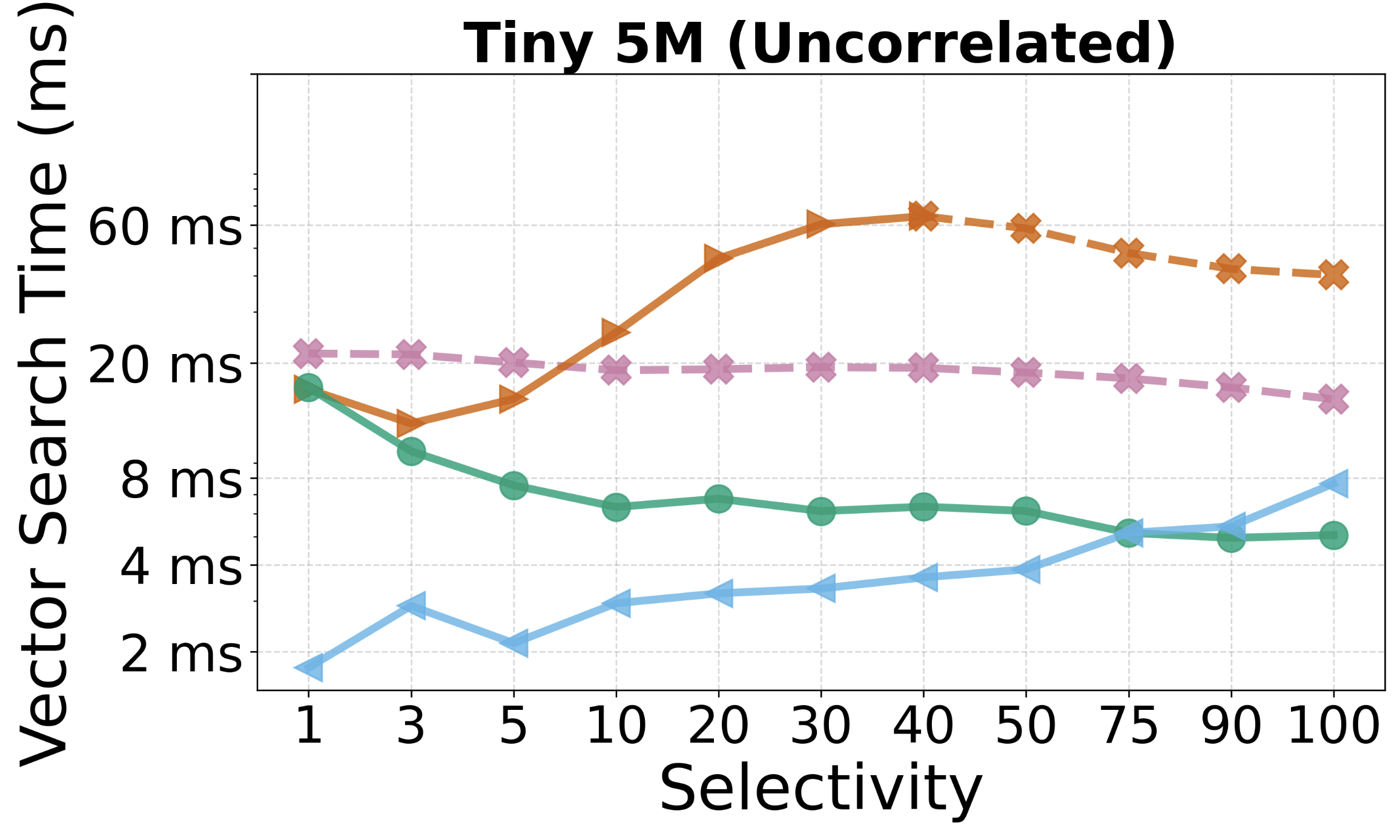}
  \end{subfigure}
                \vspace{-10pt}
  \caption{ACORN and iRangeGraph benchmarks at 95\% recall.}
                \vspace{-10pt}
  \label{fig:acorn-irangegraph}
  \vspace{-5pt}
\end{figure}
\fi

\iflong
\subsection{iRangeGraph}
\label{subsec:irange_benchmark}
In our next set of experiments, we compare
\Indexname\ against iRangeGraph. 
iRangeGraph is a specialized vector implementation that supports a limited set of 
selectivity queries.
As such, iRangeGraph is not predicate-agnostic but can be more optimized than \Indexname\
on the range filtering.
Specifically, iRangeGraph constructs segmented proximity graphs as sub-indices for different ranges and merges them during search time to support arbitrary ranges.

Since iRangeGraph is implemented using
the in-memory HNSWLib~\cite{hnswlib}, 
we used FAISS-Navix as a comparison point. Since iRangeGraph supports range selection sub-queries
and our uncorrelated workloads are based on range filters, we repeated
our experiments on uncorrelated workloads. iRangeGraph
only supports L2 distance, so we used GIST and Tiny workloads, which use L2 metric. 
For reference, Table~\ref{tab:indexing_time} shows the indexing time of iRangeGraph on these datasets,
which is 13.3x and 10.8x slower than FAISS-Navix. Part of this is also due to the fact
iRangeGraph creates a larger index. Specifically,
on GIST, the size of the iRangeGraph index is 1.2 GB, 
while FAISS-Navix's size is 0.3 GB excluding the size of actual vectors. On Tiny,
the index sizes of iRangeGraph and FAISS-Navix are, respectively, 6.8 GB and 1.3 GB. 

Figure~\ref{fig:acorn-irangegraph} shows our results. 
Overall, iRangeGraph is more performant
than Faiss-Navix at low selectivity
levels. This is expected since iRangeGraph efficiently prunes sub-indices to search only those with overlapping ranges, which reduces the number of smaller sub-indices that need to be searched as selectivity decreases. Therefore,
unlike all baselines, whose performance degrades 
as selectivities decrease, iRangeGraph's performance gets better. FAISS-Navix's performance becomes more competitive as selectivities increase and at very high selectivity levels, as expected, iRangeGraph loses its performance advantage over FAISS-Navix.
\fi

\iflong
\subsection{PGVectorScale and VBase}
\label{subsubsec:postfiltering_baselines}
In this section, we compare \Indexname\ against postfiltering-based systems. 
We begin by studying the behavior of these systems in Wiki uncorrelated 
workload. Our results are shown in Figure~\ref{fig:postfilter_baselines}. Our observation for this experiment is similar on the Arxiv workload (see Appendix~\ref{app:arxiv_post}). However, for Tiny and GIST, both postfiltering systems were unable to reach 95\% target recall. PGVectorScale achieves 39.65\% and 39.39\%, whereas VBase achieves 74.09\% and 81.26\% for Tiny and GIST workloads respectively. Therefore, we omit these experiments.
We can break the cost of postfiltering approaches into three
main components:
\begin{squishedlist}
    \item {\em Preprocessing cost}, if any, to prepare any intermediate tables that are needed to
    check if a streamed vector is in $S$ or not.
    \item {\em Vector search cost} of streaming nearest neighbors from closest to furthest to $k$. This is determined by the likelihood of vectors in the nearest neighbors of $v_Q$ being in $S$. The major factor that determines this is the selectivity level.
    \item {\em Verification cost} of checking if a streamed vector
    is in $S$ or not.
\end{squishedlist}

In our uncorrelated workloads, observe that 
at high-selectivity levels, postfiltering approaches have an advantage over prefiltering appraoches. This is the ideal case for postfiltering:
there is no preprocessing cost and the verification cost is cheap, a lookup to run a simple predicate on the object ID of the vector tuple.
At these selectivity levels, prefiltering approaches, specifically
\Indexname\, pays the upfront cost of running the predicate on all
vectors. Therefore postfiltering systems outperform \Indexname\ above 50\% selectivity.

As the selectivity decreases, vector search cost increases, as
more tuples need to be streamed, so cumulative verification cost
also increases. Therefore the performance of postfiltering systems degrades significantly. For example, PGVectorScale's latency
numbers degrade from about 10.22 ms to 334.42 ms. A similar
degradation also appears in prefiltering approach, e.g., \Indexname\, but to a lesser
extent because vector search is aware of filtering which helps it to only consider selected nodes to converge faster.
Thus, \Indexname\ latency only degrades from 8.26 ms to 53.14 ms.

We next used our correlated workloads. Our results are shown
in Figure~\ref{fig:postfilter_baselines}.
Although our correlated queries have joins of Person and Chunks, 
PGVectorScale and VBase plan still do not perform any preprocessing.
Their plans stream a Chunk tuple \texttt{v(cID)} from a vector search operator. Then an IndexNestedLoopJoin (INLJ) operator joins
$v$ with PersonChunk on \texttt{cID}. If a tuple \texttt{pc(pID,cID)} is found,
a predicate on the \texttt{pID} is executed. 
The primary difference
is that now there is a more expensive verification cost. However,
the general patterns exist. As expected, queries are broadly
easier on positively correlated scenarios than negatively correlated
scenarios for all systems, i.e., latencies generally go down. 
For example, PGVectorScale's runtimes in the 1\% to 22.9\% range
are 331.38 ms to 2052.44 ms and 28.67 ms to 524.54 ms respectively,
in the negatively and positively correlated cases.
\Indexname\ is more robust to selectivity changes and its latency numbers are between 
71.86 ms to 65.7 ms and 38.86 ms to 29.66 ms. 

We also extended our experiments with an additional 2-hop join benchmark
on Wiki that represents a graph RAG application \cite{linkedinGraphRag,langchainGraphRag} use case. In this workload selection subqueries are of the following structure:
\begin{lstlisting}[language=SQL,style=mystyleSQL]
MATCH (p:persons)-[pr:PersonResource]->(r:Resource)
      (r)-[rc:ResourceChun k]->(c:Chunk)
WHERE p.birth_date >= {s_date} AND p.birth_date < {e_date}
PROJECT GRAPH S(c)
CALL QUERY_HNSW_INDEX(S, HNSWIndex, v$_Q$, k) RETURN c.cID
\end{lstlisting}
To write an equivalent SQL version, we create a temporary table
using SQL's WITH statement. Unlike our previous experiments, PGVectorScale and VBase generate plans that have preprocessing
costs. Specifically, they first materialize the 2-hop join
in a temporary table \texttt{Tmp}. Then, for each streamed vector $v$, they 
check if $v$ joins with any tuple in \texttt{Tmp} in a NestedLoopJoin operator. For vector queries, we used our previous queries in the correlated workloads. Our results are shown in Figure~\ref{fig:prefiltering_baselines}. 
First, these queries are harder than prior queries because the
selection subquery is harder. Therefore each system's performance
degrades. Also, observe that every system's curve is now an upward
curve. This is because as selectivity levels increase, even though vector search gets faster, the preprocessing cost for both post and
prefiltering systems increases. This is because the size of the 2-hop
join that needs to be computed increases, which offsets
the costs. 




\subsubsection{Vector Search with Filtering and two-hop traversal}
This benchmark evaluates vector search performance with multi-hop joins and date range filtering. Multi-hop joins, in this case, 2-hop traversal are particularly useful in graph RAG applications. This benchmark also helps identify how postfiltering and prefiltering systems behave in complex query scenarios. Again, we will use the date range filtering to reduce the selectivity of this query.
\begin{lstlisting}[language=SQL,style=mystyleSQL]
-- CYPHER
MATCH (p:persons)-[r]->(:resources)-[m]->(e:embeddings)
WHERE p.birth_date >= {start_date} AND p.birth_date < {end_date}
CALL ANN_SEARCH(e.embedding, {vector}, 100, {efsearch}) 
RETURN e.chunk_id;


-- SQL
with t1 as (
	SELECT r.wiki_id AS r_wiki_id
	FROM persons AS p
	JOIN person_resources AS pr ON p.wiki_id = pr.person_wiki_id
	JOIN resources AS r ON pr.resource_wiki_id = r.wiki_id
	WHERE p.birth_date >= {start_date} AND p.birth_date < {end_date}
	GROUP BY r.wiki_id)
SELECT e.chunk_id
FROM embeddings as e, t1
WHERE e.wiki_id = t1.r_wiki_id
ORDER BY e.embedding <=> '{vector}'
LIMIT 100;
\end{lstlisting}

\fi

\begin{figure}[h]
  \centering
    \begin{subfigure}[b]{0.23\textwidth}
    \centering
    \includegraphics[width=\textwidth]{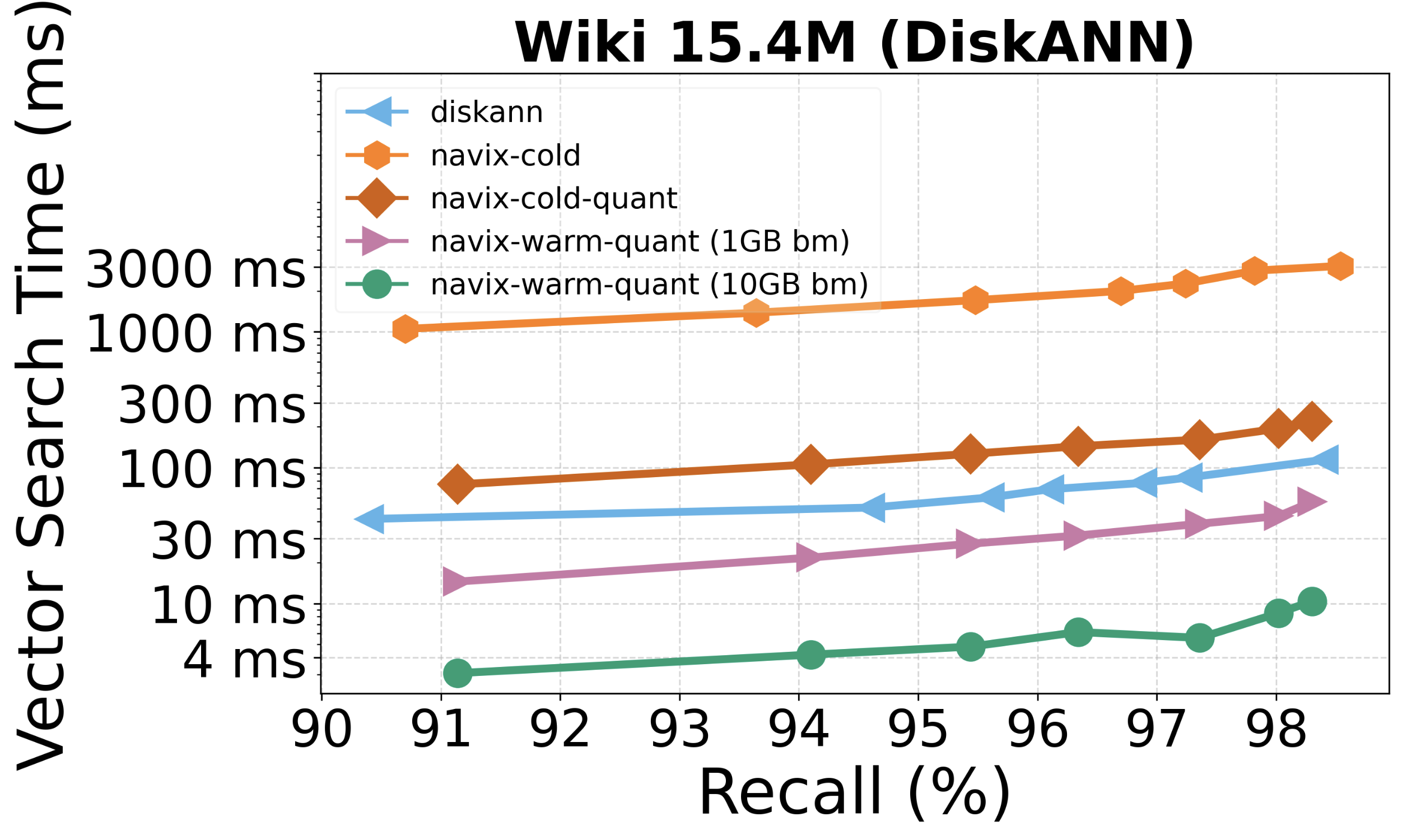}
  \end{subfigure}
  \begin{subfigure}[b]{0.23\textwidth}
    \centering
    \includegraphics[width=\textwidth]{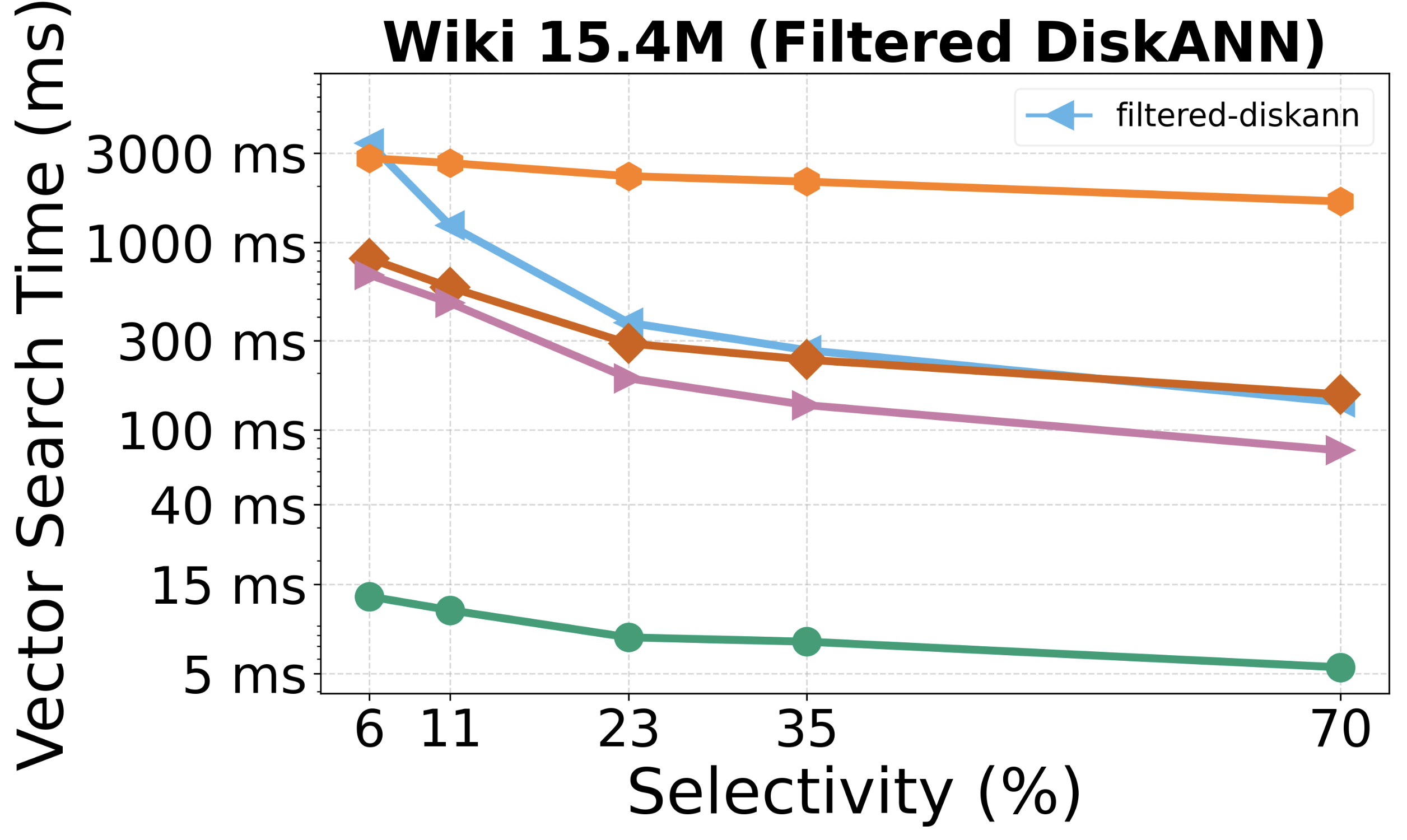}
  \end{subfigure}
                \vspace{-10pt}
  \caption{DiskANN and Filtered-DiskANN benchmarks at 95\% recall.}
                \vspace{-10pt}
  \label{fig:disk-experiments}
\end{figure}

\subsection{DiskANN and Filtered-DiskANN}
\label{subsec:disk}
\subsubsection{DiskANN comparisons}
We next compare \Indexname\ with Disk\-ANN~\cite{diskann} and FilteredDiskANN~\cite{filteredDiskANN}, which are two disk-based 
proximity graph-based vector indices.
Our goal is to evaluate the disk-based performance of \Indexname. 
DiskANN is a proximity graph-based index optimized for search on SSDs. It stores each vector and its corresponding adjacency list in a single 4KB page while keeping quantized vectors~\cite{ivf_pq}, which are compressed versions of the vectors, in memory. During search, it uses the quantized vectors and performs disk I/O only to read the index graph. 
After search, it reads the actual vectors and reranks them using actual distances.

Since DiskANN and Navix have very different designs, it is difficult to have very controlled experiments. However, to focus on benchmarking the disk performance of Navix,
we ran \Indexname\ in the following configurations:
\begin{squishedlist}
    \item \Indexname-cold: We start up the system and just measure the first cold run of each query. This forces all accesses to both the vectors and neighborhoods to do disk I/Os.
    \item \Indexname-cold-quant: We quantized the vectors and stored them in an in-memory data structure. We modified our search algorithm to read the quantized vectors from this structure instead of Kuzu's buffer manager. This gives us a Kuzu configuration that, similar to DiskANN, performs I/Os only to read adjacency lists. 
    \item \Indexname-warm-quant-1GB: We give a small amount buffer manager space to \Indexname-cold-quant. Before we run each query $q$, we run 1000 random queries, so the adjacency lists are partially cached. We run $q$ only once unlike our previous experiments, because running it multiple times would cache all of the adjacency lists accessed when evaluating $q$ in Kuzu's buffer manager.
    \item \Indexname-warm-quant-10GB: We give enough buffer manager space to Kuzu to cache most or all of the adjacency lists.
\end{squishedlist}
\noindent We expect each Kuzu configuration was to outperform the previous one, with performance improvements revealing how different scan operations affect the purely disk-based runtime.
We used $R=64$, $LBuild=200$ as DiskANN index configurations. 
Since Kuzu currently lacks asynchronous I/O support, we set DiskANN's asynchronous I/O count to 1 as well. Before running each Navix benchmark, we flushed the file system cache. %

Since DiskANN does not support filters, we ran experiments on our largest dataset Wiki
using the uncorrelated workload queries without filters. Our results are shown in Figure~\ref{fig:disk-experiments}.
We observe that scanning of vectors is the major contributor
to the runtime. This is because the largest performance difference is between 
\Indexname-cold and \Indexname-cold-quant. 
Observe that DiskANN, which only performs disk I/O on scanning adjacency lists
outperforms \Indexname-cold by a major factor, between 25.07x and 26.72x. However, when we also cache the vectors in memory in quantized way,
\Kuzu-\Indexname-cold-quant closes the performance gap significantly to between 1.79x and 1.92x.
DiskANN is still more performant than \Indexname-cold-quant,
which we attribute to DiskANN's more optimized I/O path. Specifically, Kuzu requires two random I/Os to read each adjacency list: one to read the metadata i.e. offset and size of its CSRs, and another to read the actual adjacency list. 
In addition, while DiskANN performs direct I/O, Kuzu always scans through the operating system.

We next observe that even with a small amount of buffer manager cache, 
\Kuzu-\Indexname-warm-quant-1GB outperforms DiskANN. This is primarily because we store adjacency lists and vectors separately, thus within a single 4KB page we can pack more adjacency lists and cache more effectively.
Finally, by caching more of the index in
\Indexname-warm-quant-10GB, we can obtain run times 
that match our previous experiments in Figure~\ref{fig:navix_vs_adaptive_g} (e.g., around 5ms at 95\%
recall), which are between 13.9x and 10.9x faster than DiskANN (e.g., DiskANN takes 60.27ms at 95\% recall). These results highlight one advantage of implementing a vector index natively inside a DBMS. Specifically, since DBMSs already have caching mechanisms, the performance of the vector index automatically improves with additional memory resources in the system.

\subsubsection{FilteredDiskANN comparisons}
\label{subsubsec:filtereddiskann}
We next compared the performance of the same \Indexname\ configurations against
FilteredDiskANN. Briefly, FilteredDiskANN is similar to DiskANN except it supports single-label filtering on a low-cardinality label set (up to 5,000 unique labels). It modifies the proximity graph by creating extra edges between nodes with the same label. 

We generated 200 unique random labels and assigned them to each vector in our Wiki dataset with zipf distribution using
FilteredDiskANN's synthetic label generation tool. We fixed 
the recall rate to 95\% as before and used the same index building parameters as in the DiskANN benchmark. FilteredDiskANN
can only answer queries that ask for one label, which limits the selectivities we 
can use. We varied the labels from least selective query, which had a selectivity
of 70\% to 6\%. Our results are shown in Figure~\ref{fig:disk-experiments}.
Our results are similar to our DiskANN results except even 
\Indexname-cold-quant now outperforms FilteredDiskANN. We attribute this to
\Indexname's \adaptivel\ being a more efficient filtered search algorithm
than FilteredDiskANN. The inefficiency of  
FilteredDiskANN has also been observed in previous work~\cite{acorn}. 


\section{Related Work}
\label{sec:rw}
Our work is related to prior work from several areas: \\
\noindent \textbf{Approximate Nearest Neighbours Indices:} Existing 
indices that are used for approximate kNN queries can be broadly categorized into two: (i) clustering-based indices~\cite{clustering_q_index,soarclustering,spannclustering,ivf_pq,lhs}; and (ii) graph-based indices~\cite{nsg,diskann,rnndescent,hnsw}. 
\Indexname\ adopts HNSW, which is a graph-based index.
Broadly, clustering based indices 
partitions the n-dimensional space into different partitions
around a set of centroid vectors. Then given a query vector $v_Q$,
one or more of these partitions are searched to find kNNs of $v_Q$.
IVF \cite{ivf_pq,faiss} is one of the most used clustering-based approaches. Locality sensitive hashing \cite{lhs} is another approach to cluster the data based on hashing.
 In contrast, graph-based indices form a proximity graph
across vectors. 
RNN-Descent \cite{rnndescent}, Vamana \cite{diskann}, and HNSW \cite{hnsw} are graph-based indices which rely on proximity graphs,
where vectors that are close to each other are connected with each other to form a graph. 
Graph-based indices are empirically shown to be superior 
in terms of recall and performance compared to clustering-based approaches~\cite{annbenchmarks}.


\noindent \textbf{Predicate-agnostic Vector Search:} Several prior DBMSs, vector databases, or vector search systems are capable of evaluating predicate-agnostic vector search queries.  These systems adopt pre- or postfiltering or a mix of these approaches. They also adopt graph or clustering-based indices, or a mix of these. We covered Weaviate, Milvus, PGVectorScale, and VBase in prior sections. We cover Acorn and AnalyticDB-V here.

AnalyticDB-V~\cite{analyticaldb} is a distributed analytical RDBMS that
implements a mix of HNSW index and a compressed clustering-based index called
{\em Voronoi Graph Product Quantization} (VGPQ) index. The primary index is VGPQ. The newly inserted vectors are first inserted into an HNSW index and
periodically merged into the VGPQ index. Since the primary index is 
VGPQ, the HNSW search is not the bottleneck in predicate-agnostic search queries. On the VGPQ index, AnalyticDB-V performs 
a mix of post- and prefiltering approach based on the estimated selectivity
of the selection subquery.

\noindent \textbf{Specialized indices for predicates on low cardinality attributes:} Several prior works have developed graph-based indices 
that can evaluate {\em predicate-aware} vector search queries. These approaches assume a set of predicates, or the structures of these predicates, are known apriori and construct indices 
whose edges contain extra edges that can be used to evaluate these
predicates. 
We covered Filtered-DiskANN \cite{filteredDiskANN} in Section~\ref{subsec:disk}. 
NHQ \cite{nhq} and HQANN \cite{hqannlowcard} are approaches that encode the attribute to filter on directly into the vector as a new dimension and use
this during index construction to create extra edges between similar attribute vectors. This method works on small cardinality attributes and on simple filtering operations. 
HQI\cite{hqi} partitions the index based 
on attributes whose cardinality can be at most 20. This allows searching
within a one or a set of partitions to evaluate equality or range queries on
this attribute domain. 

Finally, SeRF\cite{serf} and iRangeGraph\cite{irange}, which we
also compared against
\iflong
in Section~\ref{subsec:irange_benchmark},
\else
in the longer version of our paper~\cite{longerpapernavix},
\fi
support arbitrary range filtering queries on an attribute in the context of graph-based vector indices. At a high level, both approaches construct segmented sub-indices for different ranges and merge them together to support arbitrary ranges.
However, since these approaches require constructing many sub-indices, they are significantly slower during index construction. 

Overall, these specialized indices are generally highly limited in terms
of the predicates or the structures of their selection subqueries. For example, they are not designed to support selection subqueries that includes
join or filters on string-based predicates. 

\noindent \textbf{Native Vector Indices in DBMS:}
We compared \Indexname\ against PGVectorScale and VBase in
\iflong
Section~\ref{subsubsec:postfiltering_baselines}.
\else
the longer version of our paper~\cite{longerpapernavix}.
\fi
Several other
DBMSs or DBMS extensions also support vector indices. PGVector~\cite{pgvector} and PASE~\cite{pase} are vector index extension of PostgresSQL. Finally DuckDB and Neo4j also support vector indices.
 DuckDB uses the USearch~\cite{usearch} library and Neo4j uses Apache Lucene~\cite{lucene}. Unlike our approach, which integrates a native vector index, these systems
use separate indices and do not support predicate agnostic search.



\section{Conclusions}
\label{sec:conclusions}

We presented the design and implementation of \Indexname, 
a native vector index designed for GDBMSs that can evaluate predicate agnostic vector search queries. \Indexname\ leverages
some of the core capabilities of the underlying GDBMS, such as its disk-based adjacency list storage and querying capabilities to compute selected vectors in filtered vector search queries. We also introduced an optimization
to perform distance computation more efficiently inside a DBMS. Our optimization runs the distance computations directly inside the DBMS's buffer manager cache. For filtered vector search,
\Indexname\ adopts the prefiltering approach and uses a novel search heuristic called  \adaptivel, which utilizes
the selectivity information in the selected subset $S$ to decide through
simple rules which exploration heuristic to pick per candidate vector. To capture
possible correlations of $S$ with a query vector $v_Q$, \adaptivel\ 
uses the local selectivity of the neighbors of each candidate vector it explores during search. Our results show that \Indexname\ is efficient
and robust and is competitive with or outperforms both existing prefiltering and postfiltering based systems and heuristics under a variety of query workloads with different selectivities and correlations.



\bibliographystyle{ACM-Reference-Format}
\bibliography{references}
\balance

\iflong
\appendix
\section{Appendix}
\label{sec:appendix}

\begin{figure*}[t]
  \centering
    \begin{subfigure}[b]{0.33\textwidth}
    \centering
    \includegraphics[width=\textwidth]{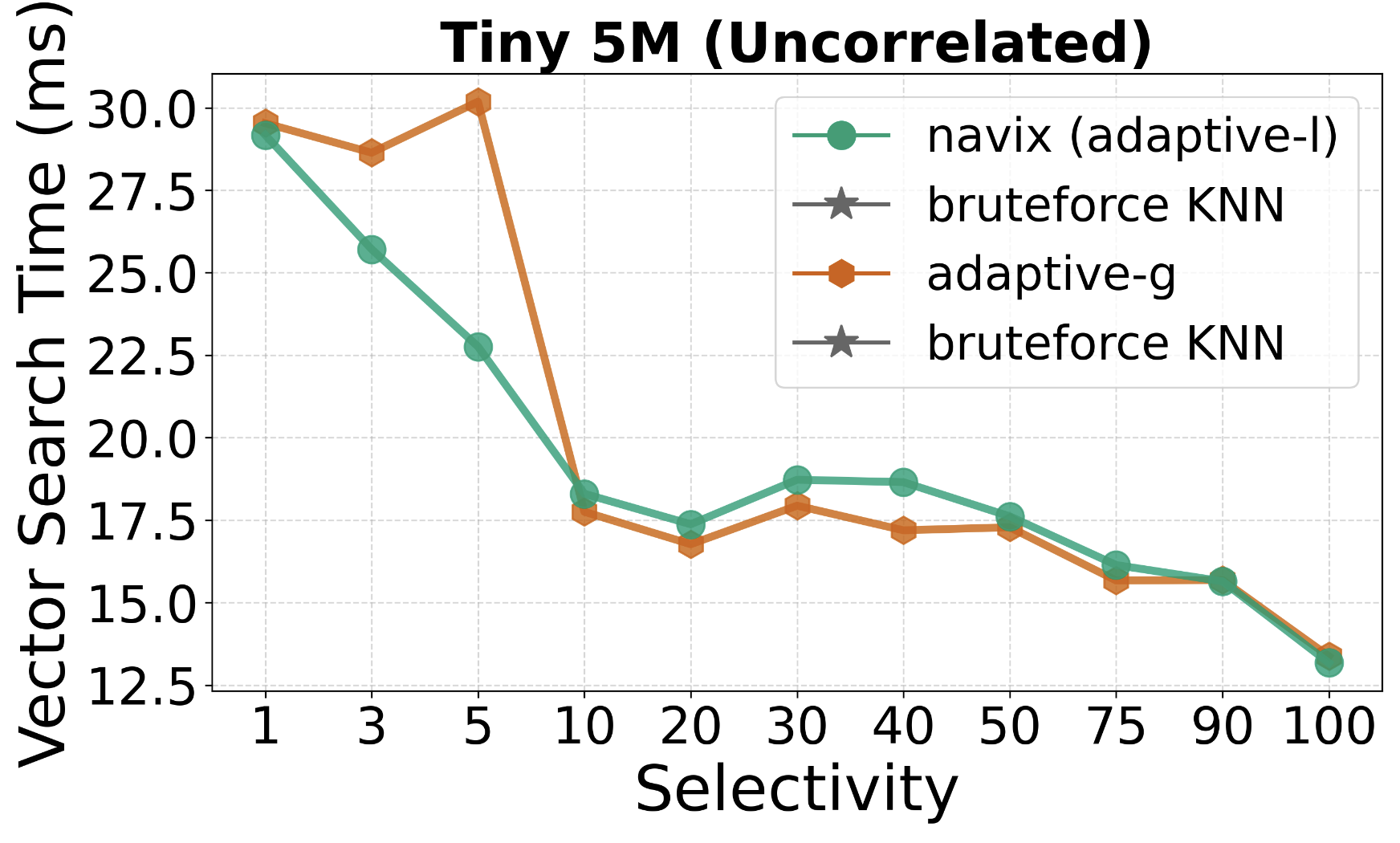}
  \end{subfigure}
  \begin{subfigure}[b]{0.33\textwidth}
    \centering
    \includegraphics[width=\textwidth]{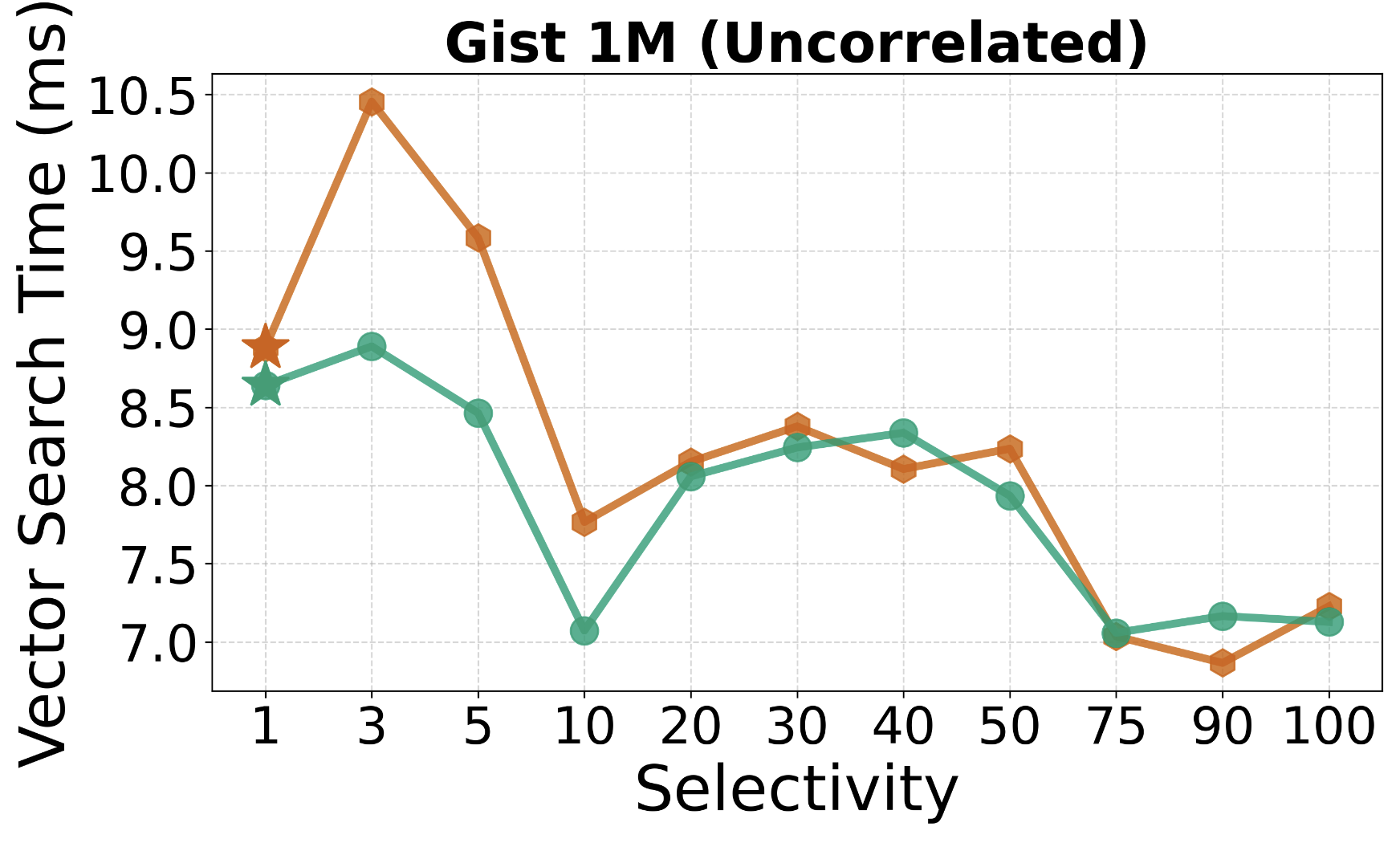}
  \end{subfigure}
   \begin{subfigure}[b]{0.33\textwidth}
    \centering
    \includegraphics[width=\textwidth]{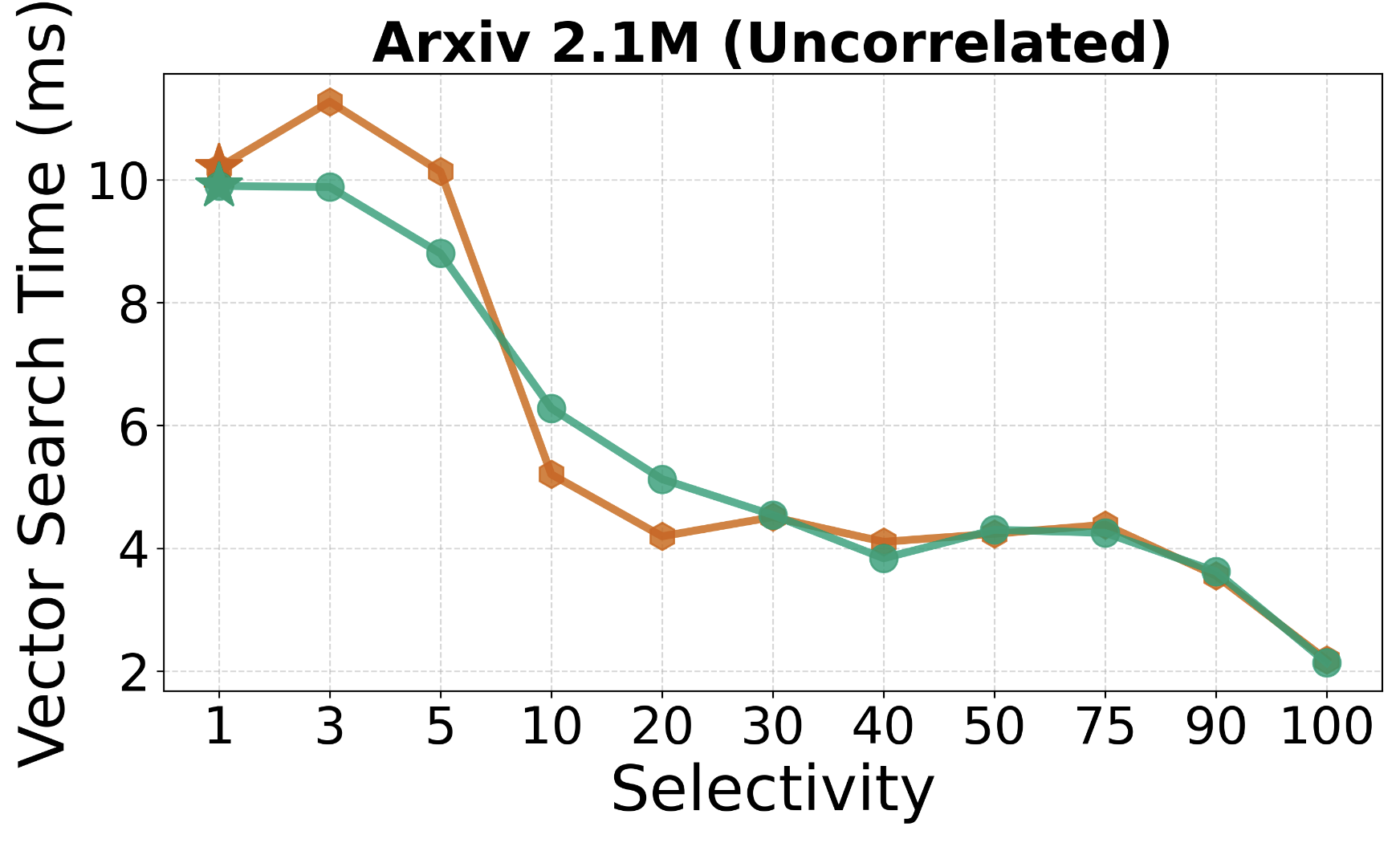}
  \end{subfigure}
  \caption{Vector search time vs selectivity for NaviX and adaptive-g within 95\% to 95.5\% recall}
  \label{fig:adaptive_g_navix_other_datasets}
\end{figure*}

\subsection{Evaluation of Adaptive-G and Navix For Other Uncorrelated Workloads:}
\label{app:adaptive_g_navix}
As mentioned in Section \ref{sec:adaptiveg-navix}, we present vector search time for \Indexname\ and \adaptiveg\ for other uncorrelated workloads i.e. Tiny, GIST, and Arxiv. Figure \ref{fig:adaptive_g_navix_other_datasets} demonstrates vector search time across different selectivities. \Indexname\ and \adaptiveg\ perform similarly to each other except at lower selectivities for Tiny and GIST, where \Indexname\ outperforms \adaptiveg\ by up to 30\%. This performance advantage occurs because \Indexname\ can make better decisions when switching between different heuristics based on the local selectivity of the index traversal region.

\begin{figure}[H]
\centering
  \includegraphics[keepaspectratio, height=4.5cm]{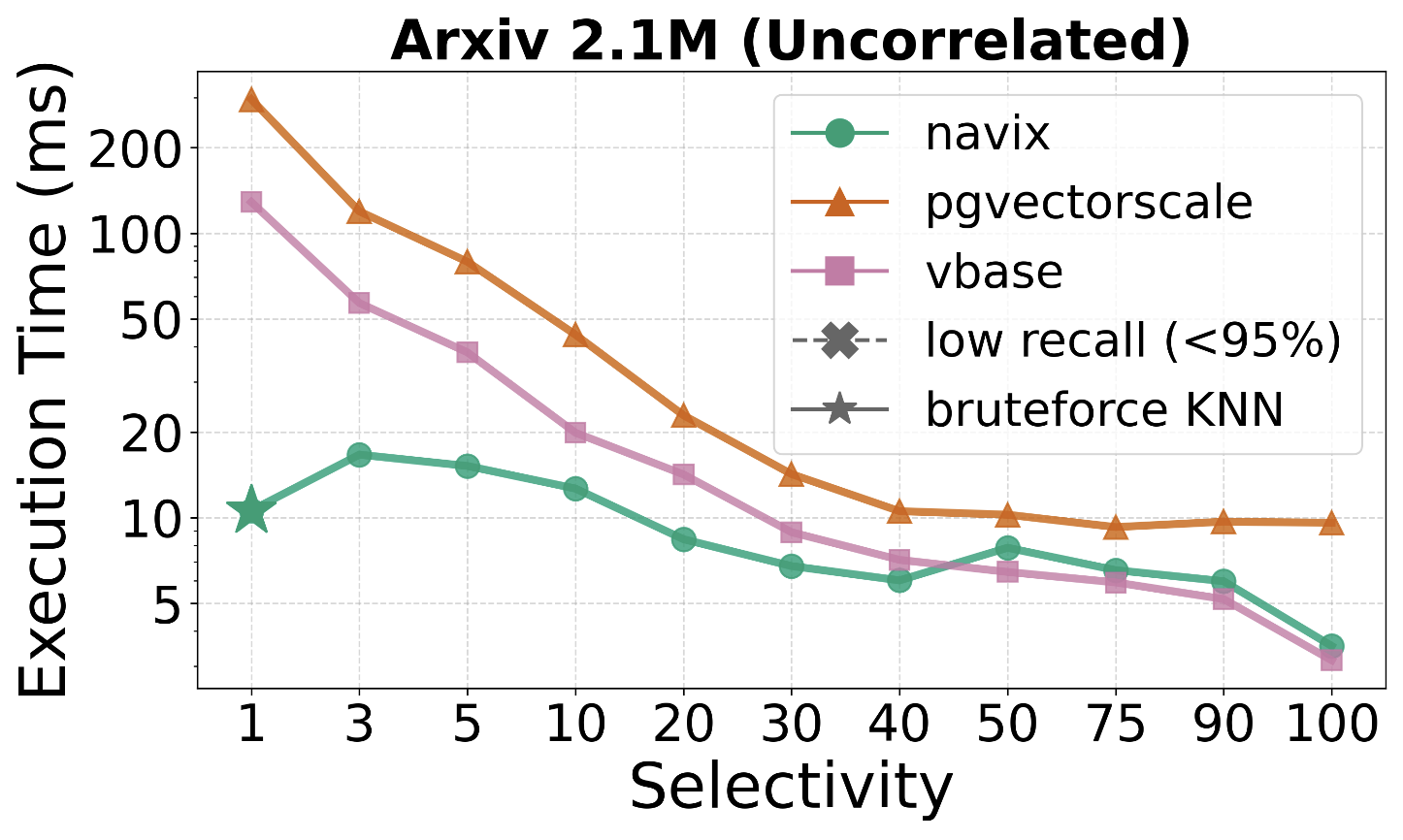}
  \caption{Execution Time vs Selectivity for postfiltering baselines with >95\% recall and within 1\% of each other}
  \label{fig:postfilterin-arxiv}
\end{figure}

\subsection{Evaluation of Postfiltering Baselines for Arxiv Uncorrelated Workload:}
\label{app:arxiv_post}

We extend the experiments in Section \ref{subsubsec:postfiltering_baselines} with the Arxiv uncorrelated workload in Figure \ref{fig:postfilterin-arxiv}. The results are similar to the Wiki uncorrelated workload in Section~\ref{subsubsec:postfiltering_baselines}: both postfiltering systems degrade significantly in mid to lower selectivities where \Indexname\ outperforms both of them. For higher selectivities, the difference is not as significant. However, \Indexname\ slightly outperforms PGVectorScale since the Arxiv dataset is smaller, thus the majority of the cost is due to vector search time rather than prefiltering time.

\begin{figure}[H]
\centering
\includegraphics[keepaspectratio, height=3.2cm]{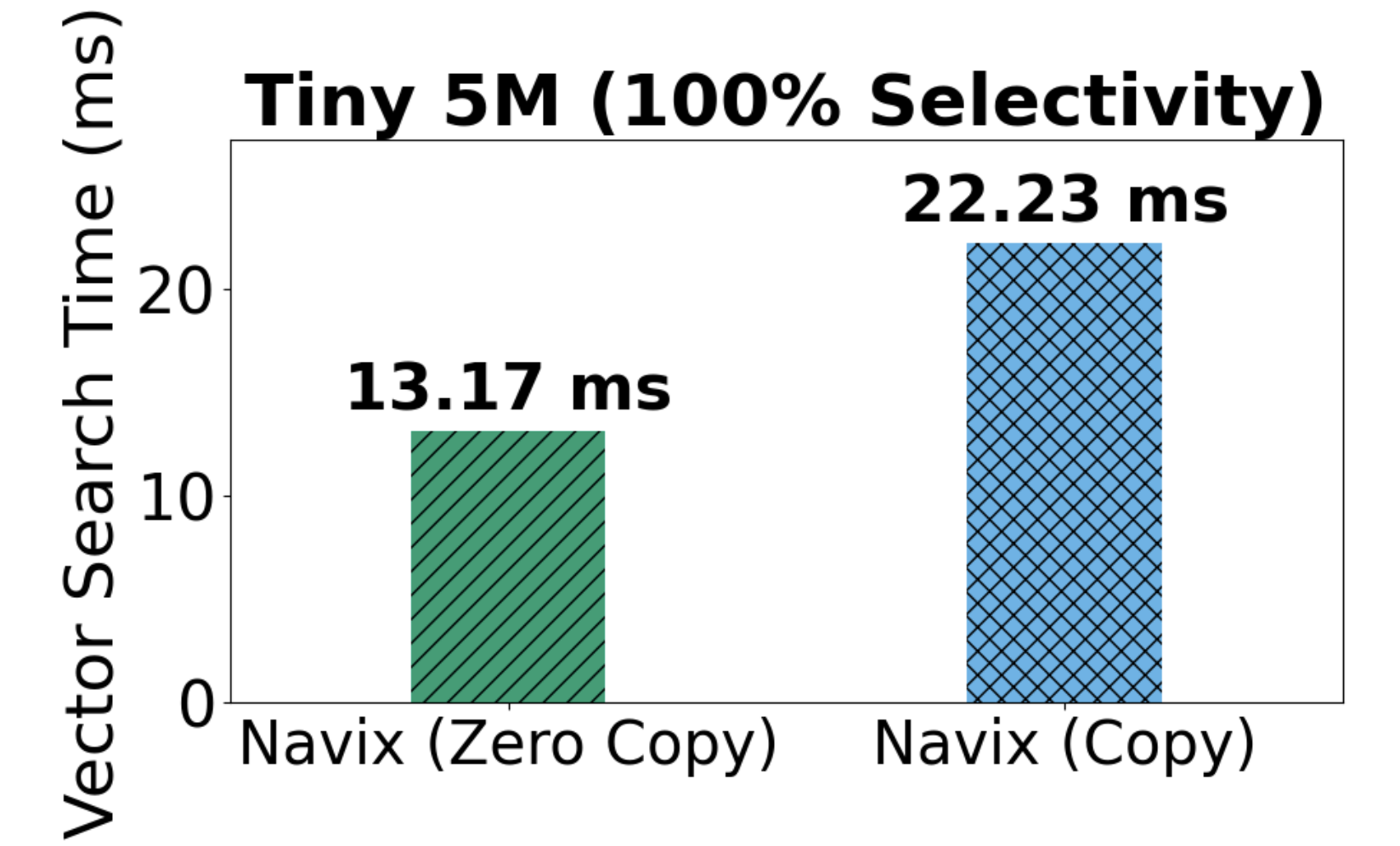}
  \caption{Vector Search Time with and without zero-copy optimization for 95\% recall}
  \label{fig:zero_copy_bench}
\end{figure}


\subsection{Evaluation of In-BM Distance Calculations:}
\label{app:in_bm_distance}

In this evaluation, we do an ablation study on how much performance benefits our in-bm distance calculation optimization provides. As shown in Figure \ref{fig:zero_copy_bench},
we used the Tiny dataset at 100\% selectivity.
We turned the in-bm distance calculation optimization
off in \Indexname\. We refer to this configuration as \Indexnamecopy.
Figure~\ref{fig:zero_copy_bench} shows our results.
We see that the in-bm optimization improves the performance of \Indexname\ by 1.6x, demonstrating its importance
for DBMSs that use a BM to scan their vectors into their vector search operators.




\begin{figure*}[t!]
  \centering
             \begin{subfigure}[b]{0.24\textwidth}
    \centering
    \includegraphics[width=\textwidth]{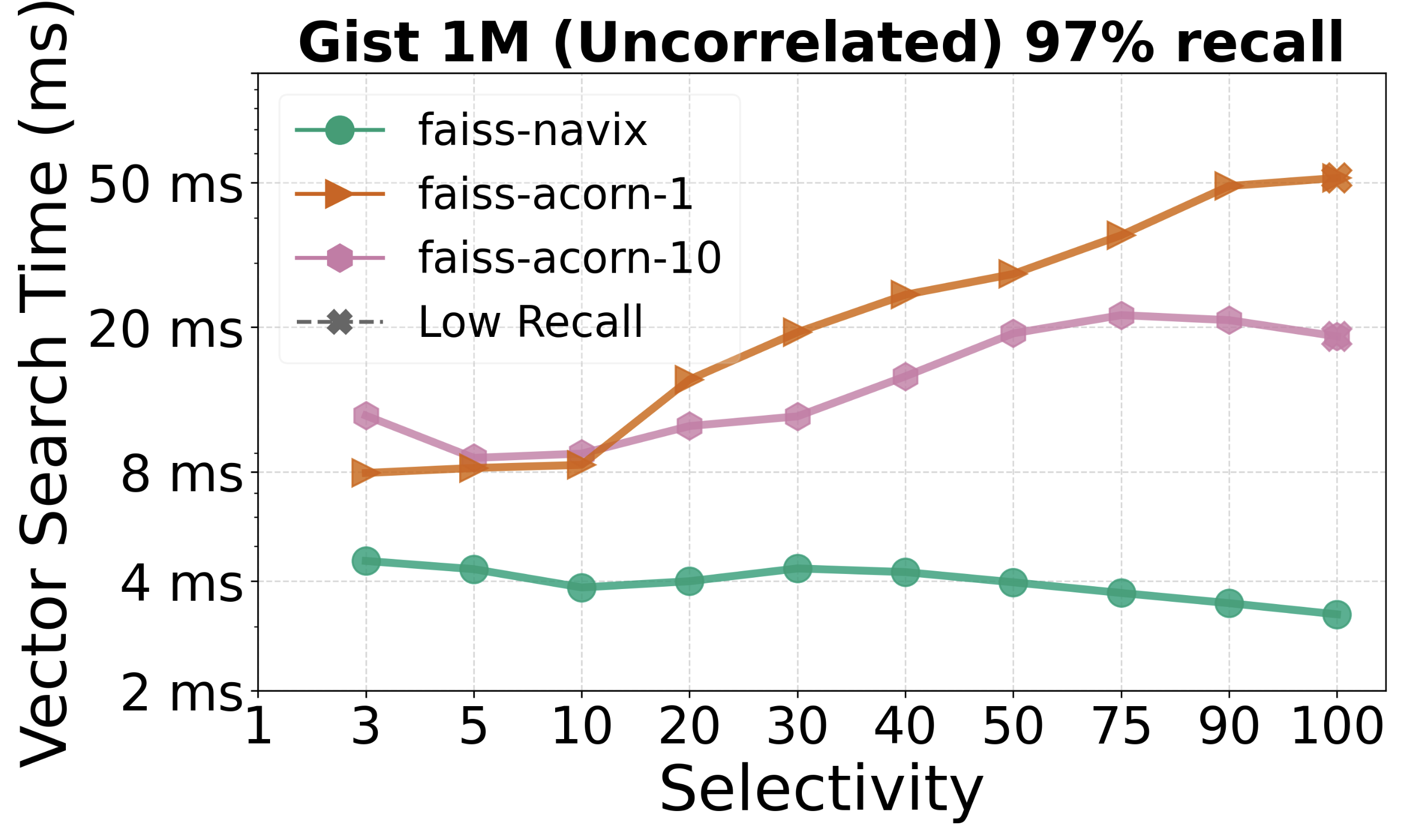}
      \end{subfigure}
           \begin{subfigure}[b]{0.24\textwidth}
    \centering
    \includegraphics[width=\textwidth]{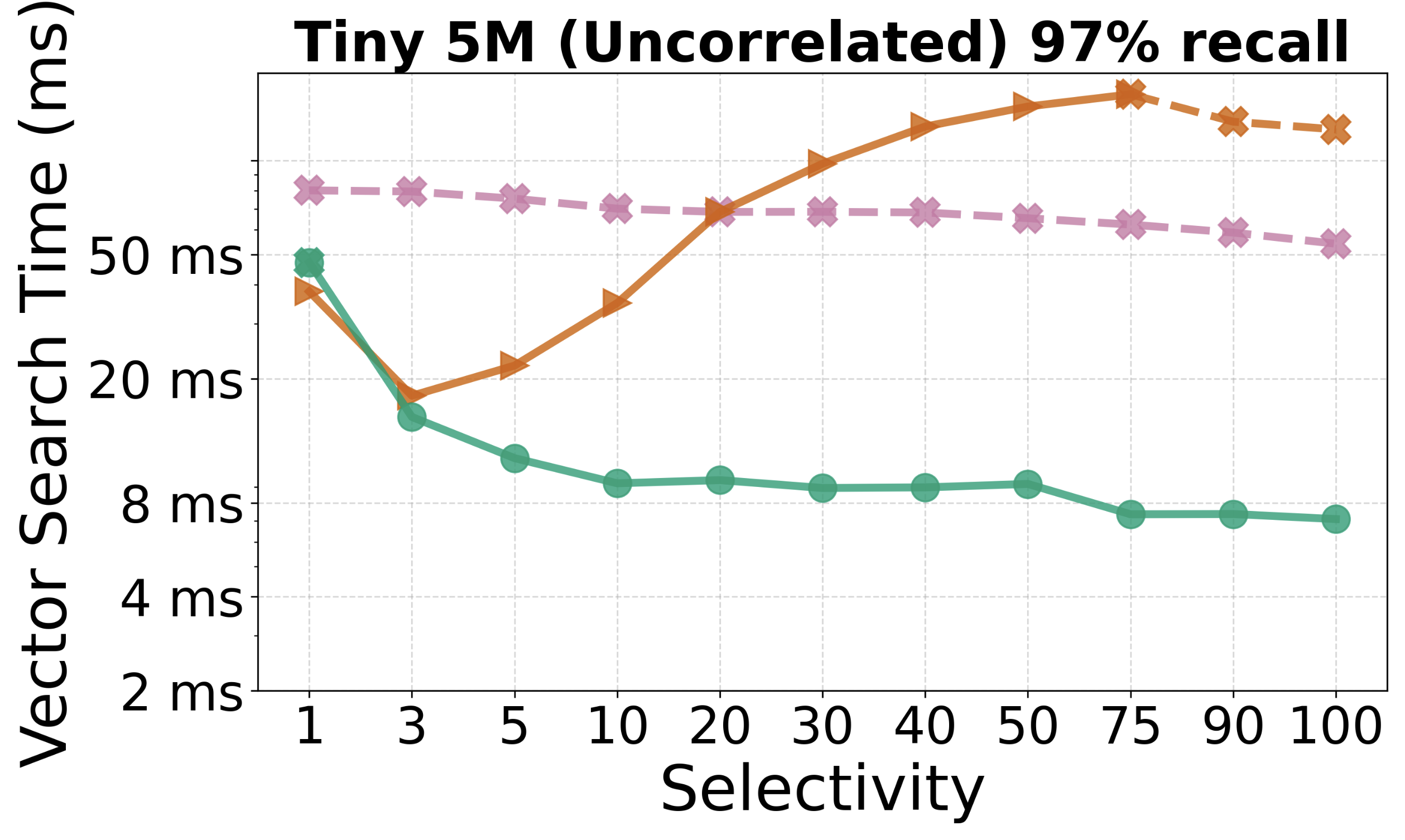}
      \end{subfigure}
   \begin{subfigure}[b]{0.24\textwidth}
    \centering
    \includegraphics[width=\textwidth]{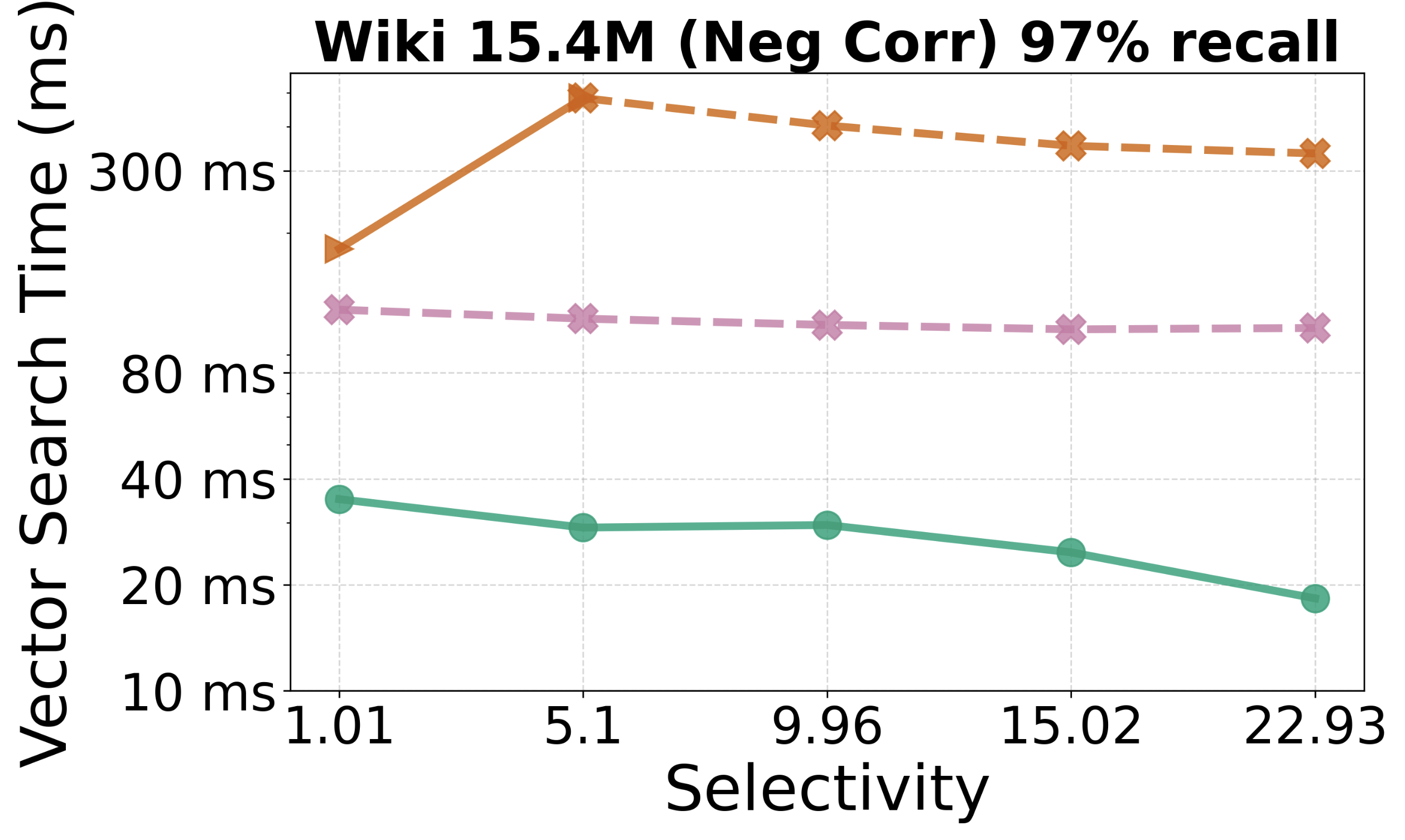}
  \end{subfigure}
     \begin{subfigure}[b]{0.24\textwidth}
    \centering
    \includegraphics[width=\textwidth]{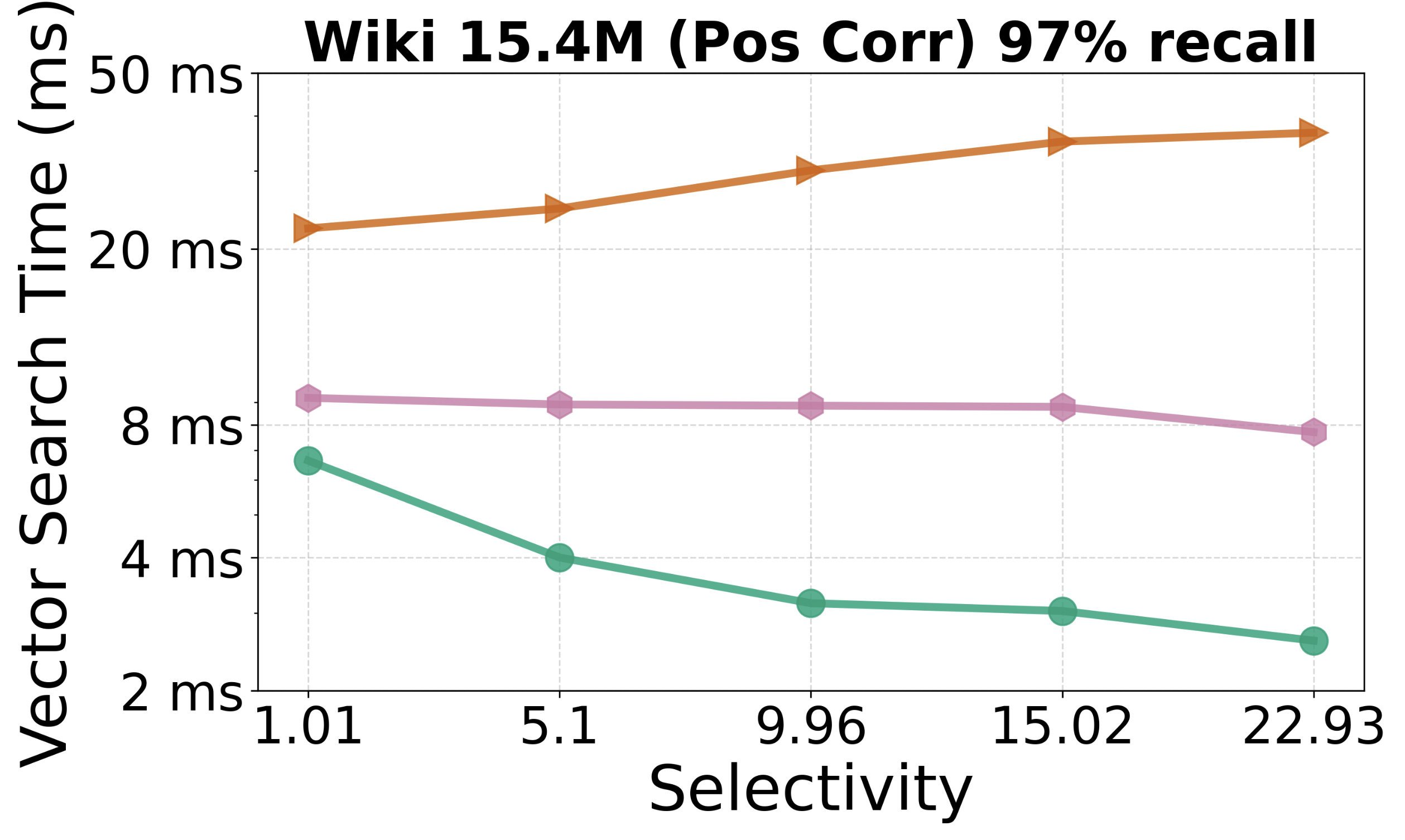}
      \end{subfigure}

    \begin{subfigure}[b]{0.23\textwidth}
    \centering
    \includegraphics[width=\textwidth]{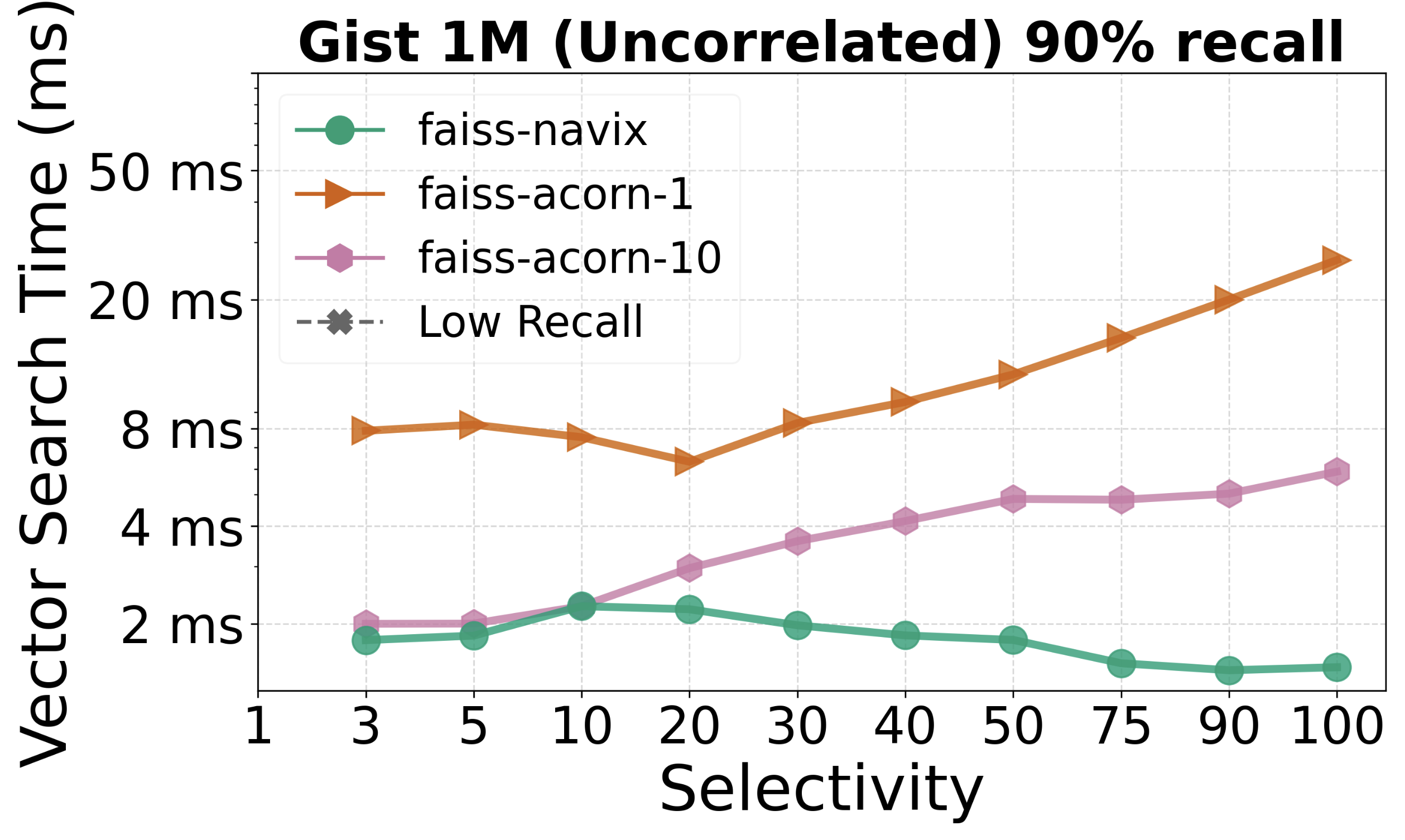}
  \end{subfigure}
  \begin{subfigure}[b]{0.23\textwidth}
    \centering
    \includegraphics[width=\textwidth]{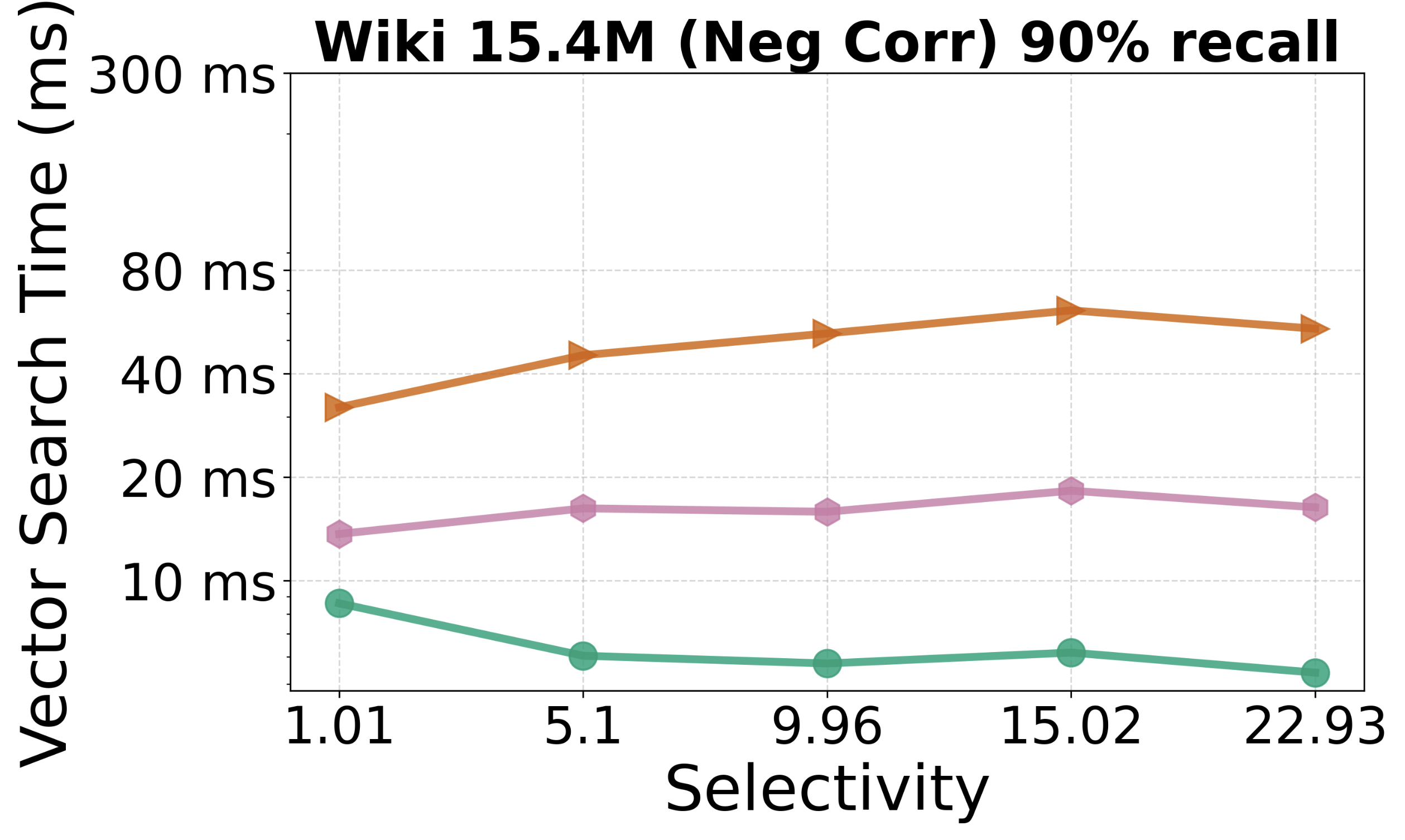}
  \end{subfigure}
    \begin{subfigure}[b]{0.23\textwidth}
    \centering
    \includegraphics[width=\textwidth]{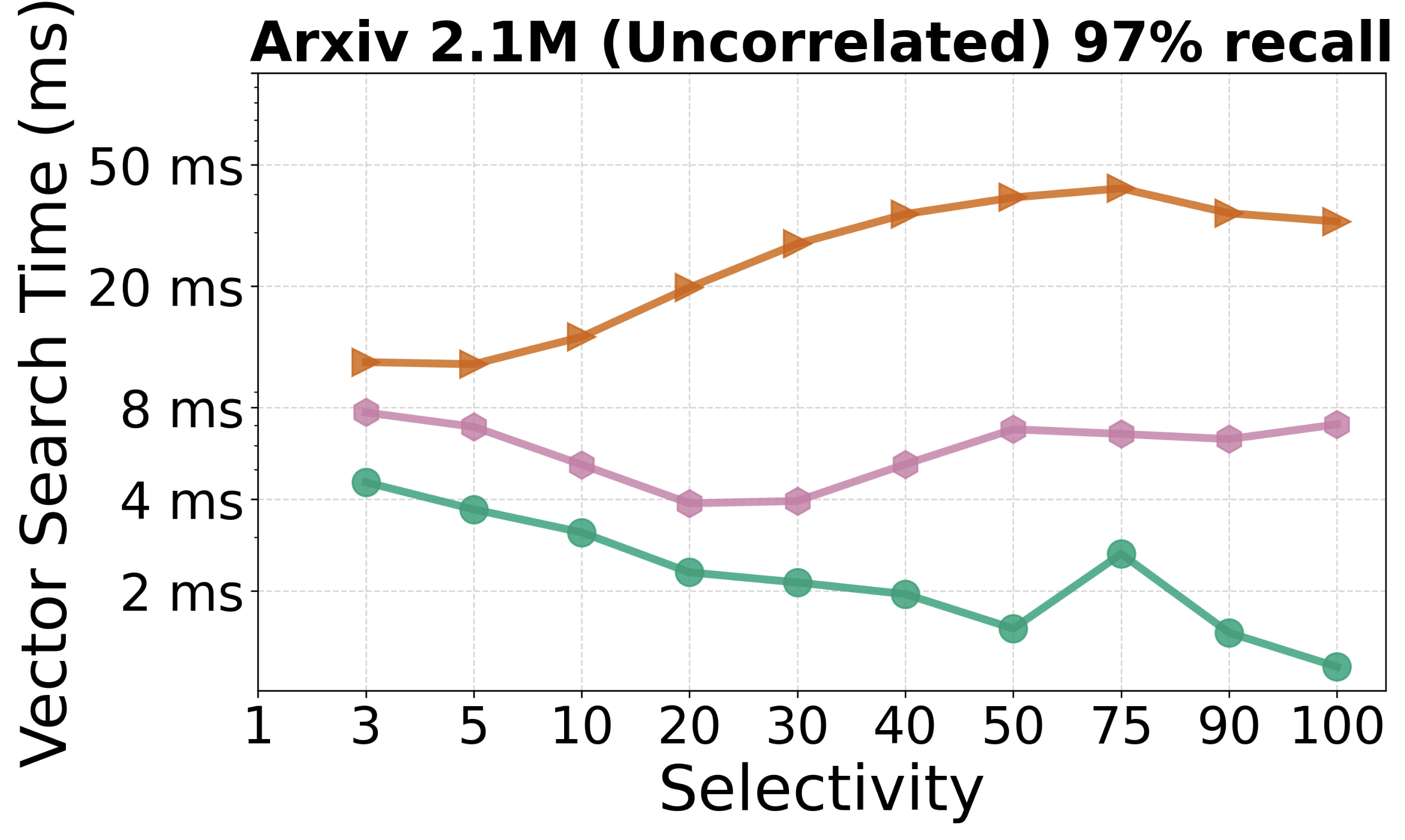}
  \end{subfigure}
  \caption{Vector Search Time vs Selectivity for ACORN and Faiss-Navix baselines at different recalls}  \label{fig:acorn_all_diff_recall}
\end{figure*}

\subsection{Comparison Against ACORN at Different Recalls}
\label{appsubsec:acorn_different_recall}
As mentioned in Section~\ref{subsec:acorn-ex}, Figure~\ref{fig:acorn_all_diff_recall} present remaining results comparing Faiss-\Indexname\ and Acorn at 90\% and 97\% recall levels on our workloads. The findings align with our previous results where Faiss-\Indexname\ consistently outperforms both Acorn configurations i.e. Acorn-1 and Acorn-10.

\begin{figure*}[t!]
  \centering
    \begin{subfigure}[b]{0.23\textwidth}
    \centering
    \includegraphics[width=\textwidth]{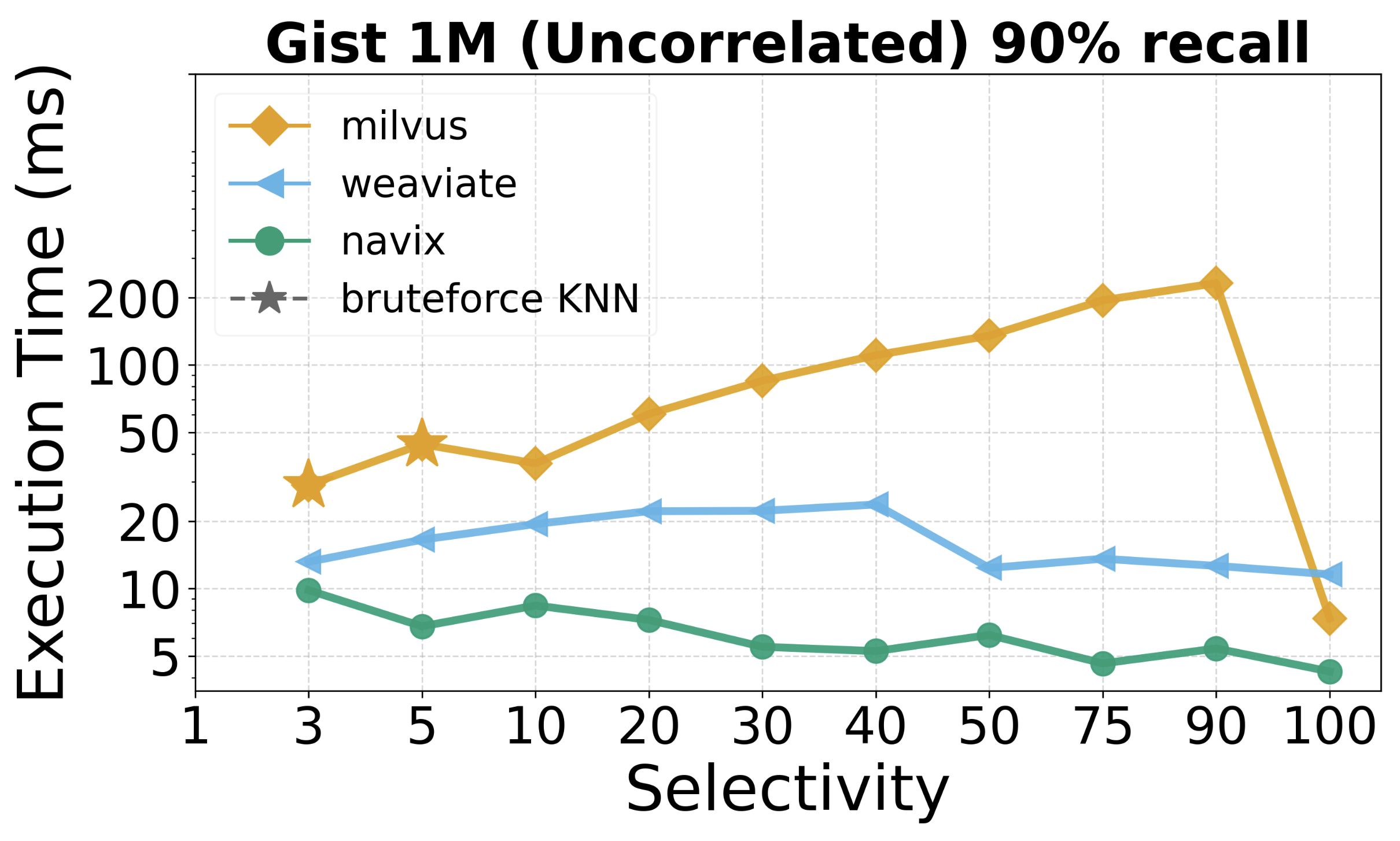}
  \end{subfigure}
  \begin{subfigure}[b]{0.23\textwidth}
    \centering
    \includegraphics[width=\textwidth]{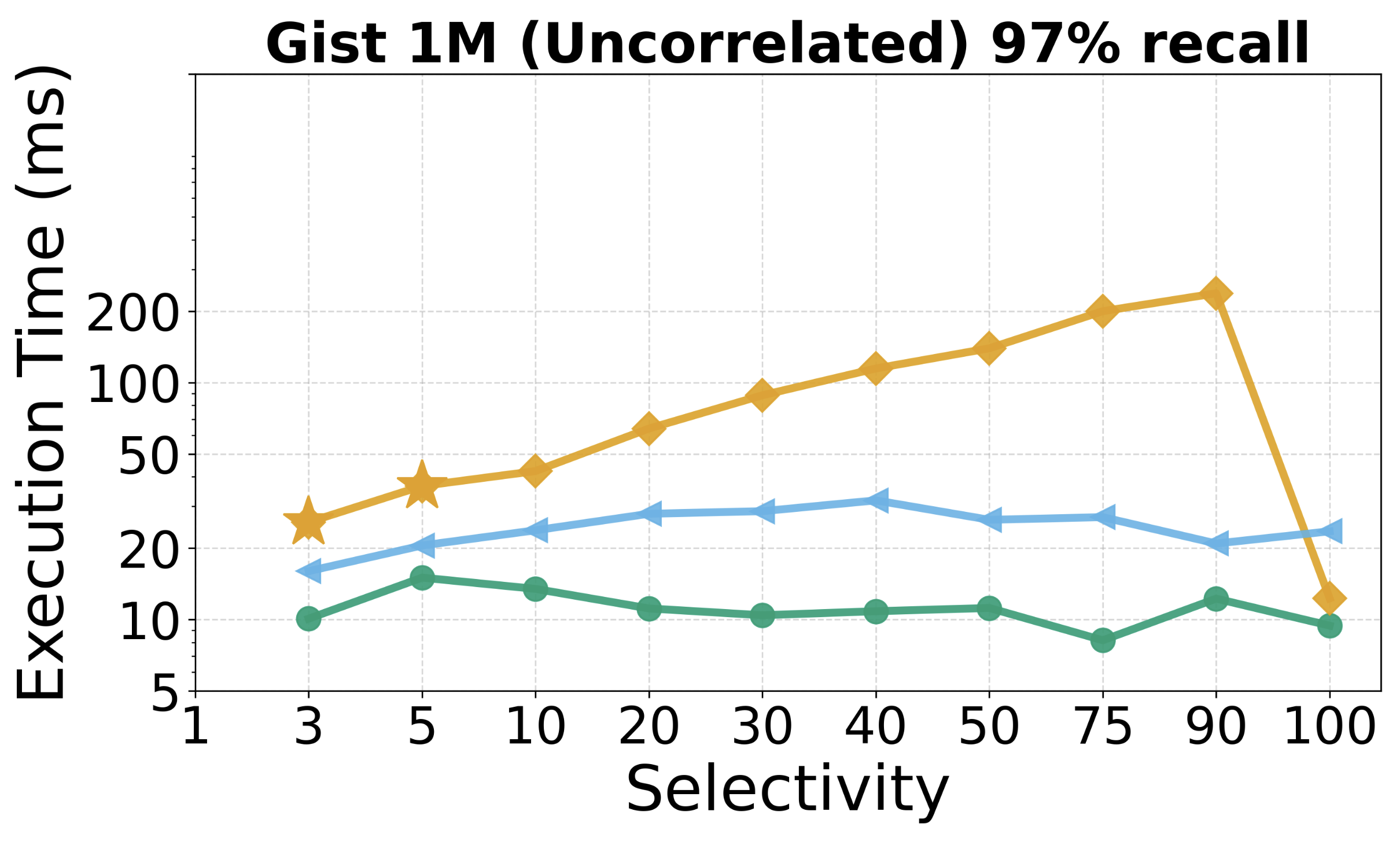}
  \end{subfigure}
   \begin{subfigure}[b]{0.24\textwidth}
    \centering
    \includegraphics[width=\textwidth]{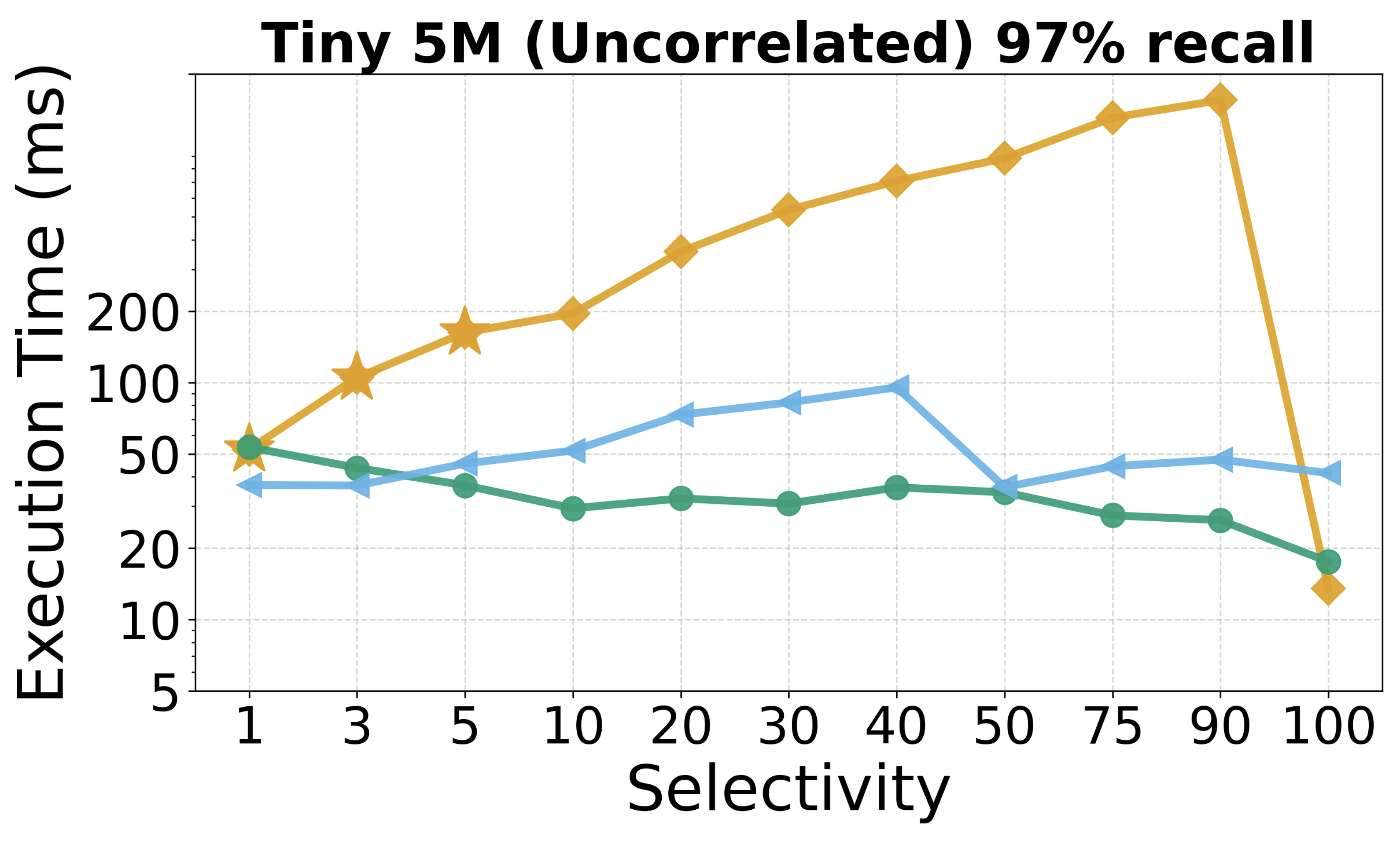}
  \end{subfigure}
     \begin{subfigure}[b]{0.24\textwidth}
    \centering
    \includegraphics[width=\textwidth]{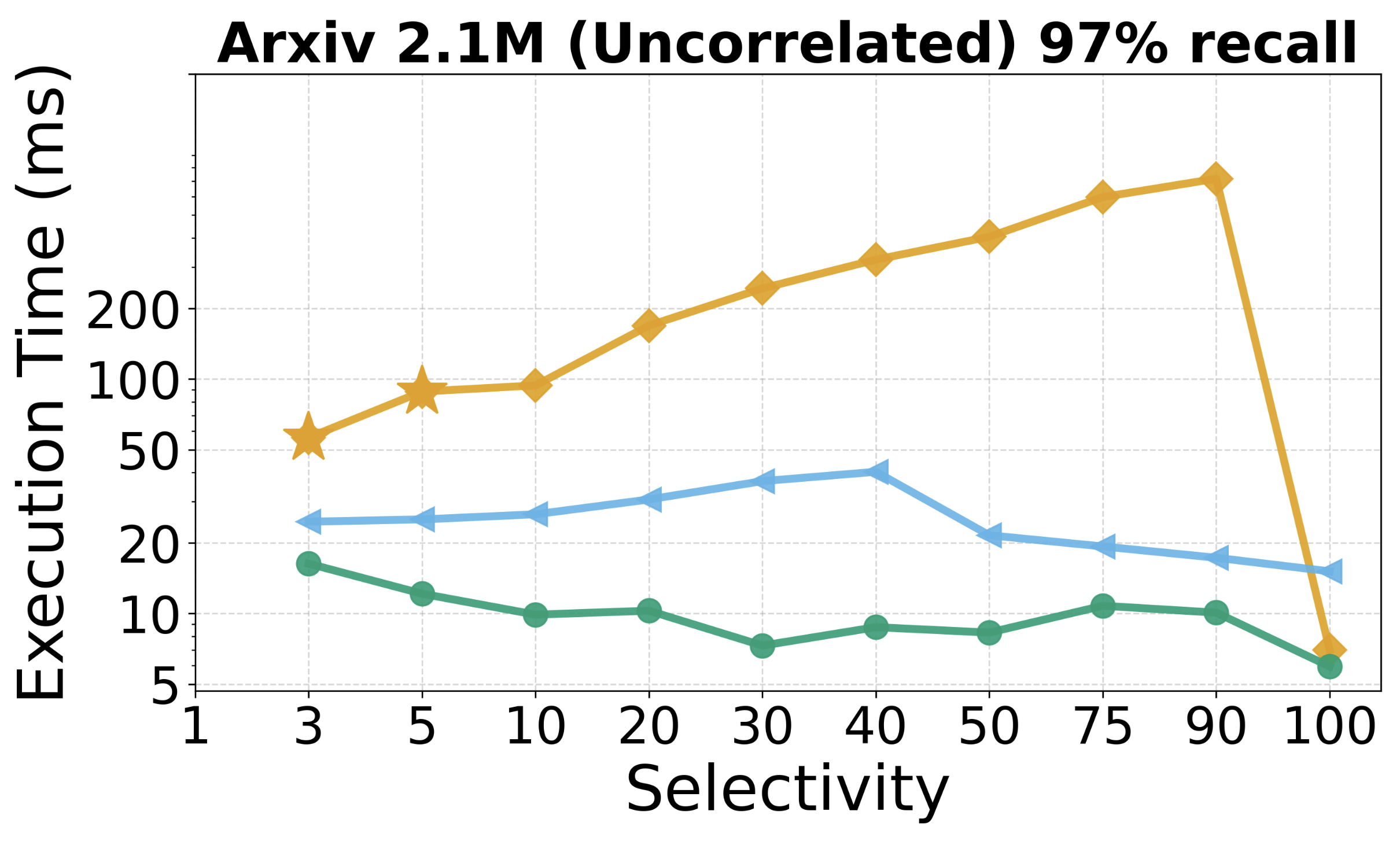}
      \end{subfigure}
  \caption{Execution Time vs Selectivity for Weaviate and Milvus baselines at different Recalls}  \label{fig:prefiltering_baseline_all_recall}
\end{figure*}

\subsection{Comparison Against Weaviate and Milvus at Different Recalls}
\label{appsubsec:weaviate_different_recall}

As mentioned in Section~\ref{subsec:weaviate_milvus}, Figure~\ref{fig:prefiltering_baseline_all_recall} present remaining results comparing \Indexname\ with Weaviate and Milvus at 90\% and 97\% recall levels on our workloads. The findings align with our previous results where \Indexname\ is competitive or outperforms both Weaviate and Milvus at different selectivities.

\fi

\end{document}